\documentclass[11pt,a4paper]{article}
\usepackage[margin=22mm]{geometry}
\usepackage{fontspec}
\usepackage{amsmath,amssymb,graphicx,longtable,booktabs,array,caption,fancyhdr}
\usepackage[unicode,colorlinks=true,linkcolor=blue,urlcolor=blue]{hyperref}
\newlength{\tableunit}
\title{Tracing dependency networks explains how origins shape downstream vulnerability}
\author{Ruiwu Niu, Muhammad Azhar, Michaël Antonie van Wyk, Xiaoqun Wu}
\date{}
\usepackage{placeins}
\hypersetup{pdftitle={Tracing dependency networks explains how origins shape downstream vulnerability},pdfauthor={Ruiwu Niu, Muhammad Azhar, Michaël Antonie van Wyk, Xiaoqun Wu}}
\begin{document}
\begin{center}
{\LARGE Tracing dependency networks explains how origins shape downstream vulnerability\par}\medskip
{\large \mbox{Ruiwu Niu\textsuperscript{1}}, \mbox{Muhammad Azhar\textsuperscript{1}}, \mbox{Michaël Antonie van Wyk\textsuperscript{2}}, \mbox{Xiaoqun Wu\textsuperscript{3,*}}\par}\medskip
\begin{minipage}{0.98\linewidth}\centering\fontsize{9.5}{11.5}\selectfont
\textsuperscript{1}Department of Data Science and Digital Innovation, Hong Kong Shue Yan University, Hong Kong SAR, China\par\smallskip
\textsuperscript{2}School of Electrical and Information Engineering, University of the Witwatersrand, Johannesburg, South Africa\par\smallskip
\textsuperscript{3}College of Computer Science and Software Engineering, Shenzhen University, Shenzhen, China\par\smallskip
\medskip\textsuperscript{*}Correspondence to Xiaoqun Wu (\href{mailto:xqwu@szu.edu.cn}{xqwu@szu.edu.cn}).\par\smallskip
Ruiwu Niu (\href{mailto:rniu@hksyu.edu}{rniu@hksyu.edu}); Muhammad Azhar (\href{mailto:azhar@hksyu.edu}{azhar@hksyu.edu})\par
Michaël Antonie van Wyk (\href{mailto:Anton.VanWyk@wits.ac.za}{Anton.VanWyk@wits.ac.za})\par
\end{minipage}\par
\end{center}\medskip

\FloatBarrier\section*{Abstract}
\addcontentsline{toc}{section}{Abstract}

Many systems sustain outcomes through chains of prerequisites, yet these dependencies are often partially observed. We trace backward from sampled outcomes to recover dependency structure and measure edge importance by the fraction of outcomes lost after an edge fails. Across 1,620 synthetic networks, revealing 90\% of origins required tracing 7.24\% of outcomes on average. We define pair thickness as the number of origin and outcome pairs connected through an edge. Under strict dependency, where every input is necessary, thickness counts each affected outcome once per upstream origin, whereas loss counts it once. Edges therefore form origin groups with distinct thickness and loss relations, also seen in software, citation and river topologies. Within a network, greater origin diversity corresponds to lower strict loss at equal complete thickness. We develop a growth model that predicts how edges populate these groups and how losses are distributed under partial observation, tested on 2,560 independent networks. Separating origin counts from downstream reach improves identification of edges causing the greatest losses among unsampled outcomes in strict synthetic tests. Substitutable inputs reduce losses and change which importance scores perform best. Edge importance thus measures a dependency’s contribution to sustaining later outcomes, guiding the protection of critical connections.

\FloatBarrier\section*{Introduction}
\addcontentsline{toc}{section}{Introduction}

Many systems work only while their prerequisites hold. Biochemical reactions need substrates before further compounds can form \textsuperscript{\hyperlink{ref:1}{1}}, regulators shape gene expression \textsuperscript{\hyperlink{ref:2}{2}}, essential input shortages spread along supply chains \textsuperscript{\hyperlink{ref:3}{3}}, and software packages declare the dependencies they require \textsuperscript{\hyperlink{ref:4}{4},\hyperlink{ref:5}{5}}. Behavioral dependencies between urban places also predict how economic disruption spreads \textsuperscript{\hyperlink{ref:6}{6}}. A local dependency can determine whether downstream outcomes are realized or maintained.

Reconstructed COVID-19 trees connect earlier infections to later cases \textsuperscript{\hyperlink{ref:7}{7}}, motivating transmission tracing and control \textsuperscript{\hyperlink{ref:8}{8}}. River tributaries combine upstream contributions, and downstream environmental DNA helps reconstruct species distributions \textsuperscript{\hyperlink{ref:9}{9}}. Regulatory perturbations affect expression outputs in context dependent ways \textsuperscript{\hyperlink{ref:2}{2}}. These systems share an origin to outcome structure but require different failure rules.

Network function depends on both topology and the interactions represented by its links, as illustrated by synchronization in multiplex networks \textsuperscript{\hyperlink{ref:10}{10}}. Here, we treat dependency networks as abstractions of how outcomes are generated or sustained by prerequisites. Edges point from prerequisites to the nodes they support. Roots are externally sustained origins, intermediate nodes relay support, and designated terminals are the outcomes assessed. Any finite directed acyclic graph can be equipped with input requirement rules to define such a model. Its acyclic structure admits a formation order, while edge meanings and input rules supply the causal interpretation. Under strict dependency every original input is required; under tolerant dependency remaining inputs can sustain function. Dependency links in cascading failure models likewise encode support requirements \textsuperscript{\hyperlink{ref:11}{11},\hyperlink{ref:12}{12}}. Threshold cascades, interdependent failures, multiple suppliers and operational resilience show how these requirements reshape robustness \textsuperscript{\hyperlink{ref:13}{13},\hyperlink{ref:14}{14},\hyperlink{ref:15}{15},\hyperlink{ref:16}{16}}.

We measure downstream vulnerability by terminal loss, the fraction of designated outcomes that fail when one edge is removed under the specified rule. An early prerequisite can thus be important despite having few immediate dependents. This interpretation connects edge ranking to the preservation of later outcomes and the assessment of interventions. A completed network permits retrospective assessment; prediction beyond its observation period requires a formation model and time horizon. Several origins may reach the same outcomes through an edge. Counting origin and outcome pairs therefore differs from counting affected outcomes, and substitutable inputs can prevent reachable outcomes from failing.

Dependencies are rarely observed in full. Tracing sampled outcomes can reveal their origins while missing other outcomes that share those dependencies. Sampling distorts apparent topology \textsuperscript{\hyperlink{ref:17}{17},\hyperlink{ref:18}{18}}, and reconstruction methods infer missing links from measurement models \textsuperscript{\hyperlink{ref:19}{19},\hyperlink{ref:20}{20},\hyperlink{ref:21}{21}}. We query sampled outcomes' prerequisites and ask what they reveal beyond the sample.

Influence maximization and reverse reachability estimate downstream reach \textsuperscript{\hyperlink{ref:22}{22},\hyperlink{ref:23}{23},\hyperlink{ref:24}{24}}. Citation analysis counts pairs and paths \textsuperscript{\hyperlink{ref:25}{25},\hyperlink{ref:26}{26}}, including path centrality and intermediacy \textsuperscript{\hyperlink{ref:27}{27},\hyperlink{ref:28}{28}}, while betweenness counts shortest paths \textsuperscript{\hyperlink{ref:29}{29}}. Software impact sets and vulnerability measures track necessary dependencies and defect exposure \textsuperscript{\hyperlink{ref:30}{30},\hyperlink{ref:31}{31}}. Infrastructure studies \textsuperscript{\hyperlink{ref:32}{32},\hyperlink{ref:33}{33}}, supply network reconstruction \textsuperscript{\hyperlink{ref:34}{34},\hyperlink{ref:35}{35}} and country specific exposures \textsuperscript{\hyperlink{ref:36}{36}} also relate structure to losses. We ask how edge failure consequences vary with origin multiplicity when only some outcomes are observed.

We address this question by separating the origins that reach an edge from the outcomes that depend on it. Our tracing experiments show that many origins can be recovered while edge rankings remain uncertain. Under strict dependency and complete observation, pair thickness counts each affected outcome once per upstream origin, while terminal loss counts it once. This distinction explains why edges follow different thickness and loss relationships according to origin count. We then examine how sampling and input substitution alter these relationships and their value for edge prioritization. A growth model predicts root group composition and loss distributions under partial observation.

\FloatBarrier\section*{Results}
\addcontentsline{toc}{section}{Results}

\subsection*{Backward tracing, pair thickness and terminal loss}

\begin{figure}[!htbp]\centering\begin{minipage}{\linewidth}\phantomsection\label{fig:1}\hypertarget{figure-1}{}\addcontentsline{toc}{subsection}{Figure 1. Backward tracing, pair thickness and terminal loss on a constructed example and the complete Bioconductor network}
{\centering
\includegraphics[width=\linewidth,height=0.63\textheight,keepaspectratio]{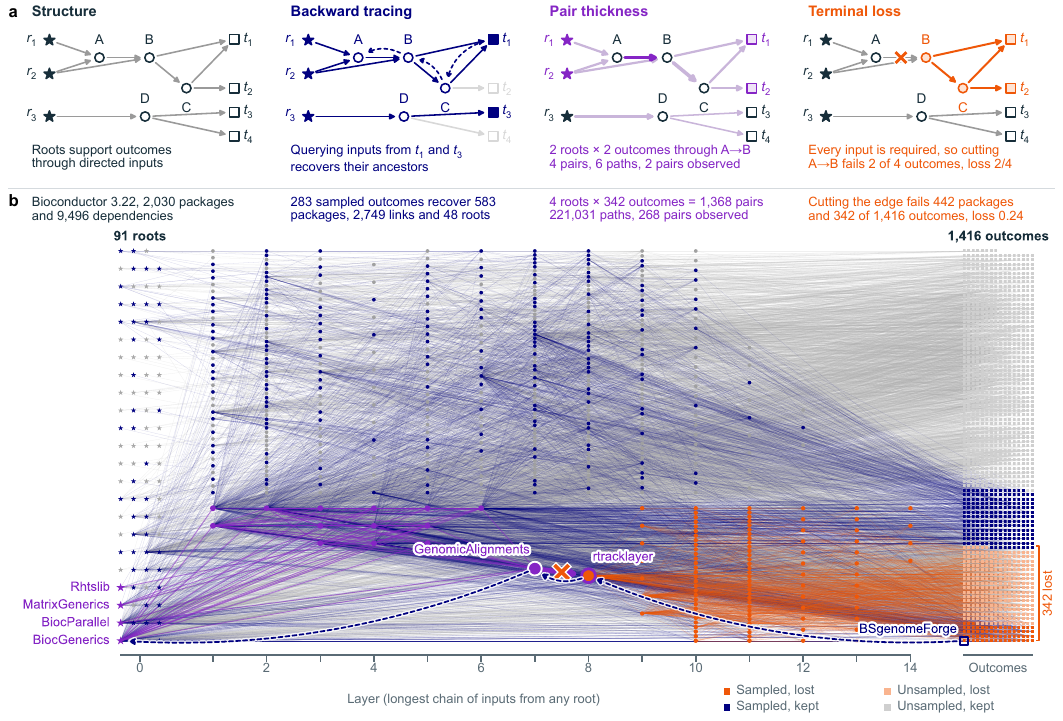}\par}\smallskip
{\fontsize{9}{11}\selectfont \textbf{Figure 1. Backward tracing, pair thickness and terminal loss on a constructed example and the complete Bioconductor network.} \textbf{a}, A constructed network introduces the four concepts. Stars mark roots, circles intermediate nodes and squares designated outcomes, and edges point from input to dependent. Querying inputs from \(t_1\) and \(t_3\) recovers their ancestors (blue), with dashed arrows showing reverse queries. Edge A to B joins two roots to two outcomes through four pairs and six paths, and the sample observes two of these pairs (purple). Under the strict rule, cutting it fails B, C, \(t_1\) and \(t_2\), a terminal loss of 2/4 (orange). \textbf{b}, The same steps on all 2,030 packages and 9,496 dependencies of Bioconductor 3.22. Roots occupy the first column. Intermediate packages are arranged by the length of the longest input chain from any root. The 1,416 designated outcomes fill the final column, with one cell per package. The 283 sampled outcomes of the primary observation recover 583 packages, 2,749 links and 48 of the 91 roots (blue). The dashed route follows one chain of reverse queries from the sampled outcome BSgenomeForge through the focal edge to the root BiocGenerics. The focal edge from GenomicAlignments to rtracklayer connects four roots to 342 outcomes, giving 1,368 pairs and 221,031 paths, and 268 pairs are observed because 67 of these outcomes are sampled. Cutting it fails 442 packages, including the 342 outcomes, a terminal loss of 0.24 (orange). Outcome cells show loss and sampling status. Away from the focal edge, blue and gray separate recovered from unrecovered packages and links. Vertical positions within a layer only reduce crossings. Supplementary Note 1.5 gives the fixed rule that selects the focal edge and the route.\par}
\end{minipage}\end{figure}

Let \(G=(V_G,E_G)\) be a finite, simple, unweighted directed acyclic graph (DAG) with node set \(V_G\) and edge set \(E_G\), all nodes initially active. An edge \(e=(u\to v)\in E_G\) supplies one input from \(u\) to \(v\). Roots \(\mathcal R_G\subseteq V_G\) have no incoming edges and are sustained externally. The nonempty set \(\mathcal T_G\subseteq V_G\) contains the designated outcomes, or terminals, which may also support other nodes. \hyperref[fig:1]{Figure~1a} illustrates these roles on a small constructed network, and \hyperref[fig:1]{Figure~1b} shows them on all 2,030 packages and 9,496 dependencies of the Bioconductor reference used below.
\par\par
To observe the network, we select a nonempty set of terminals \(\mathcal T_S\subseteq\mathcal T_G\) and query the direct predecessors of every discovered node until none remain unqueried. The result, \(S=(V_S,E_S)\), is the ancestor closure of the sample, containing every node and edge that leads to it. Tracing back from the 283 sampled terminals in \hyperref[fig:1]{Figure~1b} recovers the blue subgraph of 583 packages and 2,749 links, with dashed arrows marking reverse queries. This is the deterministic backward search used in reverse influence sampling (RIS) \textsuperscript{\hyperlink{ref:23}{23},\hyperlink{ref:24}{24}}. We keep the recovered structure to study individual edge failures.

For either graph \(H\in\{G,S\}\), let \(R_H=|\mathcal R_H|\ge1\) and \(T_H=|\mathcal T_H|\ge1\) count roots and terminals, and write \(x\leadsto y\) when a directed path, possibly of length zero, connects two nodes. For edge \(e=(u\to v)\), define

\[
\begin{aligned}
n_H(e)&=|\{r\in\mathcal R_H:r\leadsto u\}|,\qquad
U_H(e)=|\{t\in\mathcal T_H:v\leadsto t\}|,\\
C_H(e)&=n_H(e)U_H(e),\qquad
\bar C_H(e)=\frac{C_H(e)}{R_HT_H}.
\end{aligned}
\tag{1}
\]

The integer \(n_H\in\{1,\ldots,R_H\}\) counts upstream roots and \(U_H\in\{0,\ldots,T_H\}\) counts reachable terminals. These measure upstream and downstream reach restricted to roots and designated terminals. Their product, which we call pair thickness, is the number of distinct root and terminal pairs connected through the edge, so \(0\le C_H\le R_HT_H\) and \(0\le\bar C_H\le1\). A pair joined by several paths counts once. The purple edge in \hyperref[fig:1]{Figure~1b} links four roots to 342 terminals, giving 1,368 distinct pairs although 221,031 paths traverse it. Only 67 of these terminals are sampled, so the raw observed thickness is 268.

Conventional node pair projection count (NPPC) allows every node as an endpoint \textsuperscript{\hyperlink{ref:25}{25},\hyperlink{ref:26}{26}}. Restricting endpoints to roots and designated outcomes preserves this counting principle while distinguishing origins from outcomes. Inserting an intermediate step into an edge leaves its pair thickness and the loss of original terminals unchanged. Supplementary Note 5.1 proves this consistency under process refinement. Classical pair and path scores also multiply upstream and downstream counts (Supplementary Note 5.5).

For a nonroot target, let \(d_v\ge1\) be its original number of inputs and \(m_v\in\{0,\ldots,d_v\}\) the number made unavailable by a cut or an upstream failure. A real tolerance \(\tau\in[0,1]\) sets the failure rule

\[
v\text{ fails}\quad\Longleftrightarrow\quad
m_v\ge\max\{1,\lceil\tau d_v\rceil\},\qquad
I_e(\tau)=\frac{|\mathcal F_e(\tau)\cap\mathcal T_G|}{T_G}.
\tag{2}
\]

Here \(\lceil\cdot\rceil\) rounds up, \(\mathcal F_e(\tau)\subseteq V_G\) is the final set of failed nodes and \(I_e(\tau)\in[0,1]\) is the terminal loss. Every cut starts from the intact graph. At zero tolerance any missing input causes failure, and at full tolerance one surviving input suffices. Cutting the purple edge in \hyperref[fig:1]{Figure~1b} fails 442 packages, including 342 of the 1,416 terminals, a terminal loss of 0.24. Here failure means that a package can no longer satisfy its required inputs (Methods).

Complete backward tracing sees the two sides of an edge differently. It finds every upstream root, so \(n_S(e)=n_G(e)=n(e)\), but it can miss downstream terminals. We write \(m=T_S\in\{1,\ldots,T_G\}\) for the sample size and \(k_e=U_S(e)\in\{0,\ldots,m\}\) for the sampled coverage, and call an edge discovered when \(k_e>0\). Tracing to a fixed depth behaves differently, because apparent roots may still have unqueried predecessors.

\subsection*{Backward tracing reveals origins before reliable edge rankings}

The different information recovered upstream and downstream also affects how much sampling is needed. Tracing reveals origins before it reliably recovers thickness rankings. In a planted experiment with 1,620 independent networks across 27 conditions, each with 100 roots, revealing 90 roots required observing 7.24\% of terminals on average. A stable thickness ranking over all reference edges required 87.7\%, with rank correlation staying at 0.90 or above at that and every larger tested sample size. Discovering origins and estimating their downstream reach require different levels of observation. \hyperlink{figure-S1}{Supplementary Fig. 1} compares observed and complete normalized pair thickness across six network families.

\begin{figure}[!htbp]\centering\begin{minipage}{\linewidth}\phantomsection\label{fig:2}\hypertarget{figure-2}{}\addcontentsline{toc}{subsection}{Figure 2. Recovery of pair thickness rankings in synthetic networks}
{\centering
\includegraphics[width=\linewidth,height=0.63\textheight,keepaspectratio]{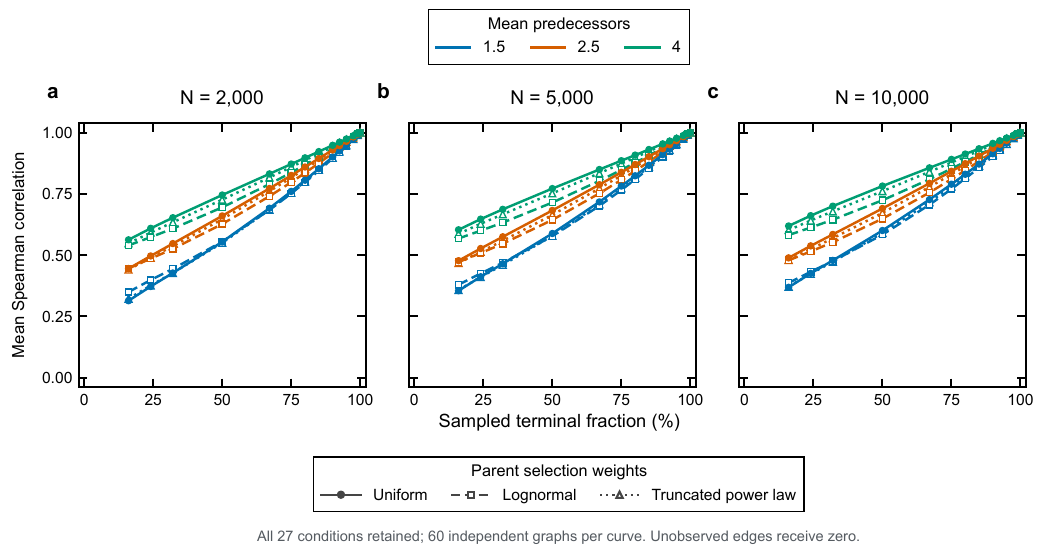}\par}\smallskip
{\fontsize{9}{11}\selectfont \textbf{Figure 2. Recovery of the complete reference ranking as more terminals are sampled, across 27 synthetic conditions.} Panels a, b and c show networks with \(N=2{,}000\), \(5{,}000\) and \(10{,}000\) nodes, respectively. Blue, orange and green mark mean predecessor counts of \(1.5\), \(2.5\) and \(4\), and solid lines with filled circles, dashed lines with open squares and dotted lines with open triangles mark uniform, lognormal and truncated power law parent selection weights. These labels specify the propensities used to select parents. Each panel shows all nine combinations of predecessor count and selection weight. The horizontal axis is the sampled fraction of designated terminals, and the vertical axis is the Spearman correlation between observed and complete pair thickness rankings over every reference edge, with unobserved edges scored zero. Observations are first averaged within each graph, and each curve then gives equal weight to the 60 independent graphs in that condition.\par}
\end{minipage}\end{figure}

\hyperref[fig:2]{Figure~2} shows the gap for networks of 2,000 to 10,000 nodes with three mean predecessor counts and three parent selection weights. All nine combinations of predecessor count and selection weight require extensive observation for reliable ranking, with mean stable success reaching 95\% only between 80\% and 92.5\% sampling. Even at 90\% sampling, mean Spearman correlation was 0.991 among discovered edges but 0.925 over all reference edges once undiscovered edges were scored zero. The undiscovered edges account for this difference.
\par\par
\begin{figure}[!htbp]\centering\begin{minipage}{\linewidth}\phantomsection\label{fig:3}\hypertarget{figure-3}{}\addcontentsline{toc}{subsection}{Figure 3. Recovery within the OpenAlex reference network}
{\centering
\includegraphics[width=\linewidth,height=0.63\textheight,keepaspectratio]{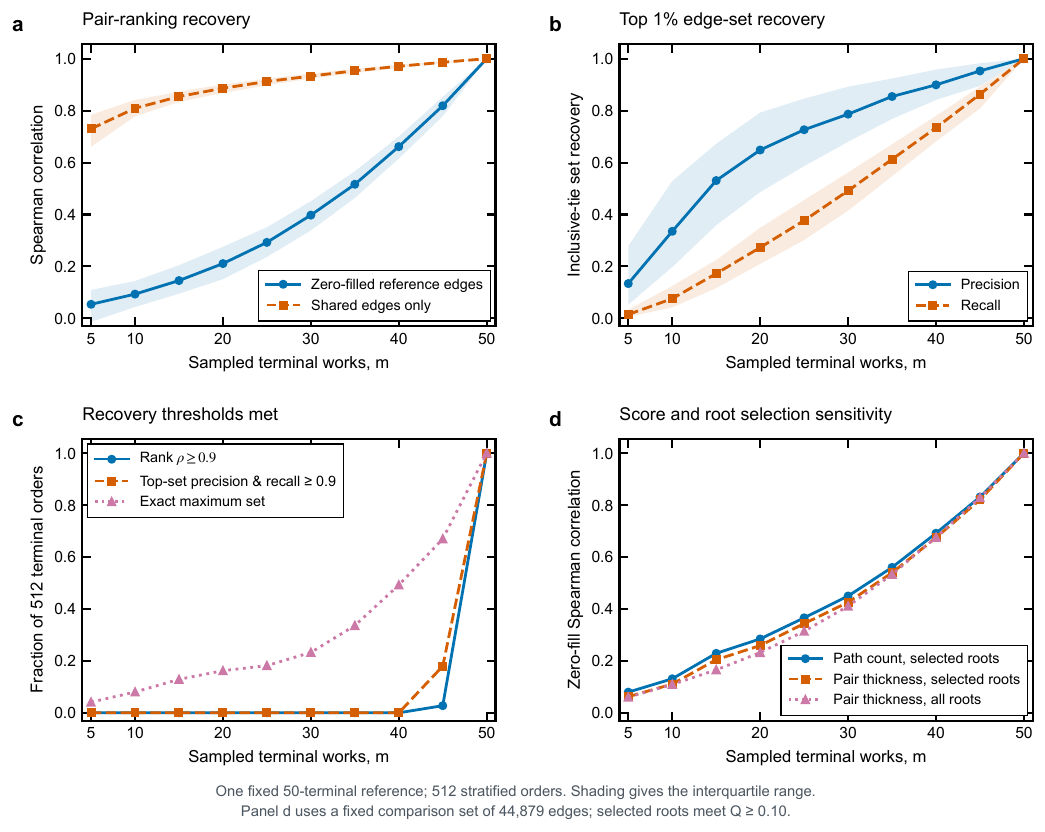}\par}\smallskip
{\fontsize{9}{11}\selectfont \textbf{Figure 3. Recovery within one fixed OpenAlex reference of 50 designated terminal works.} \textbf{a}, Ranking correlation over shared edges compared with correlation over all reference edges, with unseen edges assigned a score of zero. \textbf{b}, Precision and recall of the top 1\% edge set, keeping ties at the cutoff. Precision is the share of selected edges that belong to the reference top set, and recall the share of that set recovered. \textbf{c}, Fractions of the 512 sampling orders that reach a ranking correlation of at least 0.9, reach both precision and recall of at least 0.9, or recover the exact maximum set. \textbf{d}, Sensitivity to score and root selection on a fixed comparison set of 44,879 reference edges. Path counts and pair thickness with restricted roots are compared with pair thickness using all DAG roots. Restricted roots reach at least 10\% of designated terminals and are selected separately in each scored graph. Shading shows interquartile ranges across sampling orders stratified by publication year. The orders are repeated observations of one reference. At all 50 terminals, every sampling order gives the same observation.\par}
\end{minipage}\end{figure}

The same gap appears in a real citation network. We fixed an OpenAlex reference \textsuperscript{\hyperlink{ref:37}{37}} by tracing two steps back from 50 works published during 2021 to 2025, giving 51,384 publication nodes and 87,326 edges, together with a backbone of 13,948 nodes that support at least 10\% of those works. \hyperref[fig:3]{Figure~3} tracks rank recovery among discovered and all reference edges, precision and recall for the top 1\% of scores, and the frequency of meeting recovery thresholds. The final panel checks sensitivity to earlier counting conventions and reference sets. The 512 sampling orders are repeated observations of this one reference.
\par\par
With 25 sampled works, median rank correlation was 0.912 among discovered edges but only 0.293 over the reference, and mean backbone precision and recall were 0.946 and 0.568. At 40 works, median reference correlation reached 0.661 and mean backbone recall 0.834. Even at 45 works, only 14 of 512 orders reached a reference correlation of 0.90. A well recovered subgraph can coexist with substantial uncertainty about the wider network (Supplementary Note 1).

\subsection*{Origin counts explain root groups in thickness and loss}

To relate thickness to failure, we use 30 Growth, 30 Erdős-Rényi (ER) derived and 30 Barabási-Albert (BA) derived networks, Bioconductor software dependencies, HEP-TH citations and a binary tree. Growth uses staged attachment, and the ER and BA graphs are oriented acyclic controls. A Rio Grande reference adds 3,372 reaches, 3,411 links and 81 gauge outcomes, yielding 2,836 candidate cuts. Bioconductor declares necessary inputs, so strict loss measures counterfactual failure to install or rebuild at fixed versions, with external dependencies available and no repair or replacement \textsuperscript{\hyperlink{ref:4}{4},\hyperlink{ref:38}{38}}. Citation, river and transmission networks combine empirical topology with equation (2).

\begin{figure}[!htbp]\centering\begin{minipage}{\linewidth}\phantomsection\label{fig:4}\hypertarget{figure-4}{}\addcontentsline{toc}{subsection}{Figure 4. Observed thickness, model loss and reference coverage}
{\centering
\includegraphics[width=\linewidth,height=190mm,keepaspectratio]{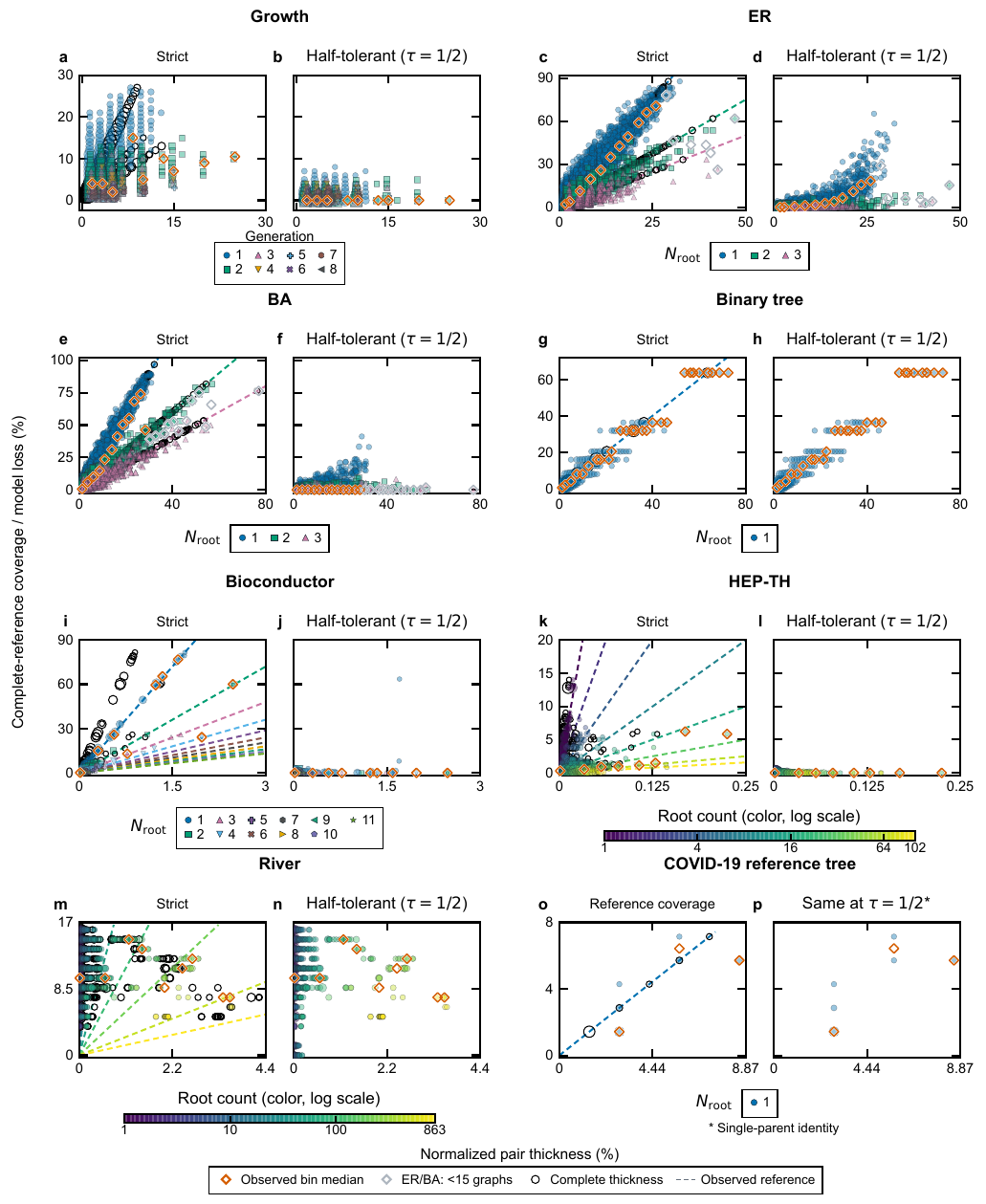}\par}\smallskip
{\fontsize{9}{11}\selectfont \textbf{Figure 4. Pair thickness and terminal loss under strict dependency and tolerance \(\tau=1/2\).} Paired panels show Growth (a,b), ER (c,d), BA (e,f), the binary tree (g,h), Bioconductor (i,j), HEP-TH (k,l), Rio Grande (m,n) and a COVID-19 reference tree (o,p). Both axes are percentages, with observed normalized pair thickness horizontally and complete reference coverage or model terminal loss vertically. In strict panels, black hollow circles show the same edges at complete thickness and unchanged loss; each thickness uses its own root and terminal denominator. Dashed rays give the observed graph's strict response scale \(R_S/n\). Marker area shows capped multiplicity. Growth panels use generation labels and small horizontal offsets; other panels encode upstream root count. Diamonds give medians within fixed bins; gray ER and BA diamonds have fewer than 15 supporting graphs. Synthetic families pool 30 graphs and the binary tree 30 observations. Empirical panels use primary observations, including 2,836 river candidates and 50 from the largest eligible COVID reference. COVID coordinates are sampled coverage \(k_e/m\) and reference leaf coverage \(U/T\); panel p follows the identity for one parent per nonroot node. Losses and reference coverage on empirical topologies are calculated under the specified dependency rule.\par}
\end{minipage}\end{figure}

\hyperref[fig:4]{Figure~4} shows strict responses on the left and responses at half tolerance on the right. Under strict dependency the edges split into groups by upstream root count. Black hollow circles place each edge at its complete thickness and colored points at its observed thickness on a shared loss axis, exposing the effect of incomplete observation. Dashed rays mark the response within the observed graph. In Growth, relationships within groups are positive, yet the median pooled rank correlation is \ensuremath{-}0.029.
\par\par
\begin{minipage}{\linewidth}
\textbf{Proposition 1. Pair counts and strict terminal loss.} In the finite DAG defined above, with fixed roots and designated terminals, complete observation and strict dependency give

\[
I_e(0)=\frac{U_G(e)}{T_G}
=\frac{R_G}{n(e)}\bar C_G(e).
\tag{3}
\]
\end{minipage}\par

Endpoint counting \textsuperscript{\hyperlink{ref:26}{26}} combined with strict failure propagation proves the relation (Supplementary Note 3.1). Equivalently, \(\bar C_G=(n/R_G)I(0)\), so pair thickness is outcome loss weighted by origin coverage. Edges with positive coverage lie on different rays exactly when their root counts differ, and fitting a slope within a ray also requires coverage to vary. These comparisons hold the network fixed. At a given terminal coverage, loss stays constant while pair thickness increases with root count. This difference can reverse consequence rankings. Equation (5) removes the root factor and sets the scale for the calibration below.

The empirical networks also contain multiple root groups. Among 4,305 primary candidate edges in HEP-TH, root counts take 65 values, 31.1\% of edges have several roots and \(n\) reaches 102. Bioconductor has 11 classes among 2,749 edges, with 34.4\% multiroot edges and a maximum \(n=11\), and the river has 154 classes. Multiple root classes record the convergence of origin lineages upstream of edges. Each class defines an exact ray, whose visible separation from others depends on coverage variation and sampling calibration.

Observed slopes also depend on how faithfully the sample represents complete coverage. With terminal coverage \(W_H(e)=U_H(e)/T_H\in[0,1]\), the ordinary least squares slope with a fitted intercept, within a root group of varying observed coverage, is

\[
b_n(0)=\frac{R_S}{n}\lambda_n,\qquad
\lambda_n=\frac{\operatorname{Cov}(W_S,W_G\mid n)}
{\operatorname{Var}(W_S\mid n)}.
\tag{4}
\]

The moments are computed over edges within a root group; \(b_n\) and the calibration coefficient \(\lambda_n\) are real and require positive predictor variance. Root count sets the structural scale, and calibration can shift or even reverse the order of observed slopes.

\begin{figure}[!htbp]\centering\begin{minipage}{\linewidth}\phantomsection\label{fig:5}\hypertarget{figure-5}{}\addcontentsline{toc}{subsection}{Figure 5. Slopes by root count and the single root boundary}
{\centering
\includegraphics[width=\linewidth,height=145mm,keepaspectratio]{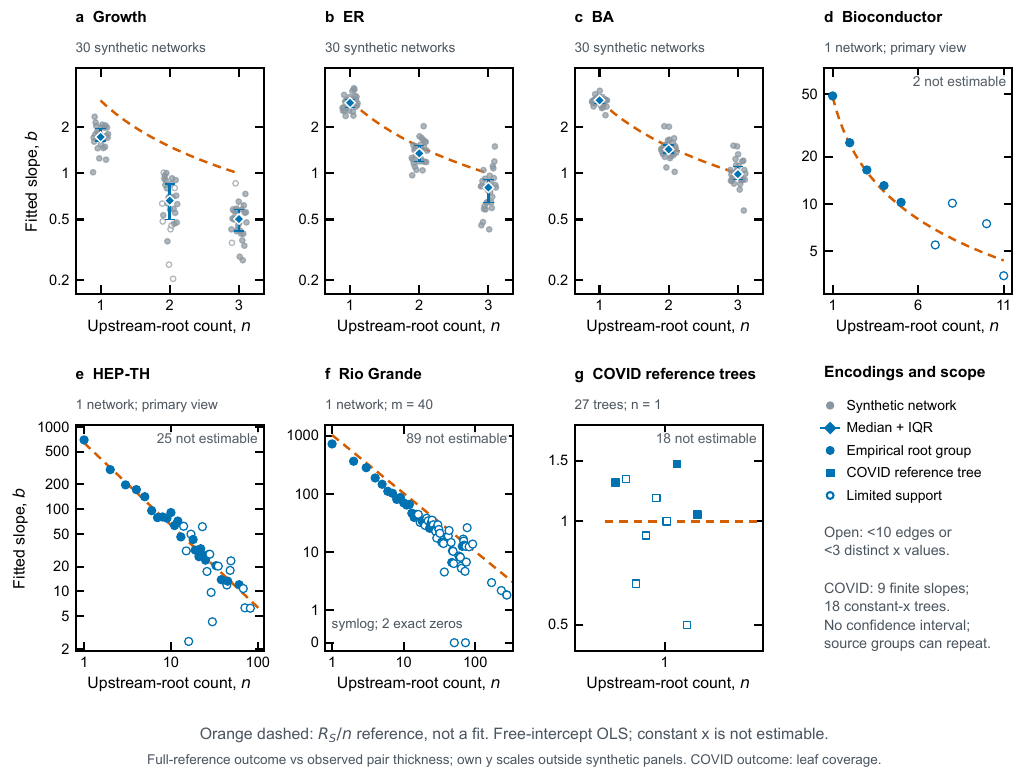}\par}\smallskip
{\fontsize{9}{11}\selectfont \textbf{Figure 5. Slopes within groups of edges that share an upstream root count.} Ordinary least squares with a fitted intercept regresses complete reference strict loss on observed normalized thickness; for COVID the response is reference leaf coverage. Panels a to c show Growth, ER and BA networks, d to f Bioconductor, HEP-TH and Rio Grande, and g the COVID references. Synthetic diamonds give network medians and interquartile ranges. Filled markers have at least 10 edges and 3 distinct predictor values, and open markers denote groups below these support thresholds. Orange dashed curves give the scale within the observed graph \(R_S/n\). All 319 original and 65 river slopes are shown, including exact river zeros at \(n=51,73\); a further 89 river groups have constant predictors and undefined slopes. The river panel uses a symmetric logarithmic scale that is linear for \(|b|\le1\), where \(b\) is the fitted slope. COVID references contribute all nine finite slopes from the primary observation, and 18 have constant predictors. All have \(n=R_S=1\), and horizontal offsets are used only to separate markers. The references share publication sources; their slopes are descriptive, with no confidence interval across independent epidemics.\par}
\end{minipage}\end{figure}

\hyperref[fig:5]{Figure~5} compares these observed slopes. In every synthetic network the one root slope exceeded the two and three root slopes, with Growth medians of 1.73, 0.664 and 0.503, and all three decreased monotonically in 24/30 Growth, 30/30 ER and 29/30 BA graphs. The river contributes 63 positive and two zero slopes, while 89 groups lack predictor variation. Its observed slopes rise between adjacent groups 18 times. These departures from a simple inverse trend are consistent with unequal calibration, as equation (4) allows. Under complete observation, the groups have a direct physical meaning. At equal complete pair thickness within a network, edges with more upstream roots reach fewer terminals and therefore cause smaller losses under strict dependency. Single root edges consequently cause the greatest loss at that thickness, as equation (3) shows. At fixed terminal coverage, loss is independent of root count because each exposed outcome is counted once, regardless of how many origins reach it.
\par\par
COVID-19 reference trees give the limiting case \textsuperscript{\hyperlink{ref:7}{7}}. Each of the 27 eligible trees has one root, so all its edges share one root class, as do networks whose roots never converge. \hyperref[fig:4]{Figure~4} shows the sampling scatter in the largest tree, and \hyperref[fig:5]{Figure~5} shows all nine estimable primary slopes; the other 18 had constant predictors.

Within a root group, complete thickness and strict loss differ only by a positive constant, so their regressions on observed thickness share the same coefficient of determination, \(R^2\in[0,1]\), whenever predictor and response vary. Incomplete observation can alter both the slope and the scatter of a root group. Dividing out origin multiplicity gives

\[
W_S(e)=\frac{R_S}{n(e)}\bar C_S(e)=\frac{k_e}{m}.
\tag{5}
\]

This normalization improved pooled linear fit and rank correlation in all 92 original synthetic and empirical analysis units. Median \(R^2\) rose from 0.027 to 0.511 in Growth, from 0.766 to 0.921 in ER and from 0.768 to 0.977 in BA (\hyperlink{figure-S6}{Supplementary Fig. 6}). It separates how many origins reach an edge from how many outcomes the edge supports.

Sparse sampling helps explain the weak pooled correlation in Growth. At \(m=20\), a median 70.3\% of discovered edges reached only one sampled terminal, so their thickness varied mainly with root count. Across positive-coverage reference edges, root count was negatively associated with complete coverage, with median Spearman correlation \ensuremath{-}0.479. The median correlation between thickness and loss rose from \ensuremath{-}0.029 at \(m=20\) to 0.799 under complete observation, and changed by different amounts when the candidate set was held fixed. Both measurement resolution and which edges are discovered therefore contribute (\hyperlink{figure-S14}{Supplementary Fig. 14} and Supplementary Note 11).

The empirical references retain stronger pooled correlations, 0.490 in HEP-TH and 0.592 in Bioconductor at their primary samples. In OpenAlex, the median across sampling orders rose from 0.640 to 0.773 as the sample grew from ten to fifty terminals with tracing limited to two steps. Across full references, root count and coverage correlated at \ensuremath{-}0.050 in HEP-TH and \ensuremath{-}0.005 in OpenAlex, and temporal associations could be examined only on dated subsets (Supplementary Note 12 and \hyperlink{figure-S15}{Supplementary Fig. 15}).

\begin{figure}[!htbp]\centering\begin{minipage}{\linewidth}\phantomsection\label{fig:6}\hypertarget{figure-6}{}\addcontentsline{toc}{subsection}{Figure 6. Exact tolerance responses and the single parent boundary}
{\centering
\includegraphics[width=\linewidth,height=200mm,keepaspectratio]{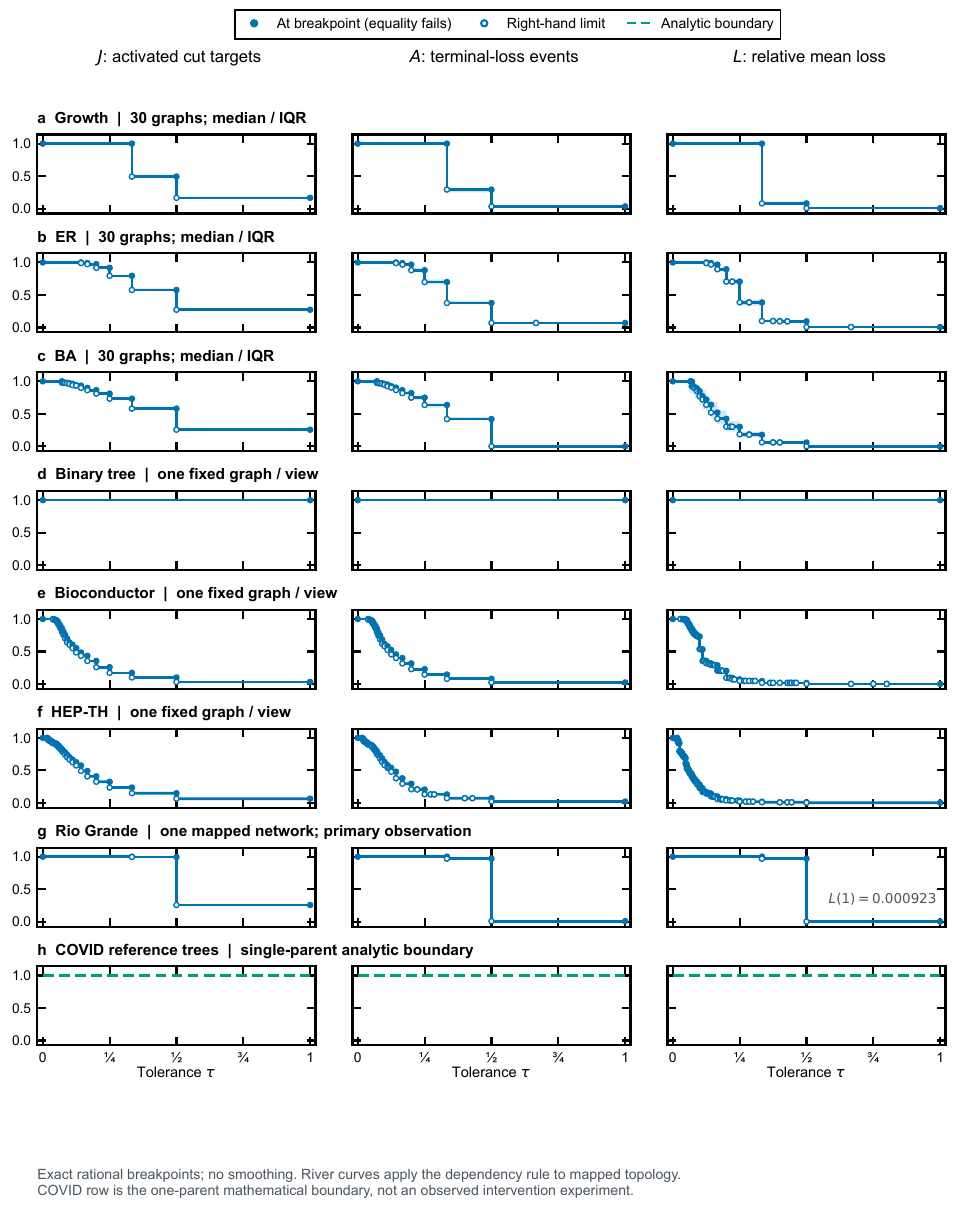}\par}\smallskip
{\fontsize{9}{11}\selectfont \textbf{Figure 6. Input tolerance separates failure initiation, propagation to outcomes and total loss.} Columns show the fraction \(J\) of cuts that fail the target, the fraction \(A\) causing any terminal loss and mean terminal loss relative to its strict value, \(L\). Rows compare network families. Synthetic curves give medians and interquartile ranges over 30 networks; empirical curves use primary observations. Filled markers show values at critical tolerances and open markers the limits immediately to their right. Equality triggers failure, so loss falls after each breakpoint. All breakpoints in the original networks and river reference are shown without smoothing. Rio Grande has 2,836 candidates and 81 designated reaches, with \(L=22/23838\) above \(\tau=1/2\). The COVID reference has \(J=A=L=1\) throughout \(0\le\tau\le1\) because each nonroot node has one parent, a property of this topology.\par}
\end{minipage}\end{figure}

The strict relation identifies outcomes exposed to an edge failure. When remaining inputs can sustain function, some of those outcomes survive, altering responses within root groups. As tolerance increases, \hyperref[fig:6]{Figure~6} tracks the fraction of cuts that fail their target, the fraction that reach any terminal and the mean loss relative to strict dependency. The curves fall in steps because nodes lose whole inputs.
\par\par
Growth keeps its strict response up to tolerance one third, as expected with at most three inputs per node, and the binary tree and COVID reference trees never change because every nonroot has one input. At half tolerance, median mean loss drops to 8.08\%, 9.40\% and 5.87\% of its strict value in Growth, ER and BA, respectively. The river shows how sharp these steps can be. It retains 96.8\% of strict loss at exactly half tolerance but only 0.0923\% just above it, where only 22 of 2,836 cuts still reach a designated outcome. Equality still triggers failure, placing the jump immediately above the threshold.

For an edge with positive complete coverage, let \(q_e(\tau)\in[0,1]\) be the fraction of its reachable terminals that fail. Then

\[
I_e(\tau)=W_G(e)q_e(\tau).
\tag{6}
\]

A single cut starts a cascade only when \(\tau\le1/d_v\). Further propagation depends on joint losses of downstream inputs. We compute the whole curve by propagating the largest tolerance at which each terminal fails, with the recursion and proof in Supplementary Note 3.4. The local aggregation has the established discrete Sugeno form \textsuperscript{\hyperlink{ref:39}{39}}. The calculation captures thresholds that a handful of chosen tolerance values could miss.

At full tolerance, a terminal survives whenever any path from an original root survives. Let \(D_G(e)\in\{0,\ldots,U_G(e)\}\) count the terminals whose every such path uses edge \(e\). Then

\[
\frac{D_G(e)}{T_G}=I_e(1)\le I_e(\tau)\le I_e(0)=\frac{U_G(e)}{T_G}.
\tag{7}
\]

This is the graph theoretic dominator condition \textsuperscript{\hyperlink{ref:40}{40}}. Edges with \(D_G=U_G\) stay equally consequential at every tolerance, and some have several upstream roots. Many other strict response lines are selectively suppressed, as the right panels of \hyperref[fig:4]{Figure~4} show. Cross validation shows that both groupwise and pooled linear predictions can become inaccurate as tolerance rises (\hyperlink{figure-S2}{Supplementary Fig. 2}). Input substitution depends on the arrangement and requirements of downstream support.

\subsection*{Counting distinct outcomes improves strict edge prioritization}

\begin{figure}[!htbp]\centering\begin{minipage}{\linewidth}\phantomsection\label{fig:7}\hypertarget{figure-7}{}\addcontentsline{toc}{subsection}{Figure 7. Edge importance across seven network families}
{\centering
\includegraphics[width=\linewidth,height=125mm,keepaspectratio]{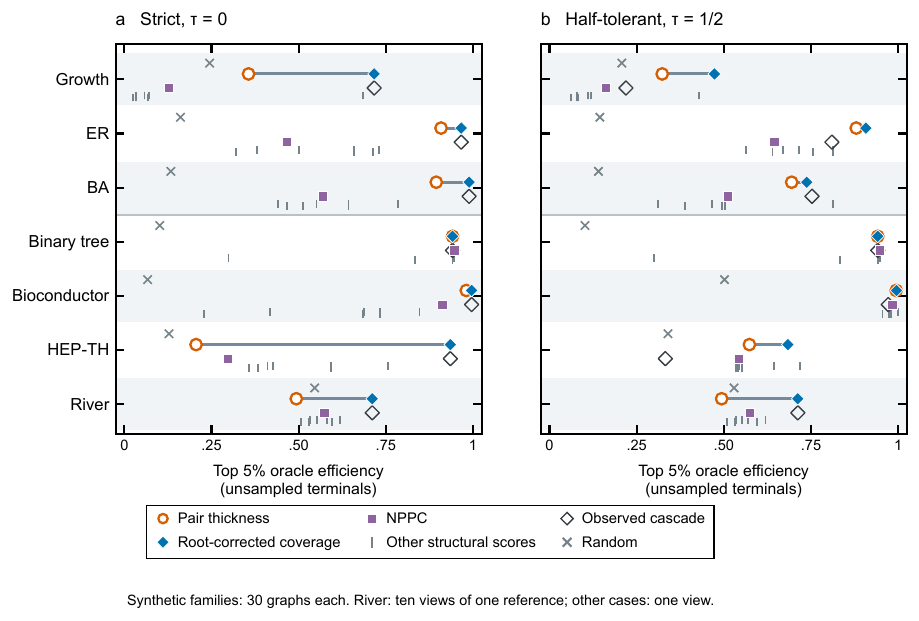}\par}\smallskip
{\fontsize{9}{11}\selectfont \textbf{Figure 7. Predicting loss among unsampled terminals under strict support (a) and tolerance \(\tau=1/2\) (b).} Top 5\% oracle efficiency divides mean loss from independent cuts of selected edges by the greatest mean loss attainable at the same budget with complete information. Rows compare networks on a common efficiency axis. Orange circles show pair thickness and blue diamonds coverage corrected for upstream root count; connecting lines highlight their difference. Purple squares denote node pair projection count (NPPC), gray ticks the seven other structural comparators, open diamonds observed cascade loss and crosses random ranking. Vertical offsets separate coincident methods. All methods share candidate edges; at half tolerance, structural scores also share the failure initiation gate. Synthetic means average 30 independent graphs per family. The tree, Bioconductor and HEP-TH are single cases; river means average ten observations of \(m=40\) terminals from one reference. Under strict support, corrected coverage equals observed cascade loss and deterministic reverse reachability coverage restricted to sampled terminals. \hyperlink{figure-S3}{Supplementary Fig. 3} lists every comparator and numerical value.\par}
\end{minipage}\end{figure}

The distinction between origins, exposed outcomes and failed outcomes determines what an edge score must predict. We define importance as terminal loss under the specified support rule, and \hyperref[fig:7]{Figure~7} tests scores against that consequence. Every score uses the same observed graph to predict losses among unsampled outcomes. Performance is the mean loss from independently cutting the highest ranked 5\% of edges, divided by the best achievable mean at that budget. This oracle efficiency lies between zero and one whenever some candidate causes loss. Connecting segments compare pair thickness with corrected coverage, and further markers show NPPC, path counts, betweenness, target outdegree and random ranking.
\par\par
Corrected coverage equals observed strict loss and the deterministic reverse reachability coverage of designated terminals \textsuperscript{\hyperlink{ref:23}{23},\hyperlink{ref:24}{24}}. This equality lets us test how accurately sampled coverage predicts losses among unsampled outcomes.

Under strict dependency, correction raised mean efficiency from 0.356 to 0.716 in Growth, from 0.908 to 0.966 in ER and from 0.895 to 0.989 in BA, and every synthetic graph improved. In Growth, NPPC and the search path count (SPC), search path link count (SPLC) and search path node pair count (SPNP) scored 0.128, 0.0325, 0.0243 and 0.0712, respectively, while target outdegree came closer to coverage at 0.684. In Bioconductor and HEP-TH, efficiency rose from 0.981 to 0.996 and from 0.206 to 0.935. In ten views of the Rio Grande, strict coverage efficiency averaged 0.711, against 0.493 for thickness and 0.619 for SPNP. In the single root tree, correction changes no ranking and some classical scores slightly outperform coverage, at 0.947 versus 0.941. \hyperlink{figure-S3}{Supplementary Fig. 3} lists all comparator values.

Classical pair and path scores multiply a count of upstream nodes or paths by a corresponding downstream count \textsuperscript{\hyperlink{ref:26}{26}}. Removing the upstream factor from each of six such scores improved strict efficiency in all 90 synthetic graphs at the primary 5\% budget (\hyperlink{figure-S16}{Supplementary Fig. 16} and Supplementary Note 13). The benefit is therefore shared by several downstream measures. In Growth, descendant counts and path counts to all nodes reached 0.760 and 0.762, above terminal coverage at 0.716, and forward path scoring has prior precedent \textsuperscript{\hyperlink{ref:41}{41}}. The empirical results were less uniform, and coverage underperformed thickness in one of ten river views.

Alternative inputs change which scores perform well, as the second panel of \hyperref[fig:7]{Figure~7} shows. At half tolerance, every structural score receives the same filter that excludes cuts unable to initiate failure. In BA, target outdegree scores 0.812 against 0.738 for coverage. A separate comparison in the Rio Grande at full tolerance gave an efficiency of 0.103 for coverage and 0.739 for betweenness. Alternative support shifts the ranking target from exposed outcomes to outcomes that lack a surviving route.

\subsection*{Growth and sampling predict root group composition and responses}

The exact relation gives the response within each root group, but leaves the sizes of those groups unspecified. Growth creates multiple root groups when new nodes take parents from branches descending from different roots. Later dependencies inherit the merged ancestry, while subsequent births build up terminal coverage. Modeling these two processes together lets us predict group composition and the response distributions seen after sampling.

The growth model starts with an integer number of roots \(R_0\ge1\) in generation zero. Generations \(g=1,\ldots,L\), with integer \(L\ge1\), each add a positive integer number of nodes \(N_g\), forming a cohort born at the same stage. Each newborn draws a positive integer number of parents \(J_g\) from a probability mass function \(\eta_g\), supported within the number of available older nodes, and newborns choose independently from the pool fixed before their generation.

At each draw without replacement, an eligible node \(i\) born in generation \(b_i\in\{0,\ldots,g-1\}\) is chosen with probability

\[
p_i(g,\mathcal A;\beta)=
\frac{\exp[-\beta(g-b_i)]}{\sum_{j\in\mathcal A}\exp[-\beta(g-b_j)]},
\qquad \beta\in\mathbb R.
\tag{8}
\]

The nonempty set \(\mathcal A\) holds the remaining eligible parents, and \(p_i\in(0,1]\) sums to one over it. The dimensionless age preference \(\beta\) sets the favored parent age. Negative values favor older nodes, zero gives uniform choice and positive values favor recent nodes. Birth generations are known formation stages, distinct from layers inferred from topology. We build on models of attachment with age preference and ancestry in random acyclic graphs \textsuperscript{\hyperlink{ref:42}{42},\hyperlink{ref:43}{43},\hyperlink{ref:44}{44},\hyperlink{ref:45}{45}} to predict dependency responses under partial observation.

The final \(T=N_L\) nodes are terminals. Exact recursions track how roots combine upstream and descendants accumulate downstream. For a tail generation \(a\), the law \(\pi_a(n;\beta)\) gives root counts \(n\in\{1,\ldots,R_0\}\); for a target generation \(b\), \(\rho_b(u;\beta)\) gives complete coverage \(u\in\{0,\ldots,T\}\), with \(0\le a<b\le L\). Combining these laws with expected edge counts and uniform sampling of \(m\in\{1,\ldots,T\}\) terminals gives \(\mathcal J_{\beta,m}(n,k,u)\), the nonnegative expected number of edges with \(n\) upstream roots, \(k\in\{0,\ldots,m\}\) sampled terminals and \(u\) reachable terminals in the complete network. Impossible combinations have zero expected count. For a discovered pair \((n,k)\) with \(k>0\) and positive expected count,

\[
p_{\beta,m}(u\mid n,k)=
\frac{\mathcal J_{\beta,m}(n,k,u)}{\sum_{z=0}^{T}\mathcal J_{\beta,m}(n,k,z)},\qquad
\mu_n(k)=\sum_{u=0}^{T}\frac{u}{T}p_{\beta,m}(u\mid n,k).
\tag{9}
\]

The probability \(p_{\beta,m}\) describes complete coverage among edges that share the observed counts. The corresponding mean response \(\mu_n(k)\in[0,1]\), variance and prediction intervals give the position and spread of the response within each root group. We evaluate the exact recursion numerically over a finite set of states. Supplementary Notes 8 and 9 give the assumptions and derivations.

Given an edge and the birth generations of its endpoints, upstream ancestry and future terminal coverage arise from independent random choices. At fixed birth conditions, root count carries no additional loss information. Mixing birth conditions couples root count to coverage (Supplementary Note 3.5). Removing the association between ancestry and coverage while preserving their marginals worsened joint count predictions (Supplementary Note 7 and \hyperlink{figure-S9}{Supplementary Fig. 9}).

We tested how merging origin lineages creates root groups in 1,024 matched network pairs. Parent selection was either restricted to each root branch or allowed across branches, with each network generating its own subsequent outcomes. At 30 sampled terminals, restricting parents to their own branch produced only single root edges, whereas cross branch selection produced on average 46.4\% multiroot edges and all three root classes in every tested graph (Supplementary Note 7 and \hyperlink{figure-S9}{Supplementary Fig. 9}).

\begin{figure}[!htbp]\centering\begin{minipage}{\linewidth}\phantomsection\label{fig:8}\hypertarget{figure-8}{}\addcontentsline{toc}{subsection}{Figure 8. Growth and sampling predict response distributions within root groups}
{\centering
\includegraphics[width=\linewidth,height=130mm,keepaspectratio]{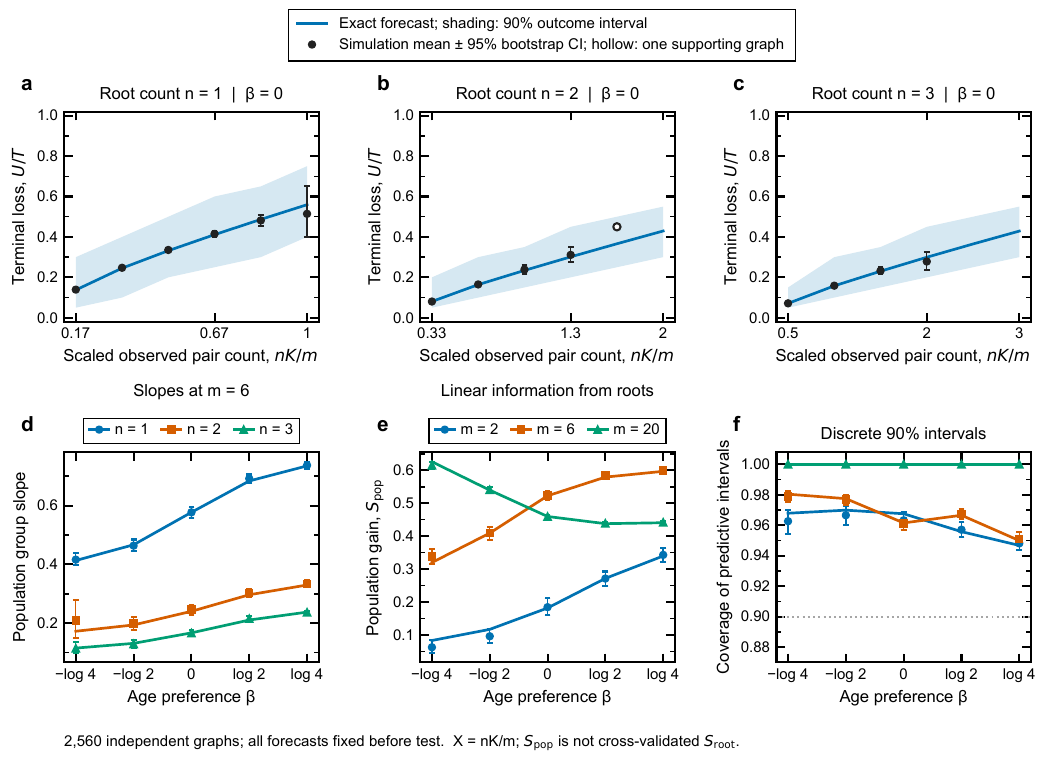}\par}\smallskip
{\fontsize{9}{11}\selectfont \textbf{Figure 8. Network formation and sampling predict response distributions across root groups.} Predictions preceded simulations of 512 independent networks at each of five age preferences, 2,560 in total. Each network has three roots, five cohorts of 20 nodes and 20 terminals in the final cohort. For a discovered edge, \(n\) counts roots reaching its tail, \(K\) counts sampled terminals reached from its head and \(U\) counts all such terminals. With \(m\) of \(T\) terminals sampled, the coordinate \(X=nK/m\) omits \hyperref[fig:4]{Fig.~4}'s random observed root denominator. \textbf{a to c}, Conditional means of \(U/T\) and discrete central 90\% prediction bands at \(\beta=0,m=6\), compared with simulated means. Shading shows outcome variation; black whiskers give pointwise 95\% uncertainty intervals for the mean from 1,000 whole graph bootstrap resamples. Hollow points have one supporting graph. \textbf{d}, Population response slopes within root groups at \(m=6\). \textbf{e}, Population gain \(S_{\mathrm{pop}}=(L_{\mathrm{pool}}-L_{\mathrm{group}})/\operatorname{Var}(U/T)\), where each \(L\) is its linear predictor's minimum mean squared error. \textbf{f}, Predicted probability mass and empirical coverage of the discrete intervals, whose mass exceeds the nominal 0.90. Lines are predictions and markers simulations. Resampling preserves whole graphs and nested samples; tables retain sparse and unobserved cells. Population gain weights edges by their expected counts and differs from the average cross validated predictive gain within individual networks examined in \hyperref[fig:9]{Fig.~9}.\par}
\end{minipage}\end{figure}

\hyperref[fig:8]{Figure~8} tests predictions frozen before generation against 2,560 independent test networks, 512 per preference, each with three roots, five generations of 20 nodes and samples of 2, 6 or all 20 terminals. The panels compare conditional means and 90\% prediction bands, group slopes, the benefit of grouping and prediction band coverage across preferences and sample sizes.
\par\par
These predictions concern a population of networks, with edges weighted by their expected abundance across that population. We use \(X=nK/m\) and \(Y=U/T\), where the random counts \(K\) and \(U\) range over sampled and complete terminal coverage, so that \(0\le X\le R_0\) and \(0\le Y\le1\). \hyperref[fig:4]{Figure~4} additionally normalizes its horizontal coordinate by the observed root count. For variable \(Y\), the population gain \(S_{\mathrm{pop}}=(L_{\mathrm{pool}}-L_{\mathrm{group}})/\operatorname{Var}(Y)\in[0,1]\) compares the nonnegative minimum mean squared errors \(L_{\mathrm{pool}}\) and \(L_{\mathrm{group}}\) of pooled and grouped linear predictors. This differs from the cross validated gain within individual graphs examined below.

The simulations agreed with the population predictions. At the primary sample of six terminals, all 15 predicted group slopes fell within pointwise bootstrap intervals around the simulated estimates, and predicted population gains of 0.320 to 0.597 matched simulated gains of 0.338 to 0.599. The largest difference between prediction band coverage and its calculated probability under partial observation was approximately 0.524 percentage points, and because terminal counts are discrete, central bands can contain more than their nominal 90\% probability. Rare groups remained poorly resolved under strong preference for old nodes with only two sampled terminals (\hyperlink{figure-S12}{Supplementary Fig. 12}). The growth and sampling calculations predict mean losses and their variation within root groups across the network population.

\par\begin{minipage}{\linewidth}
Agreement across a population leaves a further question of whether root groups improve prediction within a single observed network. To test this, we compare one regression line fitted to the supported edges with separate lines for their root groups. Both fits predict the same withheld edges. Their gain in predictive accuracy is

\[
S_{\mathrm{root}}=R^2_{\mathrm{CV,group}}-R^2_{\mathrm{CV,pooled}}
=\frac{\mathrm{SSE}_{\mathrm{CV,pooled}}-\mathrm{SSE}_{\mathrm{CV,group}}}{\mathrm{SST}}.
\tag{10}
\]

Here CV denotes cross validation, SSE is the nonnegative sum of squared prediction errors and SST is the positive sum of squared deviations of observed losses from their mean. Each predictive \(R^2=1-\mathrm{SSE}/\mathrm{SST}\le1\) can be negative. The real gain \(S_{\mathrm{root}}\) is positive when root groups help and negative when separate fitting hurts. We report grouped accuracy alongside this gain. Methods describes the withheld targets, estimability conditions and validation controls (\hyperlink{figure-S11}{Supplementary Fig. 11}).
\end{minipage}\par

\begin{figure}[!htbp]\centering\begin{minipage}{\linewidth}\phantomsection\label{fig:9}\hypertarget{figure-9}{}\addcontentsline{toc}{subsection}{Figure 9. Age preference and the predictive value of root groups}
{\centering
\includegraphics[width=\linewidth,height=160mm,keepaspectratio]{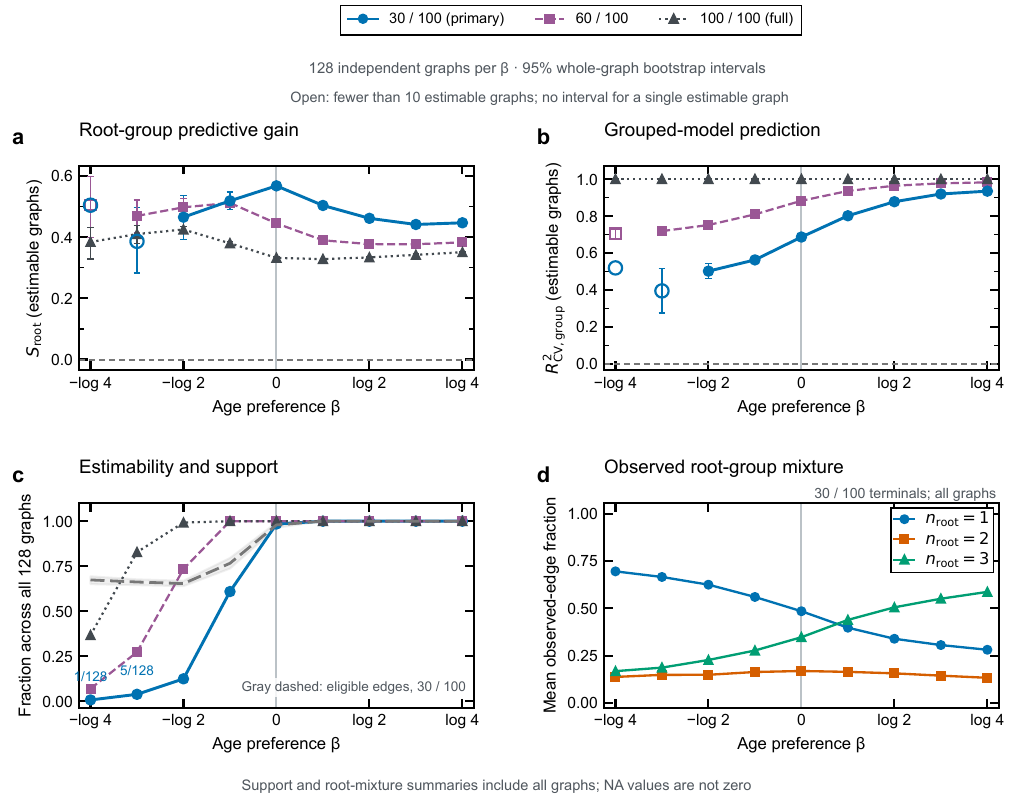}\par}\smallskip
{\fontsize{9}{11}\selectfont \textbf{Figure 9. Age preference changes root group composition and predictive value within individual networks.} Nine prespecified \(\beta\) values from \(-\log4\) to \(\log4\) were each tested on 128 independent networks. Negative \(\beta\) favors older parents, positive \(\beta\) recent parents and \(\beta=0\) uniform choice. All networks have strict support, three roots, eight generations of 100 births and the same parent count distribution. \textbf{a,b}, Mean predictive gain \(S_{\mathrm{root}}\) (a) and grouped cross validated \(R^2\) (b), using nested samples of 30, 60 or all 100 terminals. Means include only graphs with an estimable prespecified comparison and retain negative scores. \textbf{c}, Colored curves show the fraction of all 128 graphs with an estimable gain. The gray dashed curve shows the mean fraction of edges meeting support criteria at the primary sample of 30 terminals, including graphs without estimable gain. \textbf{d}, Mean observed edge fractions in each root group at the primary sample, over all graphs. Intervals in a, b and d and the gray band in c are 95\% whole graph bootstrap intervals; colored fractions in c are exact counts. Sampling budgets change discovered candidates and statistical support as well as observation error. These comparisons within individual graphs differ from the population quantity in \hyperref[fig:8]{Fig.~8}. At the two most negative preferences, only 1/128 and 5/128 graphs are estimable at \(m=30\). Conditions with fewer than ten estimable graphs use isolated open markers; the degenerate interval for one graph is retained in source data but omitted from the plot. Eligibility criteria are unchanged.\par}
\end{minipage}\end{figure}

\hyperref[fig:9]{Figure~9} examines this question in 1,152 independent graphs with three roots and eight generations of 100 nodes at nine preferences. We report the gain from grouping together with grouped accuracy, the fraction of graphs permitting estimation and the root classes present. This distinguishes a weak predictive benefit from insufficient variation to measure it.
\par\par
At 30\% observation, mean estimable \(S_{\mathrm{root}}\) declines from 0.567 under uniform choice to 0.447 at \(\beta=\log4\), even as grouped predictive \(R^2\) improves from 0.686 to 0.935. Pooled prediction improves too, leaving less for group labels to add. At \(\beta=-\log4\) all root classes remain occupied, yet only one of 128 graphs supports the required regression comparison, so estimation is limited by group size and predictor variation. \hyperref[fig:9]{Figure~9} uses larger networks and a different averaging target than \hyperref[fig:8]{Figure~8}, and predicting fluctuations within individual graphs further requires information about shared terminal sets.

\FloatBarrier\section*{Discussion}
\addcontentsline{toc}{section}{Discussion}

The origins of support and the consequences of losing it become observable at different rates. Backward tracing can recover origins while leaving pair thickness rankings uncertain. Pair thickness counts each downstream outcome once for every origin reaching an edge, whereas strict terminal loss counts that outcome once. Proposition 1 explains the resulting root groups. In Growth, sparse sampling and the association between root count and coverage can erase the pooled correlation despite strong relationships within groups. The empirical references retain stronger pooled correlations.

Converging origin lineages create edges with different root counts. The exact relation fixes group slopes under complete observation. Under strict dependency, equation (3) gives lower terminal loss for edges with more roots at equal complete pair thickness within a network. At fixed terminal coverage, reducing loss requires substitutable inputs and routes that bypass the failed edge. Growth theory predicts the abundance of edges in each group and, combined with sampling, their loss distributions. Input substitution then determines which exposed outcomes fail. Population tests assess the predicted averages; tests within individual graphs ask whether root labels improve prediction for that network.

Terminal loss defines importance for a specified support rule and outcome population. Classical product scores combine upstream participation with downstream reach, counting different endpoints or paths. Removing their upstream factors improves strict rankings in the primary synthetic tests. Terminal coverage coincides with reverse reachability coverage and exactly measures observed strict loss, although other downstream scores can better predict losses among unsampled outcomes. Information about network formation can help predict these unsampled losses. Temporal effects bias structural rankings \textsuperscript{\hyperlink{ref:46}{46}}, while partial temporal observations can predict spreading influence \textsuperscript{\hyperlink{ref:47}{47}}. Bayesian finite population methods provide a related approach to predicting unobserved outcomes from sampled data \textsuperscript{\hyperlink{ref:48}{48},\hyperlink{ref:49}{49}}. In our tests, a growth prior offers a small gain over a matched empirical prior when histories and observations are scarce (Supplementary Notes 3.6, 8.4 and 10).

We also tested structural prioritization against measured gene expression responses to transcription factor (TF) perturbations \textsuperscript{\hyperlink{ref:2}{2}}. We aggregated scores over the recorded relations associated with each TF in an independently annotated projection of Escherichia coli regulation (Supplementary Note 6). Across ten sampling views and 77 test regulators, efficiency for selecting the top 10\% by activity decrease among unsampled outputs averaged 0.706 for thickness, 0.694 for coverage, 0.688 for NPPC and 0.575 for random selection. Overall rank correlation for thickness remained weak at 0.044 (\hyperlink{figure-S8}{Supplementary Fig. 8}).

Applying the framework to software dependency security \textsuperscript{\hyperlink{ref:50}{50}}, essential input economics \textsuperscript{\hyperlink{ref:51}{51},\hyperlink{ref:52}{52}} or power grid cascades \textsuperscript{\hyperlink{ref:53}{53}} requires specifying the origins, support rules and outcomes. Historical CRAN or Bioconductor package removals could test predicted losses against observed events. River hydraulics and regulatory feedback require application specific dynamics.

The available empirical dates permit temporal comparisons on subsets of nodes. Extending the growth model to arriving roots and heterogeneous cohorts would allow more complete formation histories to be represented, while large empirical root populations require more scalable computation. Separating converging origins from dependent outcomes explains how origin structure shapes downstream vulnerability and why similar structural scores can conceal different consequences.
\FloatBarrier\section*{Methods}
\addcontentsline{toc}{section}{Methods}

\subsection*{Observation and recovery measures}

Backward tracing queried an indexed table of complete predecessor lists; only recorded empty lists identified roots. A queue and shared cache retained all returned edges with one query per node across terminals. With constant time set bookkeeping, acquisition takes \(|V_S|\) queries and \(O(|V_S|+|E_S|)\) adjacency processing, excluding query table construction and score calculation. Finite depth tracing stops before querying boundary nodes. Root and terminal sets belong to the scored graph.

Both backward tracing and RIS use local adjacency access \textsuperscript{\hyperlink{ref:23}{23}}. We retain the prerequisite subgraph of designated terminals in a fixed dependency graph; RIS uses reverse influence samples to optimize a seed set. Setting all propagation probabilities to one gives identical ancestor sets for the same terminal. Sampling designs and finite population interpretations follow below.

Root and backbone recovery used precision and recall against fixed reference sets. Shared edge rank correlation included discovered reference edges; complete reference correlation used all reference edges with positive pair counts, scoring missing edges as zero. Ties received average ranks. All planted score vectors varied, permitting ordinary correlation.

\subsection*{Pair thickness and terminal coverage}

The system boundary and designated outcome population define the counting task. In the complete dependency model \(G\), original zero indegree nodes are externally sustained roots. Designated terminals are the outcomes evaluated and may also support other nodes. Other nodes contribute through reachability. Scores and interventions share the boundary and outcome set; sampled and unsampled terminals partition this population.

For each scored graph, forward and reverse topological passes propagate sets of reachable roots and designated terminals, respectively. Set unions count each endpoint once; multiplying their sizes at an edge's tail and head gives equation (1). Conventional NPPC uses the ancestor and descendant sets of all nodes instead \textsuperscript{\hyperlink{ref:26}{26}}. Restricting endpoints focuses the count on origins and outcomes.

Dividing pair thickness by its upstream root count gives the number of distinct exposed terminals, normalized in equation (5) by the sampled terminal population. Under strict dependency, losing one necessary input fails the target and all its descendants, so normalized coverage equals observed terminal loss. Counting terminal reverse reachable (RR) sets containing the target gives exactly the same count for a matched graph, terminal set and sample. Estimating complete loss requires unsampled coverage.

\subsection*{Planted and empirical recovery designs}

The planted design crossed node counts 2,000, 5,000 and 10,000, mean predecessor counts 1.5, 2.5 and 4.0, and uniform, lognormal or truncated power law attachment propensities, with 60 independent graphs per condition. Layer zero held 100 roots, followed by five intermediate layers and a final observation layer containing 40\% of nonroot nodes. Every nonroot had a predecessor, and a direct edge from each root to one terminal ensured a terminal descendant. The edge budget was \(\operatorname{round}\{\bar d(N-100)\}\), where integer \(N\) is the total node count and positive real \(\bar d\) the prescribed mean predecessor count. Propensity weights were one, lognormal with parameters zero and one, or \(\min(1+Z,25)\) for a nonnegative Lomax random variable \(Z\) with scale one and shape 2.5.

Root recovery used 128 nested terminal orders and pair ranking used 32, with registered grids and rounding given in Supplementary Note 4. Root recovery success required at least 90 planted roots. Ranking success required a recovery score of at least 0.90; stable success required this at every subsequent registered size in the same order. Required fractions were averaged within each graph, and graphs then received equal weight. A condition crossed the recovery threshold at the first registered fraction with mean stable success of at least 0.95. The pair score analysis reused the frozen graphs and observation identities.

The OpenAlex replay used ten works from each year 2021 to 2025 and 512 sampling orders stratified by year \textsuperscript{\hyperlink{ref:37}{37}}. Each sample queried the references of the selected works and their direct predecessors, giving a two step boundary. Cleaning removed duplicate edges, self edges and identifiable temporal reversals and then condensed strongly connected components; every reference component was a singleton. Edges point from cited to citing work, and terminals need not be sinks.

The full reference contained 49,740 zero indegree roots, and its backbone comprised components reaching at least 10\% of the selected works. Each replay recomputed its own roots. Backbone comparisons used work identities, ranking comparisons used stable singleton edges and top set diagnostics kept complete tie blocks; repeated census rows were not treated as independent observations. Supplementary Notes 1 and 4 give the reference boundary and changes from historical definitions.

\subsection*{Vulnerability networks}

The vulnerability cohort was independent of the planted recovery ensemble. Growth graphs started from three roots and added eight generations of 100 nodes. Each new node chose one, two or three distinct parents, each count equally likely, uniformly from the nodes present before its generation. The final 100 nodes were terminals, and their ancestor closure defined the complete analysis graph.

The ER derived substrate had 803 nodes and independent edge probability \(1600/\binom{803}{2}\) \textsuperscript{\hyperlink{ref:54}{54},\hyperlink{ref:55}{55}}. We kept its registered largest component, selected three roots and oriented edges outward by breadth first distance from the nearest root, breaking equal distance ties with a fixed random node order and removing edges between roots. The BA derived substrate used degree proportional attachment \textsuperscript{\hyperlink{ref:56}{56}}, with three initially disconnected roots, a six node bootstrap covering their three pairs and two distinct existing parents per later node; its 803-node, 1,600-edge substrate was oriented in the same way. The binary tree was one fixed breadth first, left filled graph with 803 nodes, 802 edges and 402 terminal leaves.

The frozen Bioconductor 3.22 graph used internal Depends, Imports and LinkingTo relations, directed from prerequisite to dependent. These are documented strong dependencies, and LinkingTo supplies headers for source installation \textsuperscript{\hyperlink{ref:38}{38},\hyperlink{ref:4}{4}}. Strict loss thus represents inability to satisfy required inputs for source installation or rebuilding at fixed versions, with external dependencies available and no repair or replacement. A cut makes a required input unavailable while retaining its dependency declaration. The largest component contained 2,030 nodes, 9,496 edges, 91 roots and 1,416 designated terminals, and its primary observation 583 nodes, 2,749 edges and 48 roots. \hyperref[fig:1]{Figure~1b} draws this component in full (Supplementary Note 1.5).

The HEP-TH topology came from SNAP \textsuperscript{\hyperlink{ref:57}{57}}. We retained papers through December 1994 and citations to papers from earlier months, reversed the citation direction and selected the largest weakly connected component deterministically. The resulting DAG had 3,920 nodes, 12,349 edges, 1,070 roots and 1,334 terminals, and its primary observation had 4,305 edges.

Growth observations sampled 20 of 100 terminals, and ER, BA and empirical observations sampled approximately 20\% using frozen rounding. \hyperref[fig:4]{Figure~4} and \hyperlink{figure-S1}{Supplementary Fig. 1} pool 30 samples of 80 binary tree terminals. The tolerance scan in \hyperref[fig:6]{Figure~6} and the importance analyses use one fixed view. All vulnerability observations were complete ancestor closures, and every observed edge was a candidate for an independent cut. Supplementary Notes 11 and 12 examine formation times and sampling.

\subsection*{River topology and observation}

The Rio Grande reference was fixed around the USGS Embudo anchor (08279500) before evaluating responses. The predecessor closure in Hydrologic Fabric contained 3,372 reaches, 3,411 downstream directed links and 1,270 map boundary roots. Metadata for daily mean discharge identified 89 historical stations on 81 distinct reaches, which form the designated outcome set without requiring simultaneous flow records. The registered primary sample contains 40 outcomes, 1,025 observed roots and 2,836 candidate links. Figures 4 to 6 use this single view, and \hyperref[fig:7]{Figure~7} averages ten fixed views of 40 terminals from the same river. Sensitivity analyses use samples of 16 and 65 outcomes and an NLDI representation differing by one predecessor link. Source URLs and mapping details accompany the data (Supplementary Note 2.6). All river responses are predictions from equation (2) applied to the recorded topology.

\subsection*{COVID-19 reference trees}

We used COVID-19 reference trees from the OutbreakTrees author archive \textsuperscript{\hyperlink{ref:7}{7}}. Eligible references were simple, connected DAGs with one root, one parent per nonroot, at least ten nodes and at least four leaves, and their leaves defined the reference terminals. Ten fixed seed samples of \(m=\lceil T/2\rceil\) leaves, where \(T\) is the number of reference leaves, gave complete ancestral closures. Only topology and source identifiers were analyzed. Groups of source URLs flagged shared provenance, and repeated views and edges were not treated as independent epidemic replicates. Supplementary Note 1.4 specifies selection, estimability and the analytic boundary.

\subsection*{Exact responses and root group comparisons}

Critical tolerances were propagated as exact rational values using the recurrence in Supplementary Note 3.4, equation (S10). Curves retain both threshold values and limits immediately to their right; snapshots use \(0,1/4,1/2,3/4,1\). In this unweighted model, different cuts entering one target have identical response curves, because each removes one input and target failure causes the same downstream losses of support. Their root counts and pair scores can still differ. Original indegrees and candidate populations stay fixed.

Slopes were fitted with an intercept in root groups with at least two distinct predictor values, and groups with a constant predictor remained non-estimable. River slopes use the exact integer covariance of sampled and full coverage before normalization, and groups with a constant response have zero slope and undefined \(R^2\). Predictive comparisons required at least two supported groups. Each group needed at least ten targets, three predictor values and estimable lines in the training folds. Five target node folds were fixed by a deterministic identity hash, keeping all incoming edges of a target together, and eligibility was fixed across tolerances. A pooled line, separate lines within root groups and a prespecified pooled cubic model used the same supported candidates, with cubic centering and scaling estimated on training data only. For equation (10), both fits were scored on the same withheld target folds, pooling squared errors across folds and using a common total sum of squares about the overall mean loss of the supported edges. Scores were defined only when this denominator was positive, and negative predictive \(R^2\) values were retained. Controls with shared lines, distinct group lines, added noise and shuffled labels are given in Supplementary Note 8.2 and \hyperlink{figure-S11}{Supplementary Fig. 11}.

We separately analyzed response magnitude, zero loss frequency and shape. Relative mean loss divides the summed candidate losses by their strict sum, and shape is the relative residual norm after the best common multiplicative fit of the strict response vector (Supplementary Note 2). When all responses are zero, loss is zero and shape and predictive \(R^2\) are undefined. Synthetic summaries weight physical graphs equally; empirical cases have no interval across independent graphs.

\subsection*{Structural importance benchmark}

All scores used the same observed graph and original indegrees. NPPC counts each connected endpoint pair once. Search path count (SPC) uses roots and actual sinks as endpoints, search path link count (SPLC) uses all nodes and actual sinks, and search path node pair count (SPNP) uses all nodes at both ends. The path count between roots and terminals uses the designated terminals, which can differ from sinks. Betweenness sums the fraction of shortest directed paths through an edge over all pairs or root and terminal pairs \textsuperscript{\hyperlink{ref:25}{25},\hyperlink{ref:26}{26},\hyperlink{ref:29}{29}}. Target outdegree and uniform random ranking complete the structural comparison.

Corrected coverage is deterministic RR coverage restricted to terminals. The observed cascade score coincides with it under strict support and is retained to evaluate propagation under tolerance. When substitution could prevent failure, every static score, including random ranking, received the same filter excluding cuts that could not initiate failure. Each selected edge is cut independently from the intact graph. RIS seed set optimization instead updates scores by marginal joint coverage \textsuperscript{\hyperlink{ref:23}{23},\hyperlink{ref:24}{24}}.

Budgets were \(\lceil f|E_S|\rceil\), where \(f\in\{0.01,0.05,0.10\}\) is the selected fraction. Analyses of distinct targets used analogous target budgets and each target's maximum incoming score. Tied blocks at the cutoff received their exact expected loss under uniform tie breaking. Outcomes were measured separately on all terminals and on unsampled terminals. When all candidate losses in the evaluated terminal population are zero, oracle efficiency is undefined and absolute loss is zero; at full tolerance, BA has five graphs with nonzero full terminal loss but four with nonzero loss on unsampled terminals. Benchmark intervals are exploratory.

\subsection*{Growth model, age preference and prediction}

A newborn selects distinct older parents sequentially with probability proportional to \(\exp(\beta b_i)\), renormalizing after every selection, and different newborns choose independently from a pool fixed within their generation. Summing over ordered cohort sequences gives the probability that every parent lies in a specified subset. Conditional binomial increments propagate cohort counts for root subset ancestry or descendants. Inclusion-exclusion then gives the root count law, and the same parent selection mechanism gives the expected edge cohort weights. These quantities predict mean \((n,K,U)\) counts under hypergeometric terminal sampling. Numerical finite state sums retain the full nonlinear transitions (Supplementary Note 8.1).

The scan over age preference fixed three roots, eight cohorts of 100 nodes, uniform parent counts on one, two and three, and nine equally spaced beta values from \(-\log4\) to \(\log4\), where log denotes the natural logarithm, with 128 independent complete graphs per value. One uniform terminal permutation per graph gave nested samples of 30, 60 and 100, with 30 primary. For each sample, discovery determined candidates and predictor variation determined supported groups. We report predictive gain, with cross validation blocked by target, alongside grouped accuracy, the estimable fraction of graphs, the supported fraction of edges and root class composition. Means and 95\% percentile intervals use 2,000 whole graph resamples, and finite scores are summarized conditional on estimability. A separate experiment at five preferences used 128 fresh graphs per value with five cohorts of 20 nodes to compare exact predictions of mean counts. Configurations, seeds and predictions were frozen before formal graph generation, and graph records retain missing and degenerate states.

The uniform growth experiments cover unbiased attachment. They comprise 24,000 graphs conditional on a fixed edge, 2,048 independent graphs for prospective joint count prediction and 1,024 matched pairs testing parent choice across root branches. A separate posterior test on 100 graphs used known parameters and birth labels with samples of 10, 20 and 40 terminals, with hidden count error at 20 as the primary endpoint. Protocols, controls and uncertainty are detailed in Supplementary Notes 4 and 7 and Supplementary Figs. 4, 5, 7 and 9.

The calibrated age model comparison used 128 complete historical graphs and 512 independent test graphs for each of \(\beta=0,\log2,\log4\), with three roots and five cohorts of 20 nodes. Beta was estimated from the unordered multisets of parent birth generations by summing over possible draw orders. An empirical prior received the same historical graphs, included zero coverage and used a total pseudocount of one per birth cohort. Both priors shared test birth labels, sampled counts and the hypergeometric likelihood. Predictions were frozen before testing, with six of 20 terminals observed. Squared error in hidden counts was averaged over distinct discovered nonterminal targets and then equally over graphs. Efficiency for the top 5\% on unsampled terminals used fractional ties after scores were rounded to 12 decimal places. Paired intervals from 2,000 whole test graph resamples condition on the calibrated parameters. Supplementary Note 8.4 and \hyperlink{figure-S10}{Supplementary Fig. 10} report all comparators, separating count prediction from ranking.

\subsection*{Population prediction and comparison with limited history}

The population forecast used 512 independent networks at each of five preferences from \(-\log4\) to \(\log4\), with three roots, five cohorts of 20 nodes and uniform parent counts on one, two and three. Predictions were frozen before generation, and nested samples contained 2, 6 or 20 terminals. Conditional moments and optimal linear risks were computed from the expected edge counts restricted to discovered edges. Pointwise intervals used 1,000 whole graph resamples, and discrete predictive coverage was checked against its calculated mass (Supplementary Note 9).

The limited history test used 16 independent calibration and test blocks for each of \(\beta=-\log2,0,\log2\), with the same smaller graph schedule. Each block had 128 calibration graphs, nested histories of 16 graphs and 128 fresh test graphs, and all methods shared the histories, true birth labels and hypergeometric likelihood. The stronger empirical prior selected the shrinkage between a histogram and a beta-binomial distribution using log scores over four folds of calibration graphs. Fitted priors were frozen before test generation. The primary squared error of hidden counts weighted distinct discovered nonterminal targets within graphs, and ranking used independent cuts with fractional cutoff ties. Both budgets, all comparators and non-estimable cases were kept. Ten thousand paired block resamples within preferences captured calibration and testing variation; aggregate summaries weighted preferences equally. Intervals are descriptive and unadjusted for multiplicity (Supplementary Note 10 and \hyperlink{figure-S13}{Supplementary Fig. 13}).

\subsection*{Measured regulatory perturbations}

An independent RegulonDB annotation, projected to positive contributions between distinct operons, supplied 694 nodes and 988 edges. We compared each transcription factor (TF) relation bundle with promoter responses measured in glucose medium by pooled promoter responses to TF perturbation sequencing (PPTP-seq) \textsuperscript{\hyperlink{ref:2}{2}}. Edges were selected independently of these responses. Ten views each sampled 154 of 309 candidate outputs and left 155 unsampled, and a fixed availability screen with a deterministic split by TF identity left 77 test TFs. The outcome was the mean decrease in log base two activity among valid measurements on unsampled outputs, and the primary budget selected eight TFs. Supplementary Note 6 details ties, missingness, comparator definitions and sensitivity results. The repeated views describe sampling variation within one biological system.

\subsection*{Verification}

Frozen graph identities, seeds, source hashes and scoring universes were retained. Independent checks covered direct cascade propagation, classical scores, exact enumeration of small growth graphs and posterior forecasts. Independent statistical units were whole networks or calibration blocks. Supplementary Notes 4 and 6 to 13 give the mathematical procedures, and the archive's Verification and Provenance Record keeps check counts, reproducibility details and replay limits.

\subsection*{Use of large language models}

Claude (Anthropic) and OpenAI Codex assisted with mathematical analysis, literature searches, code development, computational checks, figure preparation, and manuscript drafting and editing. The authors retain responsibility for the scientific content and conclusions.

\FloatBarrier\section*{Data availability}
\addcontentsline{toc}{section}{Data availability}

OpenAlex, SNAP HEP-TH and Bioconductor are the public sources cited in Methods. River topology and station metadata came from \href{https://waterdata.usgs.gov/blog/nldi-intro}{USGS NLDI}, \href{https://api.water.usgs.gov/docs/fabric-pygeoapi/}{Hydrologic Fabric} and Water Data services. PPTP-seq responses and author-archived regulatory annotations are available in the \href{https://github.com/yhan0410/PPTP-seq_ECOLI}{author archive} \textsuperscript{\hyperlink{ref:2}{2}}. COVID-19 reference topology came from the \href{https://github.com/DrakeLab/taube-transmission-trees}{OutbreakTrees author archive} \textsuperscript{\hyperlink{ref:7}{7}}. The study uses frozen records and the specified transformations. Accompanying Source Data contain figure tables and panel mappings. Source provenance and dataset selections are indexed in the code and data package in the \href{https://github.com/ruiwuniu/cautree-reproducibility}{private reproducibility repository}. Editors and reviewers can request access through the corresponding author; access must be granted by the repository owner. Third-party inputs that are not redistributed are identified with their original sources in the package documentation.

\FloatBarrier\section*{Code availability}
\addcontentsline{toc}{section}{Code availability}

Generators, analysis scripts, fixed sampling identities and figure source mappings are provided in the \href{https://github.com/ruiwuniu/cautree-reproducibility}{private reproducibility repository}. Its README describes the environment, verification coverage and dependencies needed for reproduction. Editors and reviewers can request access through the corresponding author; access must be granted by the repository owner.

\FloatBarrier\section*{References}
\addcontentsline{toc}{section}{References}

\begin{samepage}\noindent\hypertarget{ref:1}{}1. Ebenhöh, O., Handorf, T. \& Heinrich, R. Structural analysis of expanding metabolic networks. \emph{Genome Informatics} \textbf{15}, 35–45 (2004). \href{https://doi.org/10.11234/gi1990.15.35}{doi:10.11234/gi1990.15.35}.\par\vspace{-6pt}\end{samepage}
\begin{samepage}\noindent\hypertarget{ref:2}{}2. Han, Y., Li, W., Filko, A., Li, J. \& Zhang, F. Genome-wide promoter responses to CRISPR perturbations of regulators reveal regulatory networks in Escherichia coli. \emph{Nature Communications} \textbf{14}, 5757 (2023). \href{https://doi.org/10.1038/s41467-023-41572-4}{\begingroup\urlstyle{same}\nolinkurl{10.1038/s41467-023-41572-4}\endgroup}.\par\vspace{-6pt}\end{samepage}
\begin{samepage}\noindent\hypertarget{ref:3}{}3. Pichler, A., Pangallo, M., del Rio-Chanona, R. M., Lafond, F. \& Farmer, J. D. Forecasting the propagation of pandemic shocks with a dynamic input-output model. \emph{Journal of Economic Dynamics and Control} \textbf{144}, 104527 (2022). \href{https://doi.org/10.1016/j.jedc.2022.104527}{doi:10.1016/j.jedc.2022.104527}.\par\vspace{-6pt}\end{samepage}
\begin{samepage}\noindent\hypertarget{ref:4}{}4. R Core Team. \emph{Writing R Extensions}, Section 1.1.3, Package Dependencies. \href{https://stat.ethz.ch/CRAN/doc/manuals/r-release/R-exts.html#Package-Dependencies}{Official manual}. Accessed 14 September 2026.\par\vspace{-6pt}\end{samepage}
\begin{samepage}\noindent\hypertarget{ref:5}{}5. CRAN Repository Maintainers. \emph{CRAN Repository Policy}, revision 6875, Source packages. \href{https://stat.ethz.ch/CRAN/web/packages/policies.html}{Official policy}. Accessed 14 September 2026.\par\vspace{-6pt}\end{samepage}
\begin{samepage}\noindent\hypertarget{ref:6}{}6. Yabe, T., García Bulle Bueno, B., Frank, M. R., Pentland, A. \& Moro, E. Behaviour-based dependency networks between places shape urban economic resilience. \emph{Nature Human Behaviour} \textbf{9}, 496–506 (2025). \href{https://doi.org/10.1038/s41562-024-02072-7}{\begingroup\urlstyle{same}\nolinkurl{10.1038/s41562-024-02072-7}\endgroup}.\par\vspace{-6pt}\end{samepage}
\begin{samepage}\noindent\hypertarget{ref:7}{}7. Taube, J. C., Miller, P. B. \& Drake, J. M. An open-access database of infectious disease transmission trees to explore superspreader epidemiology. \emph{PLOS Biology} \textbf{20}, e3001685 (2022). \href{https://doi.org/10.1371/journal.pbio.3001685}{\begingroup\urlstyle{same}\nolinkurl{10.1371/journal.pbio.3001685}\endgroup}.\par\vspace{-6pt}\end{samepage}
\begin{samepage}\noindent\hypertarget{ref:8}{}8. Ferretti, L. et al. Quantifying SARS-CoV-2 transmission suggests epidemic control with digital contact tracing. \emph{Science} \textbf{368}, eabb6936 (2020). \href{https://doi.org/10.1126/science.abb6936}{\begingroup\urlstyle{same}\nolinkurl{10.1126/science.abb6936}\endgroup}.\par\vspace{-6pt}\end{samepage}
\begin{samepage}\noindent\hypertarget{ref:9}{}9. Carraro, L., Hartikainen, H., Jokela, J., Bertuzzo, E. \& Rinaldo, A. Estimating species distribution and abundance in river networks using environmental DNA. \emph{Proceedings of the National Academy of Sciences} \textbf{115}, 11724–11729 (2018). \href{https://doi.org/10.1073/pnas.1813843115}{\begingroup\urlstyle{same}\nolinkurl{10.1073/pnas.1813843115}\endgroup}.\par\vspace{-6pt}\end{samepage}
\begin{samepage}\noindent\hypertarget{ref:10}{}10. Wu, X. et al. Synchronization in multiplex networks. \emph{Physics Reports} \textbf{1060}, 1–54 (2024). \href{https://doi.org/10.1016/j.physrep.2024.01.005}{\begingroup\urlstyle{same}\nolinkurl{10.1016/j.physrep.2024.01.005}\endgroup}.\par\vspace{-6pt}\end{samepage}
\begin{samepage}\noindent\hypertarget{ref:11}{}11. Parshani, R., Buldyrev, S. V. \& Havlin, S. Critical effect of dependency groups on the function of networks. \emph{Proceedings of the National Academy of Sciences} \textbf{108}, 1007–1010 (2011). \href{https://doi.org/10.1073/pnas.1008404108}{\begingroup\urlstyle{same}\nolinkurl{10.1073/pnas.1008404108}\endgroup}.\par\vspace{-6pt}\end{samepage}
\begin{samepage}\noindent\hypertarget{ref:12}{}12. Bashan, A., Parshani, R. \& Havlin, S. Percolation in networks composed of connectivity and dependency links. \emph{Physical Review E} \textbf{83}, 051127 (2011). \href{https://doi.org/10.1103/PhysRevE.83.051127}{\begingroup\urlstyle{same}\nolinkurl{10.1103/PhysRevE.83.051127}\endgroup}.\par\vspace{-6pt}\end{samepage}
\begin{samepage}\noindent\hypertarget{ref:13}{}13. Watts, D. J. A simple model of global cascades on random networks. \emph{Proceedings of the National Academy of Sciences} \textbf{99}, 5766-5771 (2002). \href{https://doi.org/10.1073/pnas.082090499}{\begingroup\urlstyle{same}\nolinkurl{10.1073/pnas.082090499}\endgroup}.\par\vspace{-6pt}\end{samepage}
\begin{samepage}\noindent\hypertarget{ref:14}{}14. Buldyrev, S. V., Parshani, R., Paul, G., Stanley, H. E. \& Havlin, S. Catastrophic cascade of failures in interdependent networks. \emph{Nature} \textbf{464}, 1025-1028 (2010). \href{https://doi.org/10.1038/nature08932}{\begingroup\urlstyle{same}\nolinkurl{10.1038/nature08932}\endgroup}.\par\vspace{-6pt}\end{samepage}
\begin{samepage}\noindent\hypertarget{ref:15}{}15. Di Muro, M. A. et al. Cascading Failures in Interdependent Networks with Multiple Supply-Demand Links and Functionality Thresholds. \emph{Scientific Reports} \textbf{7}, 15059 (2017). \href{https://doi.org/10.1038/s41598-017-14384-y}{\begingroup\urlstyle{same}\nolinkurl{10.1038/s41598-017-14384-y}\endgroup}.\par\vspace{-6pt}\end{samepage}
\begin{samepage}\noindent\hypertarget{ref:16}{}16. Ganin, A. A. et al. Operational resilience: concepts, design and analysis. \emph{Scientific Reports} \textbf{6}, 19540 (2016). \href{https://doi.org/10.1038/srep19540}{\begingroup\urlstyle{same}\nolinkurl{10.1038/srep19540}\endgroup}.\par\vspace{-6pt}\end{samepage}
\begin{samepage}\noindent\hypertarget{ref:17}{}17. Dall’Asta, L., Alvarez-Hamelin, I., Barrat, A., Vázquez, A. \& Vespignani, A. Exploring networks with traceroute-like probes: Theory and simulations. \emph{Theoretical Computer Science} \textbf{355(1)}, 6–24 (2006). \href{https://doi.org/10.1016/j.tcs.2005.12.009}{\begingroup\urlstyle{same}\nolinkurl{10.1016/j.tcs.2005.12.009}\endgroup}.\par\vspace{-6pt}\end{samepage}
\begin{samepage}\noindent\hypertarget{ref:18}{}18. Achlioptas, D., Clauset, A., Kempe, D. \& Moore, C. On the bias of traceroute sampling: Or, power-law degree distributions in regular graphs. \emph{Journal of the ACM} \textbf{56(4)}, 21:1–21:28 (2009). \href{https://doi.org/10.1145/1538902.1538905}{\begingroup\urlstyle{same}\nolinkurl{10.1145/1538902.1538905}\endgroup}.\par\vspace{-6pt}\end{samepage}
\begin{samepage}\noindent\hypertarget{ref:19}{}19. Kim, M. \& Leskovec, J. The Network Completion Problem: Inferring Missing Nodes and Edges in Networks. \emph{Proceedings of the 2011 SIAM International Conference on Data Mining}, 47–58 (2011). \href{https://doi.org/10.1137/1.9781611972818.5}{\begingroup\urlstyle{same}\nolinkurl{10.1137/1.9781611972818.5}\endgroup}.\par\vspace{-6pt}\end{samepage}
\begin{samepage}\noindent\hypertarget{ref:20}{}20. Peixoto, T. P. Reconstructing Networks with Unknown and Heterogeneous Errors. \emph{Physical Review X} \textbf{8}, 041011 (2018). \href{https://doi.org/10.1103/PhysRevX.8.041011}{\begingroup\urlstyle{same}\nolinkurl{10.1103/PhysRevX.8.041011}\endgroup}.\par\vspace{-6pt}\end{samepage}
\begin{samepage}\noindent\hypertarget{ref:21}{}21. Peixoto, T. P. Network Reconstruction via the Minimum Description Length Principle. \emph{Physical Review X} \textbf{15}, 011065 (2025). \href{https://doi.org/10.1103/PhysRevX.15.011065}{\begingroup\urlstyle{same}\nolinkurl{10.1103/PhysRevX.15.011065}\endgroup}.\par\vspace{-6pt}\end{samepage}
\begin{samepage}\noindent\hypertarget{ref:22}{}22. Kempe, D., Kleinberg, J. \& Tardos, É. Maximizing the spread of influence through a social network. In \emph{Proceedings of the Ninth ACM SIGKDD International Conference on Knowledge Discovery and Data Mining} 137–146 (ACM, 2003). \href{https://doi.org/10.1145/956750.956769}{\begingroup\urlstyle{same}\nolinkurl{10.1145/956750.956769}\endgroup}.\par\vspace{-6pt}\end{samepage}
\begin{samepage}\noindent\hypertarget{ref:23}{}23. Borgs, C., Brautbar, M., Chayes, J. \& Lucier, B. Maximizing Social Influence in Nearly Optimal Time. \emph{Proceedings of the Twenty-Fifth Annual ACM-SIAM Symposium on Discrete Algorithms}, 946–957 (2014). \href{https://doi.org/10.1137/1.9781611973402.70}{\begingroup\urlstyle{same}\nolinkurl{10.1137/1.9781611973402.70}\endgroup}.\par\vspace{-6pt}\end{samepage}
\begin{samepage}\noindent\hypertarget{ref:24}{}24. Tang, Y., Xiao, X. \& Shi, Y. Influence Maximization: Near-Optimal Time Complexity Meets Practical Efficiency. \emph{Proceedings of the 2014 ACM SIGMOD International Conference on Management of Data}, 75–86 (2014). \href{https://doi.org/10.1145/2588555.2593670}{\begingroup\urlstyle{same}\nolinkurl{10.1145/2588555.2593670}\endgroup}.\par\vspace{-6pt}\end{samepage}
\begin{samepage}\noindent\hypertarget{ref:25}{}25. Hummon, N. P. \& Doreian, P. Connectivity in a citation network: The development of DNA theory. \emph{Social Networks} \textbf{11}, 39–63 (1989). \href{https://doi.org/10.1016/0378-8733(89)90017-8}{doi:10.1016/0378-8733(89)90017-8}.\par\vspace{-6pt}\end{samepage}
\begin{samepage}\noindent\hypertarget{ref:26}{}26. Batagelj, V. \emph{Efficient Algorithms for Citation Network Analysis}. arXiv:cs/0309023, version 1 (2003). \href{https://arxiv.org/abs/cs/0309023v1}{Original preprint}.\par\vspace{-6pt}\end{samepage}
\begin{samepage}\noindent\hypertarget{ref:27}{}27. Sabrin, K. M. \& Dovrolis, C. The hourglass effect in hierarchical dependency networks. \emph{Network Science} \textbf{5}, 490–528 (2017). \href{https://doi.org/10.1017/nws.2017.22}{\begingroup\urlstyle{same}\nolinkurl{10.1017/nws.2017.22}\endgroup}.\par\vspace{-6pt}\end{samepage}
\begin{samepage}\noindent\hypertarget{ref:28}{}28. Šubelj, L., Waltman, L., Traag, V. \& van Eck, N. J. Intermediacy of publications. \emph{Royal Society Open Science} \textbf{7}, 190207 (2020). \href{https://doi.org/10.1098/rsos.190207}{\begingroup\urlstyle{same}\nolinkurl{10.1098/rsos.190207}\endgroup}.\par\vspace{-6pt}\end{samepage}
\begin{samepage}\noindent\hypertarget{ref:29}{}29. Brandes, U. On variants of shortest-path betweenness centrality and their generic computation. \emph{Social Networks} \textbf{30}, 136–145 (2008). \href{https://doi.org/10.1016/j.socnet.2007.11.001}{doi:10.1016/j.socnet.2007.11.001}.\par\vspace{-6pt}\end{samepage}
\begin{samepage}\noindent\hypertarget{ref:30}{}30. Abate, P., Boender, J., Di Cosmo, R. \& Zacchiroli, S. Strong dependencies between software components. In \emph{2009 3rd International Symposium on Empirical Software Engineering and Measurement} 89–99 (IEEE, 2009). \href{https://doi.org/10.1109/ESEM.2009.5316017}{\begingroup\urlstyle{same}\nolinkurl{10.1109/ESEM.2009.5316017}\endgroup}.\par\vspace{-6pt}\end{samepage}
\begin{samepage}\noindent\hypertarget{ref:31}{}31. Setó-Rey, D., Santos-Martín, J. I. \& López-Nozal, C. Vulnerability of Package Dependency Networks. \emph{IEEE Transactions on Network Science and Engineering} \textbf{10}, 3396–3408 (2023). \href{https://doi.org/10.1109/TNSE.2023.3260880}{\begingroup\urlstyle{same}\nolinkurl{10.1109/TNSE.2023.3260880}\endgroup}.\par\vspace{-6pt}\end{samepage}
\begin{samepage}\noindent\hypertarget{ref:32}{}32. Hamedmoghadam, H., Jalili, M., Vu, H. L. \& Stone, L. Percolation of heterogeneous flows uncovers the bottlenecks of infrastructure networks. \emph{Nature Communications} \textbf{12}, 1254 (2021). \href{https://doi.org/10.1038/s41467-021-21483-y}{\begingroup\urlstyle{same}\nolinkurl{10.1038/s41467-021-21483-y}\endgroup}.\par\vspace{-6pt}\end{samepage}
\begin{samepage}\noindent\hypertarget{ref:33}{}33. Chulahwat, A., Hassan, E. M., Tariverdi, M., Nunez-del-Prado, M. \& Mahmoud, H. Identifying influential assets in higher order interdependent infrastructure networks through population impact. \emph{Scientific Reports} \textbf{15}, 33351 (2025). \href{https://doi.org/10.1038/s41598-025-15824-w}{\begingroup\urlstyle{same}\nolinkurl{10.1038/s41598-025-15824-w}\endgroup}.\par\vspace{-6pt}\end{samepage}
\begin{samepage}\noindent\hypertarget{ref:34}{}34. Reisch, T., Heiler, G., Diem, C., Klimek, P. \& Thurner, S. Monitoring supply networks from mobile phone data for estimating the systemic risk of an economy. \emph{Scientific Reports} \textbf{12}, 13347 (2022). \href{https://doi.org/10.1038/s41598-022-13104-5}{\begingroup\urlstyle{same}\nolinkurl{10.1038/s41598-022-13104-5}\endgroup}.\par\vspace{-6pt}\end{samepage}
\begin{samepage}\noindent\hypertarget{ref:35}{}35. Fessina, M. et al. Inferring firm-level supply chain networks with realistic systemic risk from industry sector-level data. \emph{Scientific Reports} \textbf{16}, 19848 (2026). \href{https://doi.org/10.1038/s41598-026-47883-y}{\begingroup\urlstyle{same}\nolinkurl{10.1038/s41598-026-47883-y}\endgroup}.\par\vspace{-6pt}\end{samepage}
\begin{samepage}\noindent\hypertarget{ref:36}{}36. Chakraborty, A., Reisch, T., Diem, C., Astudillo-Estévez, P. \& Thurner, S. Inequality in economic shock exposures across the global firm-level supply network. \emph{Nature Communications} \textbf{15}, 3348 (2024). \href{https://doi.org/10.1038/s41467-024-46126-w}{\begingroup\urlstyle{same}\nolinkurl{10.1038/s41467-024-46126-w}\endgroup}.\par\vspace{-6pt}\end{samepage}
\begin{samepage}\noindent\hypertarget{ref:37}{}37. Priem, J., Piwowar, H. \& Orr, R. OpenAlex: A fully-open index of scholarly works, authors, venues, institutions, and concepts. \emph{26th International Conference on Science, Technology and Innovation Indicators (STI 2022)} (2022). \href{https://doi.org/10.5281/zenodo.6936227}{doi:10.5281/zenodo.6936227}.\par\vspace{-6pt}\end{samepage}
\begin{samepage}\noindent\hypertarget{ref:38}{}38. Su, S., Ramos, M. \& Davis, S. \emph{BiocPkgTools: Collection of Simple Tools for Learning about Bioconductor Packages}. Bioconductor reference manual, version 1.30.1 (2026). \href{https://www.bioconductor.org/packages/release/bioc/manuals/BiocPkgTools/man/BiocPkgTools.pdf}{Official manual}. Accessed 22 September 2026; cited for dependency definitions.\par\vspace{-6pt}\end{samepage}
\begin{samepage}\noindent\hypertarget{ref:39}{}39. Marichal, J.-L. On Sugeno integral as an aggregation function. \emph{Fuzzy Sets and Systems} \textbf{114}, 347–365 (2000). \href{https://doi.org/10.1016/S0165-0114(98)00116-X}{\begingroup\urlstyle{same}\nolinkurl{10.1016/S0165-0114(98)00116-X}\endgroup}.\par\vspace{-6pt}\end{samepage}
\begin{samepage}\noindent\hypertarget{ref:40}{}40. Lengauer, T. \& Tarjan, R. E. A fast algorithm for finding dominators in a flowgraph. \emph{ACM Transactions on Programming Languages and Systems} \textbf{1}, 121-141 (1979). \href{https://doi.org/10.1145/357062.357071}{\begingroup\urlstyle{same}\nolinkurl{10.1145/357062.357071}\endgroup}.\par\vspace{-6pt}\end{samepage}
\begin{samepage}\noindent\hypertarget{ref:41}{}41. Jiang, X. \& Zhuge, H. Forward search path count as an alternative indirect citation impact indicator. \emph{Journal of Informetrics} \textbf{13}, 100977 (2019). \href{https://doi.org/10.1016/j.joi.2019.100977}{\begingroup\urlstyle{same}\nolinkurl{10.1016/j.joi.2019.100977}\endgroup}.\par\vspace{-6pt}\end{samepage}
\begin{samepage}\noindent\hypertarget{ref:42}{}42. Dorogovtsev, S. N. \& Mendes, J. F. F. Evolution of networks with aging of sites. \emph{Physical Review E} \textbf{62}, 1842–1845 (2000). \href{https://doi.org/10.1103/PhysRevE.62.1842}{\begingroup\urlstyle{same}\nolinkurl{10.1103/PhysRevE.62.1842}\endgroup}.\par\vspace{-6pt}\end{samepage}
\begin{samepage}\noindent\hypertarget{ref:43}{}43. Karrer, B. \& Newman, M. E. J. Random Acyclic Networks. \emph{Physical Review Letters} \textbf{102}, 128701 (2009). \href{https://doi.org/10.1103/PhysRevLett.102.128701}{\begingroup\urlstyle{same}\nolinkurl{10.1103/PhysRevLett.102.128701}\endgroup}.\par\vspace{-6pt}\end{samepage}
\begin{samepage}\noindent\hypertarget{ref:44}{}44. Janson, S. The number of descendants in a random directed acyclic graph. \emph{Random Structures \& Algorithms} \textbf{64}, 768–803 (2024). \href{https://doi.org/10.1002/rsa.21195}{\begingroup\urlstyle{same}\nolinkurl{10.1002/rsa.21195}\endgroup}.\par\vspace{-6pt}\end{samepage}
\begin{samepage}\noindent\hypertarget{ref:45}{}45. Burghart, F. Ancestries and Descendants in a Random DAG. \emph{37th International Conference on Probabilistic, Combinatorial and Asymptotic Methods for the Analysis of Algorithms (AofA 2026)}, Leibniz International Proceedings in Informatics \textbf{381}, 7:1–7:14 (2026). \href{https://doi.org/10.4230/LIPIcs.AofA.2026.7}{\begingroup\urlstyle{same}\nolinkurl{10.4230/LIPIcs.AofA.2026.7}\endgroup}.\par\vspace{-6pt}\end{samepage}
\begin{samepage}\noindent\hypertarget{ref:46}{}46. Mariani, M. S., Medo, M. \& Zhang, Y.-C. Ranking nodes in growing networks: When PageRank fails. \emph{Scientific Reports} \textbf{5}, 16181 (2015). \href{https://doi.org/10.1038/srep16181}{\begingroup\urlstyle{same}\nolinkurl{10.1038/srep16181}\endgroup}.\par\vspace{-6pt}\end{samepage}
\begin{samepage}\noindent\hypertarget{ref:47}{}47. Mao, T., Zhang, S., Hanjalic, A. \& Wang, H. Estimating nodal spreading influence using partial temporal networks. \emph{Chaos, Solitons \& Fractals} \textbf{201}, 117163 (2025). \href{https://doi.org/10.1016/j.chaos.2025.117163}{\begingroup\urlstyle{same}\nolinkurl{10.1016/j.chaos.2025.117163}\endgroup}.\par\vspace{-6pt}\end{samepage}
\begin{samepage}\noindent\hypertarget{ref:48}{}48. Ericson, W. A. Subjective Bayesian Models in Sampling Finite Populations. \emph{Journal of the Royal Statistical Society: Series B (Methodological)} \textbf{31}, 195–224 (1969). \href{https://doi.org/10.1111/j.2517-6161.1969.tb00782.x}{\begingroup\urlstyle{same}\nolinkurl{10.1111/j.2517-6161.1969.tb00782.x}\endgroup}.\par\vspace{-6pt}\end{samepage}
\begin{samepage}\noindent\hypertarget{ref:49}{}49. Rodríguez, C. E., Walker, S. G. \& Mena, R. H. A finite population parametric Bayesian bootstrap for quick counts. \emph{The Annals of Applied Statistics} \textbf{19}, 3394–3415 (2025). \href{https://doi.org/10.1214/25-AOAS2066}{\begingroup\urlstyle{same}\nolinkurl{10.1214/25-AOAS2066}\endgroup}.\par\vspace{-6pt}\end{samepage}
\begin{samepage}\noindent\hypertarget{ref:50}{}50. Yoo, E., Craighead, C. W. \& Samtani, S. Dependency Network Structure and Security Vulnerabilities in Software Supply Chains. \emph{Journal of Management Information Systems} \textbf{42}, 1149–1176 (2025). \href{https://doi.org/10.1080/07421222.2025.2561385}{\begingroup\urlstyle{same}\nolinkurl{10.1080/07421222.2025.2561385}\endgroup}.\par\vspace{-6pt}\end{samepage}
\begin{samepage}\noindent\hypertarget{ref:51}{}51. Diem, C., Borsos, A., Reisch, T., Kertész, J. \& Thurner, S. Quantifying firm-level economic systemic risk from nation-wide supply networks. \emph{Scientific Reports} \textbf{12}, 7719 (2022). \href{https://doi.org/10.1038/s41598-022-11522-z}{\begingroup\urlstyle{same}\nolinkurl{10.1038/s41598-022-11522-z}\endgroup}.\par\vspace{-6pt}\end{samepage}
\begin{samepage}\noindent\hypertarget{ref:52}{}52. Elliott, M., Golub, B. \& Leduc, M. V. Supply Network Formation and Fragility. \emph{American Economic Review} \textbf{112}, 2701-2747 (2022). \href{https://doi.org/10.1257/aer.20210220}{\begingroup\urlstyle{same}\nolinkurl{10.1257/aer.20210220}\endgroup}.\par\vspace{-6pt}\end{samepage}
\begin{samepage}\noindent\hypertarget{ref:53}{}53. Schäfer, B., Witthaut, D., Timme, M. \& Latora, V. Dynamically induced cascading failures in power grids. \emph{Nature Communications} \textbf{9}, 1975 (2018). \href{https://doi.org/10.1038/s41467-018-04287-5}{\begingroup\urlstyle{same}\nolinkurl{10.1038/s41467-018-04287-5}\endgroup}.\par\vspace{-6pt}\end{samepage}
\begin{samepage}\noindent\hypertarget{ref:54}{}54. Erdős, P. \& Rényi, A. On random graphs. I. \emph{Publicationes Mathematicae Debrecen} \textbf{6}, 290–297 (1959). \href{https://doi.org/10.5486/PMD.1959.6.3-4.12}{doi:10.5486/PMD.1959.6.3-4.12}.\par\vspace{-6pt}\end{samepage}
\begin{samepage}\noindent\hypertarget{ref:55}{}55. Gilbert, E. N. Random graphs. \emph{The Annals of Mathematical Statistics} \textbf{30}, 1141–1144 (1959). \href{https://doi.org/10.1214/aoms/1177706098}{Original article}.\par\vspace{-6pt}\end{samepage}
\begin{samepage}\noindent\hypertarget{ref:56}{}56. Barabási, A.-L. \& Albert, R. Emergence of scaling in random networks. \emph{Science} \textbf{286}, 509–512 (1999). \href{https://doi.org/10.1126/science.286.5439.509}{doi:10.1126/science.286.5439.509}.\par\vspace{-6pt}\end{samepage}
\begin{samepage}\noindent\hypertarget{ref:57}{}57. Stanford Network Analysis Project. \emph{High-energy physics theory citation network}. \href{https://snap.stanford.edu/data/cit-HepTh.html}{Official dataset page}. Accessed 12 September 2026.\par\vspace{-6pt}\end{samepage}

\FloatBarrier\section*{Acknowledgements}
\addcontentsline{toc}{section}{Acknowledgements}

This work was supported by the National Natural Science Foundation of China under Grants 62573301 and 62503335, the Startup Grants from Shenzhen University, and the Guangdong Provincial Key Laboratory of Visual Media and Multidimensional Intelligence (2026B1212010012).

\FloatBarrier\section*{Author contributions}
\addcontentsline{toc}{section}{Author contributions}

Ruiwu Niu: Conceptualization, Data curation, Formal analysis, Investigation, Methodology, Project administration, Resources, Software, Supervision, Validation, Visualization, Writing – original draft, Writing – review \& editing.

Muhammad Azhar: Validation, Writing – review \& editing.

Michaël Antonie van Wyk: Validation, Writing – review \& editing.

Xiaoqun Wu: Validation, Resources, Writing – review \& editing.

\FloatBarrier\section*{Competing interests}
\addcontentsline{toc}{section}{Competing interests}

The authors declare no competing interests.

\clearpage
\fancyhead[R]{\small Supplementary information}

\begin{center}
{\Large\bfseries Supplementary information\par}\medskip
{\LARGE Tracing dependency networks explains how origins shape downstream vulnerability\par}\medskip
{\large \mbox{Ruiwu Niu\textsuperscript{1}}, \mbox{Muhammad Azhar\textsuperscript{1}}, \mbox{Michaël Antonie van Wyk\textsuperscript{2}}, \mbox{Xiaoqun Wu\textsuperscript{3,*}}\par}\medskip
\begin{minipage}{0.98\linewidth}\centering\fontsize{9.5}{11.5}\selectfont
\textsuperscript{1}Department of Data Science and Digital Innovation, Hong Kong Shue Yan University, Hong Kong SAR, China\par\smallskip
\textsuperscript{2}School of Electrical and Information Engineering, University of the Witwatersrand, Johannesburg, South Africa\par\smallskip
\textsuperscript{3}College of Computer Science and Software Engineering, Shenzhen University, Shenzhen, China\par\smallskip
\medskip\textsuperscript{*}Correspondence to Xiaoqun Wu (\href{mailto:xqwu@szu.edu.cn}{xqwu@szu.edu.cn}).\par\smallskip
Ruiwu Niu (\href{mailto:rniu@hksyu.edu}{rniu@hksyu.edu}); Muhammad Azhar (\href{mailto:azhar@hksyu.edu}{azhar@hksyu.edu})\par
Michaël Antonie van Wyk (\href{mailto:Anton.VanWyk@wits.ac.za}{Anton.VanWyk@wits.ac.za})\par
\end{minipage}\par
\end{center}\medskip

Supplementary Notes 1 and 2 extend the recovery and intervention results, Notes 3 to 5 give derivations, methods and comparisons with prior work, and Notes 6 to 13 report measured regulatory responses, growth predictions, temporal diagnostics and edge score comparisons. References to main figures use the numbering in the Article, and supplementary equations carry the prefix S.

\textbf{Evidence types and statistical units.} Synthetic graph ensembles test the specified mathematical models. Empirical citation, river and transmission topologies support structural calculations under an imposed dependency rule, software dependency relations support counterfactual analysis with fixed requirements, and TF perturbations provide measured activity changes for relation bundles. These evidence types answer different questions. The unit of replication in each design is a generated graph, a matched graph pair or a calibration and test block, and repeated observations of one empirical topology describe sampling variation within a single system. Figure and table descriptions state their denominators and how their intervals should be read. Verification records, source and sample identities, random seeds and execution history are kept in the accompanying archive package.

Throughout, a directed acyclic graph (DAG) is a finite network \(G=(V_G,E_G)\) with node set \(V_G\) and directed edge set \(E_G\). The root set \(\mathcal R_G\subseteq V_G\) contains all original nodes without incoming edges, which are sustained externally. The nonempty designated terminal set \(\mathcal T_G\subseteq V_G\) contains the outcomes to be assessed, which need not be sinks. Their counts are the positive integers \(R_G=|\mathcal R_G|\) and \(T_G=|\mathcal T_G|\). A nonempty sample \(\mathcal T_S\subseteq\mathcal T_G\) has integer size \(m=T_S\), with \(1\le m\le T_G\), and its complete predecessor closure is the observed graph \(S=(V_S,E_S)\). Finite query boundaries are identified explicitly wherever complete closure is not assumed.

For a scored graph \(H\in\{G,S\}\) and edge \(e=(u\to v)\in E_H\), \(n_H(e)\in\{1,\ldots,R_H\}\) counts roots reaching its tail and \(U_H(e)\in\{0,\ldots,T_H\}\) counts designated terminals reachable from its head, including paths of length zero. The distinct pair count is \(C_H(e)=n_H(e)U_H(e)\), and its normalized value \(\bar C_H(e)=C_H(e)/(R_HT_H)\) lies in \([0,1]\). For a discovered edge in a complete closure we write \(T=T_G\), \(U=U_G(e)\), \(k_e=U_S(e)\) and \(n=n_G(e)=n_S(e)\), so that \(1\le k_e\le U\). Uppercase \(K\) denotes the random sampled count and lowercase \(k\) or \(k_e\) its realized value. Terminal coverage is \(W_H=U_H/T_H\in[0,1]\).

The tolerance \(\tau\in[0,1]\) sets the support rule in main equation (2), and each cut starts from the intact graph, giving a terminal loss fraction \(I_e(\tau)\in[0,1]\). A nonroot node \(v\) has original indegree \(d_v\in\{1,2,\ldots\}\) and \(m_v\in\{0,\ldots,d_v\}\) unavailable inputs. Strict support corresponds to \(\tau=0\), and at \(\tau=1\) one surviving input suffices. Binomial coefficients in the finite sampling formulas are zero whenever the requested count lies outside the feasible integer range. Regression \(R^2\) denotes the coefficient of determination; it is undefined for a constant response and can be negative for predictions on withheld observations.

\FloatBarrier\section*{Supplementary Note 1. Recovery details and finite reference diagnostics}
\addcontentsline{toc}{section}{Supplementary Note 1. Recovery details and finite reference diagnostics}

\subsection*{1.1 Root discovery and complete ranking require different observation levels}

The planted root design, specified in Methods under “Planted and empirical recovery designs”, contains 1,620 independent DAGs, each weighted equally. Supplementary Note 4.2 gives the sampling grids, rounding and trajectory summaries.

Main \hyperlink{figure-2}{Fig.~2} compares the discovery of at least 90 planted roots with stable recovery of the complete pair ranking. Stable recovery requires a score of at least 0.90 at every subsequent registered sample size along the same order, and summaries average within graphs before averaging across graphs. Root visibility and full ranking recovery concern different target sets.

The sampling fraction at which mean stable success first reaches 95\% is a different summary from the mean first stable fraction. Shared-edge correlations omit undiscovered reference edges, whereas full-reference correlations score them zero. Main \hyperlink{figure-2}{Fig.~2} therefore shows that ordering among discovered edges can become accurate before all reference edges have been found.

\subsection*{1.2 A finite empirical reference retains a coverage gap}

The finite OpenAlex reference contains 50 works, ten from each year 2021 to 2025 \textsuperscript{\hyperlink{supp-ref:1}{1}}. Its 512 stratified terminal orders give samples of 5, 10, ..., 50 works, each replayed to a boundary two steps from the sample (Methods).

With 45 of the 50 works, mean backbone precision and recall were 1.000 and 0.891 (Supplementary Table 1), and the full cohort recovered its reference exactly. Terminal support defines membership in the backbone independently of whether a node has zero indegree.

At 45 works, median shared-edge and full-reference pair correlations were 0.986 and 0.819, and only 14 of the 512 orders reached a full-reference correlation of 0.90. The first registered point with at least 95\% stable success was the full cohort, placing the crossing in \((45,50]\) (main \hyperlink{figure-3}{Fig.~3}).

\par\begin{minipage}{\linewidth}
\textbf{Supplementary Table 1. Recovery of a finite OpenAlex reference.} Backbone precision and recall are means over 512 terminal orders, and pair ranking correlations are medians over the same orders; the statistics describe different target sets. The 50-work endpoint repeats the same complete information state across all orders.\par\smallskip

\begingroup\small
\setlength{\tableunit}{\dimexpr\linewidth-8\tabcolsep\relax}
\begin{tabular}{@{}>{\raggedright\arraybackslash}p{0.160000\tableunit}>{\raggedright\arraybackslash}p{0.210000\tableunit}>{\raggedright\arraybackslash}p{0.210000\tableunit}>{\raggedright\arraybackslash}p{0.210000\tableunit}>{\raggedright\arraybackslash}p{0.210000\tableunit}@{}}
\toprule
Sampled works & Backbone precision & Backbone recall & Shared edge pair rho & Full reference pair rho \\
\midrule
25/50 & 0.946 & 0.568 & 0.912 & 0.293 \\
40/50 & 0.969 & 0.834 & 0.971 & 0.661 \\
45/50 & 1.000 & 0.891 & 0.986 & 0.819 \\
50/50 & 1.000 & 1.000 & 1.000 & 1.000 \\
\bottomrule
\end{tabular}\endgroup
\end{minipage}\par

The full pair-ranking reference uses all 49,740 zero indegree roots and 87,326 stable edges with positive pair counts. Supplementary Note 4.3 describes the root and edge boundary controls shown in main \hyperlink{figure-3}{Fig.~3d}.

\subsection*{1.3 Observed pair thickness retains information about topology}

\hyperlink{figure-S1}{Supplementary Fig. 1} compares observed and complete normalized pair thickness on discovered edges in the vulnerability cohort (Supplementary Note 4.4). Median Spearman correlations within networks were 0.360 for Growth, 0.741 for ER and 0.822 for BA, 0.858 and 0.883 in the Bioconductor and primary HEP-TH views, and 0.710 across repeated samples of the fixed binary tree.
\par\begin{figure}[!htbp]\centering\begin{minipage}{\linewidth}\phantomsection\label{fig:S1}\hypertarget{figure-S1}{}\addcontentsline{toc}{subsection}{Supplementary Figure 1. Observed and complete pair thickness}
{\centering\textbf{Supplementary Figure 1. Observed and complete pair thickness}\par\smallskip
\includegraphics[width=\linewidth,height=190mm,keepaspectratio]{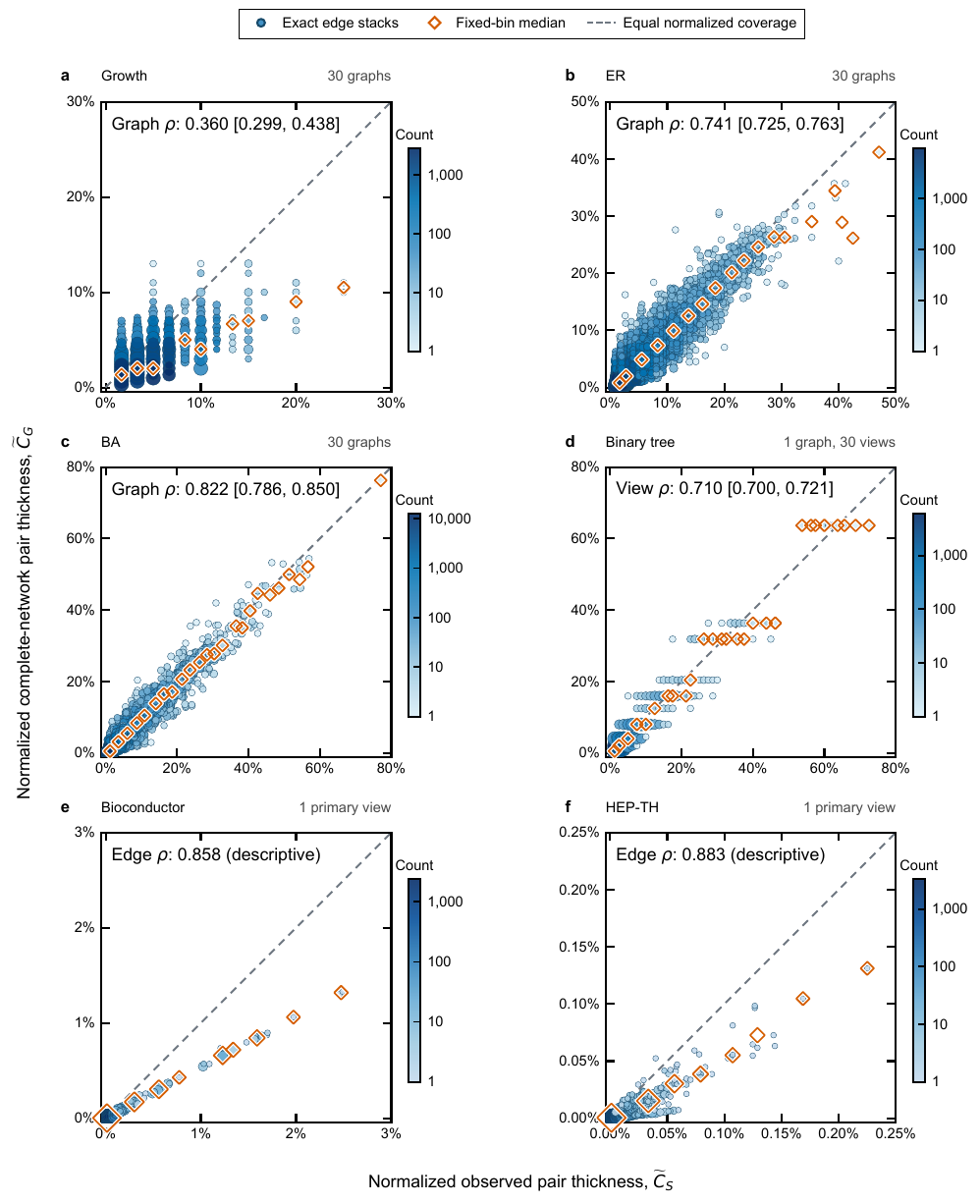}\par}\smallskip
{\fontsize{9}{11}\selectfont Panels a to c pool candidate edges from 30 independent Growth, ER and BA graphs, respectively, d pools 30 sampled terminal views of one binary tree, and e and f show one registered primary view each of Bioconductor and HEP-TH. Circles merge identical coordinates, with capped area increasing with multiplicity, and blue shading gives counts in the local plotting grid on a logarithmic scale specific to each panel. Hollow diamonds are medians within fixed bins, and dashed diagonals mark equal normalized pair thickness. Spearman coefficients for each graph or observation are summarized as median [interquartile range], and empirical edge correlations are descriptive. Each observed or complete graph is normalized by its own product of root and terminal counts.\par}
\end{minipage}\end{figure}\par
Observed thickness estimates a structural count, with accuracy that depends on topology. Differences in observed and complete root and terminal totals can change normalized scores even when rankings agree. Supplementary Note 3.2 shows how calibration within groups affects the relationship between thickness and loss.

\subsection*{1.4 COVID-19 reference trees isolate the single root sampling boundary}

The Methods criteria retained 27 COVID-19 reference trees from OutbreakTrees, in 19 groups defined by normalized source URLs \textsuperscript{\hyperlink{supp-ref:2}{2}}; overlap between groups cannot be excluded. The display tree, the largest by node count with database identity breaking ties, was selected before sampling.

Ten uniform samples without replacement of \(m=\lceil T/2\rceil\) leaves per reference gave 270 views. Complete closure keeps the single root, and with paths of length zero included it gives \(n_G=n_S=R_G=R_S=1\), \(C_G=U\), \(C_S=k_e\), \(\bar C_G=U/T\) and \(\bar C_S=k_e/m\).

The largest reference has 92 nodes, 91 edges and 70 leaves, and its primary 35-leaf sample discovers 50 edges (main \hyperlink{figure-4}{Fig.~4}). Regressing \(U/T\) on \(k_e/m\) gives slope 1.05 and \(R^2=0.689\), which describe sampling calibration; slopes in reference coordinates equal one when estimable.

Across the 27 primary views, nine observed-coordinate slopes are estimable and 18 predictors are constant (main \hyperlink{figure-5}{Fig.~5}). A fitted slope requires two edges and two predictor values. If the predictor varies but the response is constant, the slope is zero and \(R^2\) is undefined. Horizontal offsets serve only to separate markers. No pooled inference was performed.

Every nonroot has \(d_v=1\), which fixes the failure threshold at one for \(0\le\tau\le1\). Every candidate cut therefore affects at least one terminal and keeps its strict loss across the whole tolerance range (main \hyperlink{figure-6}{Fig.~6}). These results describe the published tree topologies, which may omit parts of the causal history. Disease dynamics and edge prioritization remain outside this analysis.

\subsection*{1.5 The complete Bioconductor network illustrates the four concepts}

Main \hyperlink{figure-1}{Figure~1b} draws every node and edge of the frozen Bioconductor 3.22 component used for vulnerability (Supplementary Note 4.4), 2,030 packages and 9,496 dependencies. The 91 roots occupy the first column. Intermediate packages are placed in columns according to the length of the longest input chain from any root, and the 1,416 designated terminals occupy one cell each in the final column. With this arrangement, every edge points to the right. Packages upstream of the focal edge or failing after its deletion occupy the lower band, terminal cells are grouped by loss and sampling status, and vertical order within a layer only reduces crossings. The blue subgraph is the ancestor closure of the 283 terminals in the primary observation, with 583 nodes, 2,749 edges and 48 roots.

The focal edge follows a fixed rule. Among edges with at least three upstream roots, complete coverage between 15\% and 45\% of designated terminals, and at least one but not every reachable terminal sampled, it maximizes \(n(e)U_G(e)\), with ties broken by fewer downstream nodes and then by name. Four edges qualify, and the rule selects GenomicAlignments to rtracklayer. Its four upstream roots, BiocGenerics, BiocParallel, MatrixGenerics and Rhtslib, reach 342 designated terminals, giving 1,368 pairs and 221,031 paths. Because 67 of these terminals are sampled, 268 pairs are observed. Normalized thickness is \(1368/(91\times1416)\) for the complete graph and \(268/(48\times283)\) for its observation. Strict deletion fails 442 nodes, including the 342 terminals, a loss of 342/1,416. The dashed route starts at the sampled lost terminal nearest the head of the focal edge, BSgenomeForge, and ends at the upstream root nearest its tail, BiocGenerics, with ties broken by name.

Panel a keeps the original constructed example, with three roots, four intermediate nodes, four terminals and 11 dependency edges. Sampling t1 and t3 recovers nine nodes and nine edges. The focal edge A to B connects r1 and r2 to t1 and t2, giving four pairs and six paths, with two observed pairs. Strict deletion fails B, C, t1 and t2, a loss of 2/4. These illustrative counts are separate from those of panel b, which uses the same star, circle and square symbols.

\FloatBarrier\section*{Supplementary Note 2. Extended intervention results and comparison controls}
\addcontentsline{toc}{section}{Supplementary Note 2. Extended intervention results and comparison controls}

Supplementary Notes 2.1 to 2.4 use 30 graphs from each synthetic family, the Bioconductor and HEP-TH cases and one binary tree, 93 networks and 61,488 candidate cuts in total. Each keeps its primary observed closure, candidate edges and designated terminals, and Methods gives the construction and response rules.

All thickness and loss plots use the observed predictor \(x_e=\bar C_S(e)\) and losses in the complete analysis graph. Notes 2.5 and 2.6 add an independent Growth prediction test and the river analyses, each with its own design.

\subsection*{2.1 Strict dependency produces distinct root count slope groups}

Within root groups, strict thickness and loss are positively related (Main Figures 4 and 5), a structure that a pooled correlation can obscure, most clearly in Growth.

All 319 estimable slopes across the 90 synthetic networks and two empirical views were positive, and 27 further groups had constant predictors; fitted lines included intercepts. In every synthetic graph the one-root slope exceeded both the two-root and three-root slopes, whereas the order of the latter two could reverse (main \hyperlink{figure-5}{Fig.~5}).

Supplementary Note 3 separates the root count scale from the calibration of sampled terminal coverage; together they explain the overall decline in slopes and the local reversals.

\subsection*{2.2 Increasing tolerance produces plateaus and selective loss suppression}

Tolerance was varied with topology, observed thickness, terminal set and root groups held fixed (main \hyperlink{figure-6}{Fig.~6}). With a maximum indegree of three, Growth keeps its strict response on \(0\le\tau\le1/3\), whereas higher indegrees allow earlier changes in the other families, and the binary tree never changes. Evaluation at exact critical tolerances and their right limits resolves the plateaus between displayed snapshots.

To separate response magnitude from the frequency of loss, we measured the relative mean loss \(L(\tau)\in[0,1]\), defined when the strict loss sum is positive,

\[
L(\tau)=
\frac{\sum_{e\in E_S}I_e(\tau)}
{\sum_{e\in E_S}I_e(0)}
\tag{S1}
\]

together with the fraction of candidate interventions that produce no terminal loss. At \(\tau=1/2\), the synthetic families retained 5.87\% to 9.40\% of their strict mean loss and the empirical cases 1.63\% and 0.830\% (Supplementary Table 2). At \(\tau=3/4\) and \(\tau=1\), 25 of the 30 BA networks had zero terminal loss for every candidate edge, and five kept a nonzero response.

\par\begin{minipage}{\linewidth}
\textbf{Supplementary Table 2. Response suppression and root group linear fits at intermediate tolerance.} Synthetic entries are medians of network-level quantities, with loss and zero-frequency summaries over all 30 graphs per family; empirical and tree entries describe single networks. \(R^2_{\mathrm{group}}\) is computed from pooled out-of-fold residuals for the eligible root groups in each network, using fixed target node folds. Growth fit summaries use 27 eligible networks, and ER and BA summaries use 30. A grouped comparison does not apply to the single-root tree. N/A, not applicable.\par\smallskip

\begingroup\small
\setlength{\tableunit}{\dimexpr\linewidth-8\tabcolsep\relax}
\begin{tabular}{@{}>{\raggedright\arraybackslash}p{0.160000\tableunit}>{\raggedright\arraybackslash}p{0.210000\tableunit}>{\raggedright\arraybackslash}p{0.210000\tableunit}>{\raggedright\arraybackslash}p{0.210000\tableunit}>{\raggedright\arraybackslash}p{0.210000\tableunit}@{}}
\toprule
Network & \(100L(1/2)\) & Zero loss interventions at \(\tau=1/2\) & \(R^2_{\mathrm{group}}(0)\) & \(R^2_{\mathrm{group}}(1/2)\) \\
\midrule
Growth & 8.08\% & 70.7\% & 0.565 & \ensuremath{-}0.00500 \\
ER & 9.40\% & 62.3\% & 0.924 & 0.551 \\
BA & 5.87\% & 58.1\% & 0.977 & 0.0570 \\
Bioconductor & 1.63\% & 92.0\% & 0.998 & \ensuremath{-}0.0610 \\
HEP-TH & 0.830\% & 92.9\% & 0.926 & \ensuremath{-}0.0110 \\
Binary tree & 100.00\% & 0.00\% & N/A & N/A \\
\bottomrule
\end{tabular}\endgroup
\end{minipage}\par

Similar zero-loss frequencies can hide very different nonzero tails. At \(\tau=1/2\), the largest loss from one candidate edge was 63.6\% of designated terminals in Bioconductor but 10 of 1,334 terminals, or 0.750\%, in HEP-TH, so the frequency of harmless interventions alone does not characterize the remaining vulnerability.

\subsection*{2.3 Root group lines lose descriptive power without a universal merger}

A decline in loss magnitude can move all groups toward the horizontal axis without changing their relative shape. To separate these possibilities we used the best common rescaling of the strict response vector, \(\mathbf y_0=(I_e(0))_e\). For a nonzero response \(\mathbf y_\tau\), we calculated

\[
c_\tau=\frac{\langle\mathbf y_0,\mathbf y_\tau\rangle}
{\|\mathbf y_0\|^2},\qquad
H_\tau=\frac{\|\mathbf y_\tau-c_\tau\mathbf y_0\|}
{\|\mathbf y_\tau\|}.
\tag{S2}
\]

Here the vectors contain one loss per candidate edge, \(\langle\cdot,\cdot\rangle\) and \(\|\cdot\|\) denote the Euclidean inner product and norm, \(c_\tau\in[0,1]\) is the fitted common multiplier and \(H_\tau\in[0,1]\) is the relative residual norm, which measures response shape and is distinct from the hidden terminal count defined below. The value \(H_\tau=0\) indicates a common multiplicative reduction across edges. At \(\tau=1/2\), its median was 0.872 in Growth, 0.648 in ER and 0.855 in BA, and the empirical values were 0.979 and 0.993. Thus, tolerance altered edge responses selectively rather than reducing all losses by one factor (\hyperlink{figure-S2}{Supplementary Fig. 2}).
\par\begin{figure}[!htbp]\centering\begin{minipage}{\linewidth}\phantomsection\label{fig:S2}\hypertarget{figure-S2}{}\addcontentsline{toc}{subsection}{Supplementary Figure 2. Response shape and comparison of grouped predictions}
{\centering\textbf{Supplementary Figure 2. Response shape and comparison of grouped predictions}\par\smallskip
\includegraphics[width=\linewidth,height=215mm,keepaspectratio]{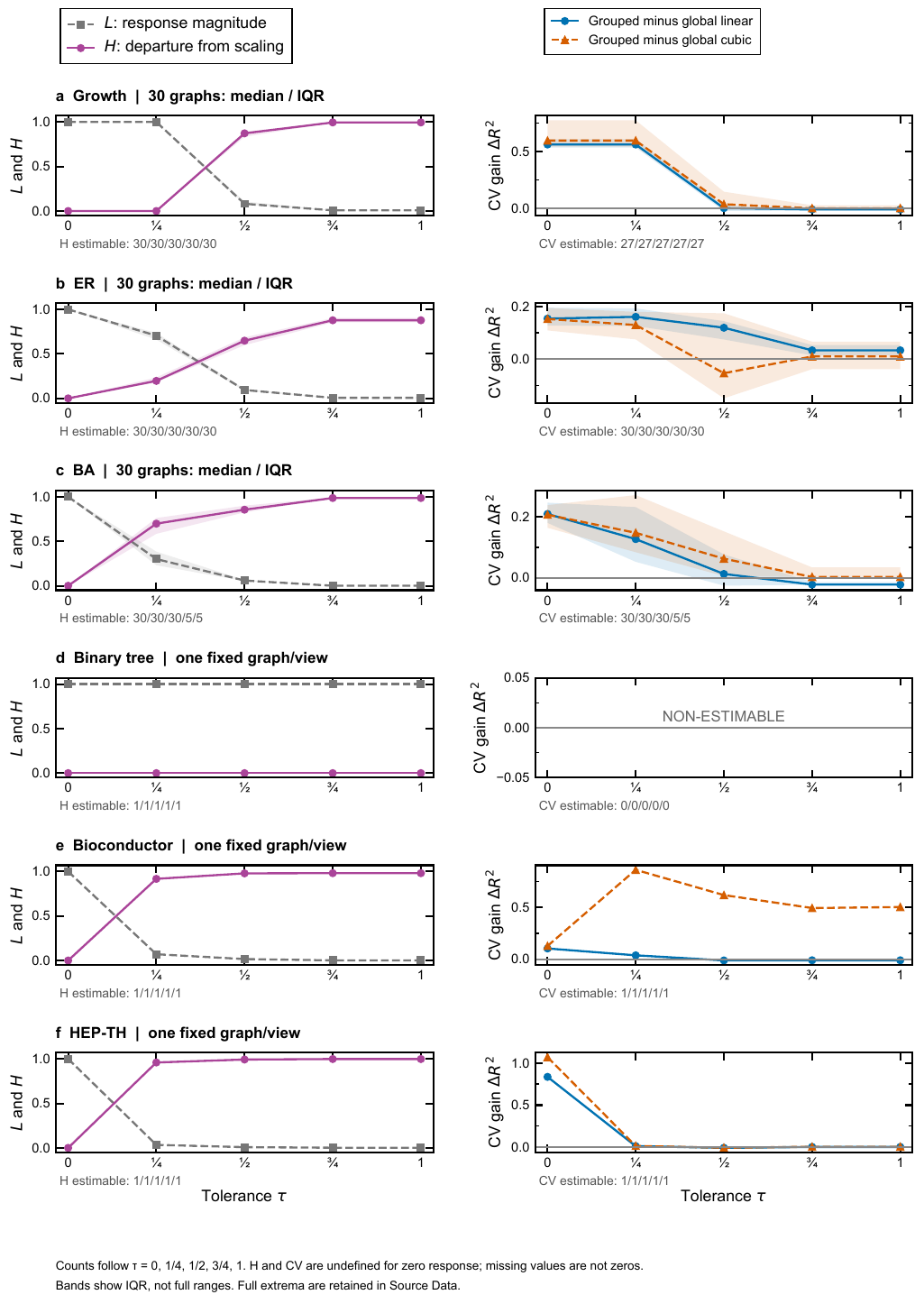}\par}\smallskip
{\fontsize{9}{11}\selectfont Left panels distinguish the reduction in mean loss from changes in the relative response across edges. The shape residual measures departure from a common rescaling of the strict response. Right panels compare separate root group lines with pooled linear and cubic predictions on the same withheld targets. Positive differences favor the grouped model. Bands show interquartile ranges across synthetic graphs; empirical cases and the tree are single observations. Undefined comparisons are omitted. At the last two tolerances, BA shape and cross-validation summaries use only the five graphs with nonzero responses, while summaries of zero-response frequency cover all 30.\par}
\end{minipage}\end{figure}\par
Grouped and pooled lines used the same five folds blocked by target, with group eligibility and fold assignment determined only by the original predictors and target identities. Both were held fixed across tolerances (Supplementary Note 4.5).

At \(\tau=1/2\), lines fitted separately by root count had little out-of-fold explanatory power in Growth, BA and the two empirical cases (Supplementary Table 2). Their median \(R^2\) advantage over a pooled line fell from 0.561 to 0.00500 in Growth and from 0.209 to 0.0120 in BA, and in Bioconductor and HEP-TH the differences changed from 0.104 and 0.837 to \ensuremath{-}0.0110 and \ensuremath{-}0.00900. The shape residuals, retained zeros and weak predictive fits show that tolerance disrupts the original linear group relationships and leaves a poor fit for a single pooled line.

ER kept a stronger grouped relationship, with median grouped \(R^2=0.551\) and a median advantage of 0.120 over the pooled line at \(\tau=1/2\). A prespecified pooled cubic model, however, beat the grouped lines in 16 of the 30 ER networks, with a median grouped minus cubic difference of \ensuremath{-}0.0540, so part of the apparent grouping advantage may reflect an overall curved relationship.

At the two highest displayed tolerances, \(H_\tau\) and predictive \(R^2\) were undefined for the 25 all-zero BA networks. These summaries use the five networks with a remaining response, while the zero response frequency still describes all 30.

\subsection*{2.4 Root correction improves edge prioritization under strict support}

Main \hyperlink{figure-7}{Fig.~7} compares structural scores at the top 5\% edge budget, and \hyperlink{figure-S3}{Supplementary Fig. 3} gives all 12 displayed methods across seven network families at strict and half tolerance. River entries average ten 40-terminal views, whereas the original empirical columns use single primary views. Methods defines independent cuts, tied selections and oracle efficiency, and evaluation on unsampled terminals tests prediction beyond the observed closure \textsuperscript{\hyperlink{supp-ref:3}{3},\hyperlink{supp-ref:4}{4},\hyperlink{supp-ref:5}{5}}.
\par\begin{figure}[!htbp]\centering\begin{minipage}{\linewidth}\phantomsection\label{fig:S3}\hypertarget{figure-S3}{}\addcontentsline{toc}{subsection}{Supplementary Figure 3. Complete numerical comparison of edge importance}
{\centering\textbf{Supplementary Figure 3. Complete numerical comparison of edge importance}\par\smallskip
\includegraphics[width=\linewidth,height=160mm,keepaspectratio]{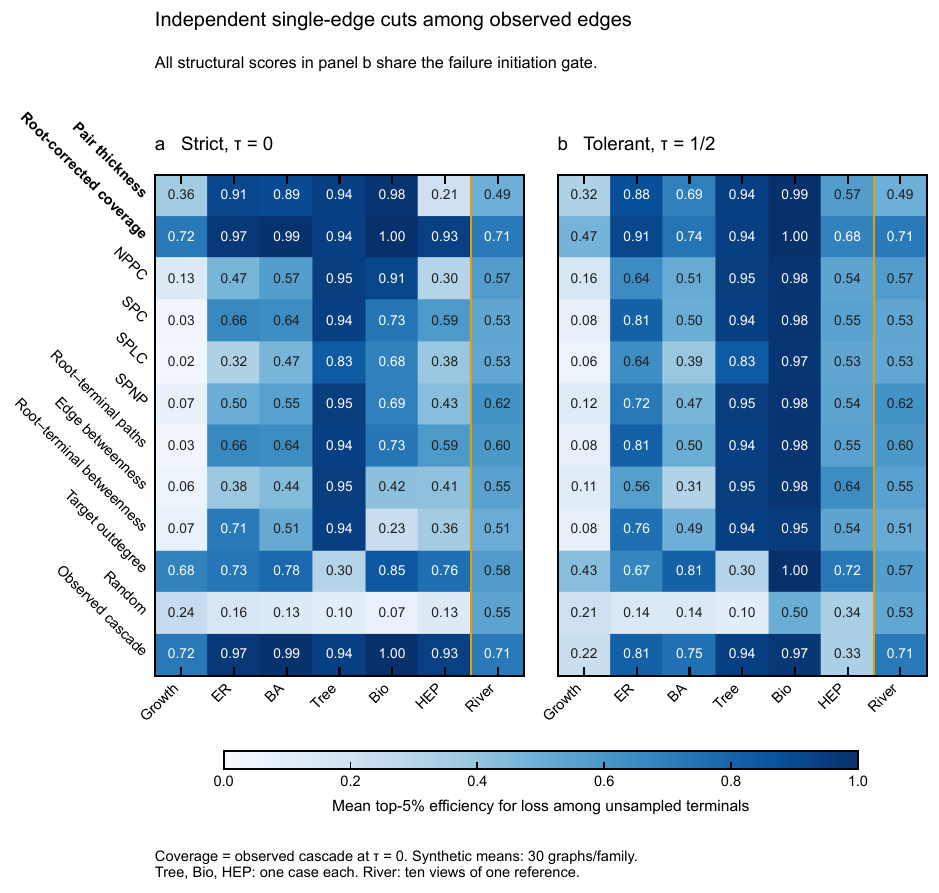}\par}\smallskip
{\fontsize{9}{11}\selectfont Mean top 5\% oracle efficiency on unsampled terminals for independent single-edge cuts under strict support (a) and half tolerance (b). All methods score the same discovered candidate edges. At half tolerance every structural score receives the same failure initiation gate, and the observed cascade uses the sampled terminal response directly. Growth, ER and BA means weight 30 independent graphs per family equally. The tree, Bioconductor and HEP-TH are single cases, and the river column averages ten views sampling 40 of 81 designated gauge reaches in one Rio Grande reference; orange outlines mark this added column. Bold labels identify pair thickness and its root correction, and corrected coverage equals deterministic reverse reachability coverage restricted to terminals. The river case extends only the structural benchmark; the posterior benchmark below uses birth information.\par}
\end{minipage}\end{figure}\par
The strict gains over raw pair thickness in main \hyperlink{figure-7}{Fig.~7} occurred in every graph of each synthetic family. Paired mean gains on unsampled terminals were 0.360, 0.0580 and 0.0940 for Growth, ER and BA, with exploratory 95\% graph-bootstrap intervals of [0.323, 0.399], [0.0440, 0.0720] and [0.0700, 0.124], and selecting distinct targets preserved these gains. In the single-root tree, pair thickness and corrected coverage rank identically, while NPPC, SPNP and all-pairs betweenness reached efficiency 0.947 against 0.941 for coverage.

Support substitution changes these comparisons. At \(\tau=1/2\), every static score received the same local condition for cascade initiation, \(\mathbf1\{\tau d_v\le1\}\). Gated coverage reached efficiencies on unsampled terminals of 0.473, 0.907 and 0.738 in Growth, ER and BA, and gated target outdegree was higher in BA (0.812), Bioconductor (1.000 versus 0.995) and HEP-TH (0.719 versus 0.684). Direct simulation of the observed cascade gave a Growth efficiency of only 0.219, although it calculates loss among sampled terminals exactly under the model. This difference separates calculation within the observation from prediction beyond it. The remaining Growth error motivates using the formation process to infer terminal coverage that backward tracing has not revealed.

\subsection*{2.5 Growth information predicts unobserved terminal coverage}

The posterior in Supplementary Note 3.6 combines the sampled count \(k_e\), the target birth generation \(b\in\{1,\ldots,L\}\), the sample size and the known Growth parameters. The positive integer \(L\) is the final generation, distinct from \(L(\tau)\), and birth is recorded by the generator. The posterior supplies the hidden count mean \(\widehat H_e\in[0,T-m]\), the complete count mean \(k_e+\widehat H_e\in[k_e,T]\) and a predictive distribution. The estimated raw pair thickness keeps the upstream factor, \(\widehat C_G(e)=n(e)(k_e+\widehat H_e)\).

The test used 100 new, unconditioned Growth graphs, each observed through samples of 10, 20 and 40 of its 100 terminals, with 20 primary. The prior, sampling plan, outcomes and comparisons were fixed before the test graphs were generated, and no parameter was fitted to test outcomes.

At the primary sample size, hidden count mean squared error fell from 21.7 for boundary-aware coverage extrapolation to 7.28 for the Growth posterior, a 66.4\% reduction (\hyperlink{figure-S4}{Supplementary Fig. 4} and Supplementary Table 3). Birth and discovery alone gave 8.45, so the observed count supplied a further 13.9\% reduction. All methods used the known terminal generation boundary, and primary errors excluded final generation targets.
\par\begin{figure}[!htbp]\centering\begin{minipage}{\linewidth}\phantomsection\label{fig:S4}\hypertarget{figure-S4}{}\addcontentsline{toc}{subsection}{Supplementary Figure 4. Predicting coverage with a uniform growth prior}
{\centering\textbf{Supplementary Figure 4. Predicting coverage with a uniform growth prior}\par\smallskip
\includegraphics[width=\linewidth,height=0.63\textheight,keepaspectratio]{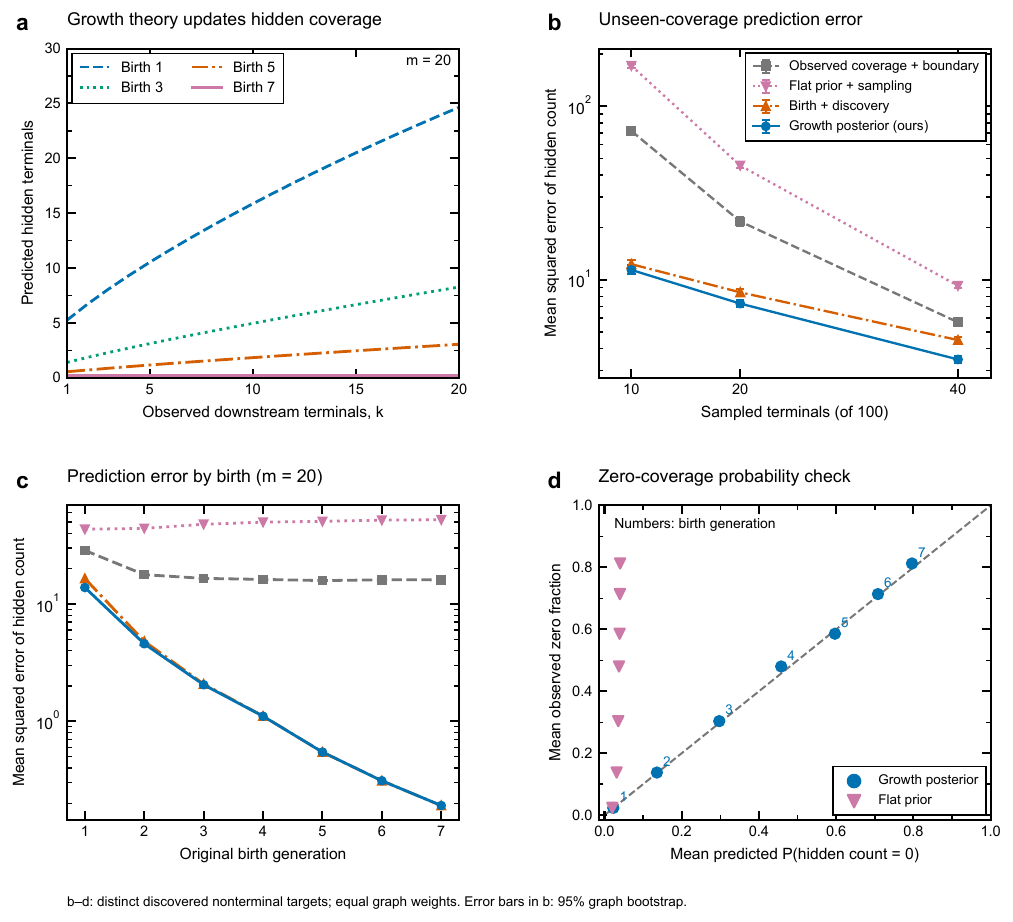}\par}\smallskip
{\fontsize{9}{11}\selectfont Theory and prediction in the fresh Growth cohort. Panel a shows the conditional mean hidden coverage against the observed count for four original birth generations at a sample of \(m=20\) terminals; rare count conditions are not frequency estimates. Panels b and c show mean squared error over distinct discovered nonterminal targets with equal weight per graph, and error bars in b are 95\% intervals from bootstrap resampling of whole graphs. Panel d compares predicted and observed probabilities of zero coverage by birth generation, again weighting graphs equally. Targets born in the final generation are excluded from b to d because, if discovered, they have no hidden terminal coverage. Colors and styles in c match b, and model parameters and birth metadata are known.\par}
\end{minipage}\end{figure}\par
\par\begin{minipage}{\linewidth}
\textbf{Supplementary Table 3. Predicting hidden terminal coverage on fresh Growth graphs.} Results use the primary sample of 20/100 terminals. Squared errors are averaged over distinct discovered nonterminal targets within each graph and then equally over the 100 graphs. Positive paired differences are baseline error minus Growth posterior error. Intervals resample physical graphs, so repeated incoming edges and sampling views add no replicates. N/A, not applicable.\par\smallskip

\begingroup\small
\setlength{\tableunit}{\dimexpr\linewidth-6\tabcolsep\relax}
\begin{tabular}{@{}>{\raggedright\arraybackslash}p{0.340000\tableunit}>{\raggedright\arraybackslash}p{0.180000\tableunit}>{\raggedright\arraybackslash}p{0.220000\tableunit}>{\raggedright\arraybackslash}p{0.260000\tableunit}@{}}
\toprule
Predictor & Mean squared error & Paired error reduction & 95\% paired interval \\
\midrule
Coverage extrapolation with boundary & 21.7 & 14.4 & [13.3, 15.6] \\
Uniform coverage prior & 45.5 & 38.2 & [36.6, 40.0] \\
Birth and discovery & 8.45 & 1.17 & [1.00, 1.34] \\
Growth posterior & \textbf{7.28} & N/A & N/A \\
\bottomrule
\end{tabular}\endgroup
\end{minipage}\par

The improvement in ranking was smaller than the improvement in count prediction (\hyperlink{figure-S5}{Supplementary Fig. 5}). At the top 5\% edge budget, mean efficiency on unsampled terminals rose from 0.684 for corrected coverage to 0.702 for the posterior, a paired improvement of 1.81 percentage points (95\% interval 1.39 to 2.21), and improvements at sample sizes 10 and 40 were 3.29 and 0.940 percentage points. For full terminal loss at the primary sampling level, posterior efficiency was 0.826, compared with 0.814 for corrected coverage. At the primary sampling level and top 5\% budget, the posterior matched the birth and discovery control in full terminal loss efficiency in all 100 graphs.
\par\begin{figure}[!htbp]\centering\begin{minipage}{\linewidth}\phantomsection\label{fig:S5}\hypertarget{figure-S5}{}\addcontentsline{toc}{subsection}{Supplementary Figure 5. Prioritization with a uniform growth prior across observation levels}
{\centering\textbf{Supplementary Figure 5. Prioritization with a uniform growth prior across observation levels}\par\smallskip
\includegraphics[width=\linewidth,height=125mm,keepaspectratio]{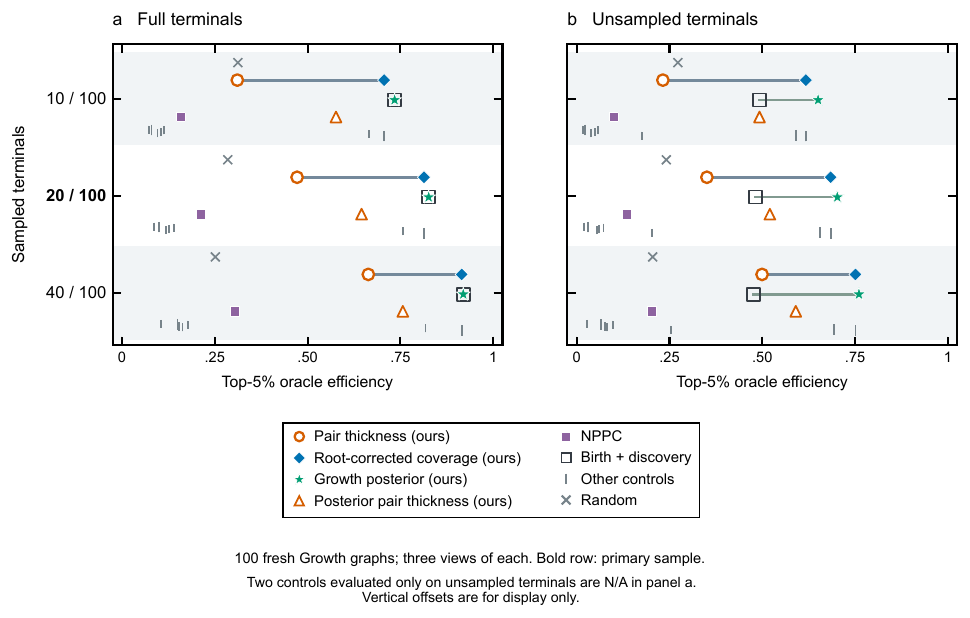}\par}\smallskip
{\fontsize{9}{11}\selectfont Mean top 5\% oracle efficiency on 100 fresh Growth graphs for full (a) and unsampled (b) terminals. Rows indicate the number of sampled terminals, with the primary sample in bold. Orange circles and blue diamonds use the \hyperlink{figure-7}{Figure~7} encoding for pair thickness and root-corrected coverage; green stars show the growth-based posterior, orange triangles posterior pair thickness, open squares birth and discovery and purple squares NPPC, while gray ticks mark the eleven other controls and crosses random ranking. Lines join pair thickness to coverage and birth and discovery to the posterior, and vertical offsets only separate methods. The two structural boundary controls evaluated only on unsampled outcomes are N/A in a. Ours denotes scores adopted or derived here. Pair thickness is NPPC restricted to designated endpoints, and corrected coverage equals both deterministic reverse reachability frequency restricted to terminals and observed strict cascade loss. Boundary controls receive the known zero hidden coverage of final generation targets. Exact ties are averaged uniformly, and sample sizes are repeated observations of the same graphs.\par}
\end{minipage}\end{figure}\par
Discrete equal-tail 95\% posterior intervals covered 97.8\% of hidden counts with a mean width of 8.93 terminals, against 95.8\% and 23.0 for the uniform prior; discreteness can make intervals conservative. The birth-resolved probability check in \hyperlink{figure-S4}{Supplementary Fig. 4} complements the interval and proper-score diagnostics, because lower error alone does not establish calibration.

\subsection*{2.6 River reference, strata and substitution}

Methods describes the frozen Rio Grande reference and station mapping. Completing the Hydrologic Fabric predecessor relations adds one reach and one link to the NLDI navigation set, so the NLDI sensitivity representation has 3,371 reaches and 3,410 links.

Ten uniform samples at each size 16, 40 and 65 use the same terminal identities in both representations. Station records were selected from daily mean discharge metadata, not flow magnitudes, and aliases on one reach count once. Each sample yields a complete predecessor closure, and the primary \(m=40\) observation has 1,778 candidate targets.

Main \hyperlink{figure-4}{Fig.~4} shows the same discovered edges in observed and full coordinates, \(\bar C_S=C_S/(1025\times40)\) and \(\bar C_G=C_G/(1270\times81)\), while retaining each edge’s loss in the full network. The black hollow circle backdrop excludes undiscovered reference edges. Dashed segments keep the strict slopes \(R_S/n\) within the observation; they are reference rays, not fitted slopes or identities for full loss against observed thickness.

Of the 154 exact root groups in main \hyperlink{figure-5}{Fig.~5}, 65 admit slopes, 63 positive and two zero, and 89 have constant predictors. The zero slopes at \(n=51,73\) arise from constant complete coverage with varying sampled coverage, so their \(R^2\) is undefined, and integer covariance avoids numerical sign errors. Slopes for \(n=1,2,3\) are 723, 364 and 282, and the eighteen upward changes between successive estimable groups reflect calibration within groups.

We resolved the complete tolerance response from exact terminal critical values, evaluating the response at every possible local threshold and immediately to its right, with original indegrees, roots and candidate edges fixed. The response is constant on \([0,1/3]\), \((1/3,1/2]\) and \((1/2,1]\). On these intervals the initiating fractions are \(1,2821/2836,725/2836\), the fractions of cuts with positive terminal loss are \(1,2749/2836,22/2836\) and the relative mean losses are \(1,23082/23838,22/23838\). Half tolerance thus retains 96.8\% of strict loss, whereas its right limit retains 0.0923\%. The denominator 23,838 sums terminal losses over separate cuts, each starting from the intact graph. The potential threshold \(2/3\) causes no further jump.

The river benchmark gives every static score the same initiation gate. Across ten primary-size views, strict efficiency on unsampled terminals averages 0.711 for coverage, 0.493 for pair thickness and 0.619 for SPNP, and 0.712, 0.494 and 0.619 at half tolerance. At full tolerance, coverage reaches 0.103 and all-pairs betweenness 0.739. The river evidence tables also keep drainage-area and stream-order comparators, while \hyperlink{figure-7}{Fig.~7} and \hyperlink{figure-S3}{Supplementary Fig. 3} show the methods shared with the original cohort.

The NLDI sensitivity changes activation, positive-loss frequency and relative mean loss by at most 0.000263, 0.0000108 and 0.0000160. These differences quantify sensitivity to the graph representation under the imposed support rule. Confluences, tributary supply, local inflow and storage can sustain downstream flow after upstream contributions are lost; equation (2) abstracts this substitution without calibrating these processes.

\FloatBarrier\section*{Supplementary Note 3. Graph and growth derivations}
\addcontentsline{toc}{section}{Supplementary Note 3. Graph and growth derivations}

\subsection*{3.1 Pair coverage determines the strict slope scale}

We prove Proposition 1 on a finite, simple, unweighted DAG with fixed original roots and a fixed nonempty designated terminal set, writing \(n(e)=n_G(e)\). For any root \(r\) reaching \(u\) and designated terminal \(t\) reachable from \(v\), concatenating an \(r\)-to-\(u\) path, the edge \(e=(u\to v)\) and a \(v\)-to-\(t\) path yields a directed root-to-terminal path through the edge, and acyclicity prevents repeated vertices. Counting each pair once gives

\[
C_G(e)=n(e)U_G(e).
\tag{S3}
\]

Under strict dependency, cutting \(e\) removes a necessary support from \(v\). The target fails, failure reaches every descendant by induction along the DAG, and nodes outside that descendant set remain supported. Consequently,

\[
I_e(0)=\frac{U_G(e)}{T_G}
=\frac{R_G}{n(e)}\bar C_G(e).
\tag{S4}
\]

Within a fixed root count, complete network thickness and strict loss therefore lie on a line of slope \(R_G/n\), and different integer root counts give distinct slope scales. This part of the grouping follows exactly from the definition of pair thickness and the strict failure rule. Pair thickness counts combinations of origins and downstream outcomes associated with an edge, whereas loss counts only the outcomes. Multiple origins therefore multiply the pair count without multiplying the number of lost outcomes.

The factorization \(\bar C_G(e)=[n(e)/R_G]I_e(0)\) separates the fraction of roots represented upstream from the fraction of terminals lost downstream. Edges with positive coverage but different root counts lie on distinct rays, although a group with constant coverage need not yield a distinct fitted line, and zero coverage places an edge at the shared origin.

For two edges \(e,f\in E_G\) with \(0<U_G(e)<U_G(f)\), raw or normalized pair thickness reverses their strict consequence order if and only if \(n(e)/n(f)>U_G(f)/U_G(e)\), which follows by dividing \(n(e)U_G(e)>n(f)U_G(f)\) by the positive factors. Root heterogeneity alone is therefore insufficient. For example, two roots feeding one intermediate node that supports three terminal children give incoming edges with \(n=1,U=3,C=3\) and outgoing edges with \(n=2,U=1,C=2\), so both scores keep the same order despite different root counts. When root counts are constant across edges, pair and loss rankings agree exactly. The argument compares existing edges within a fixed network.

\subsection*{3.2 Observation calibrates the measured slopes}

The measured plots use observed thickness, so Eq. (S4) alone does not fix their fitted slopes. When the observation \(S\) is the complete ancestor closure obtained by tracing all predecessors of its sampled terminals, every candidate edge keeps its complete upstream root set, so \(n_S(e)=n_G(e)=n(e)\). With \(W_S(e)=U_S(e)/T_S\) and \(W_G(e)=U_G(e)/T_G\), the observed predictor and strict response satisfy

\[
x_e=\frac{n(e)}{R_S}W_S(e),\qquad I_e(0)=W_G(e).
\tag{S5}
\]

For one fixed network and observation, let \(b_n(0)\in\mathbb R\) be the least-squares slope with a fitted intercept in the exact root group \(n\). When \(\operatorname{Var}(W_S\mid n)>0\), it is exactly

\[
b_n(0)=\frac{R_S}{n}\lambda_n,\qquad
\lambda_n=
\frac{\operatorname{Cov}(W_S,W_G\mid n)}
{\operatorname{Var}(W_S\mid n)}.
\tag{S6}
\]

The covariance and variance are empirical moments over the fixed edge group, and the calibration coefficient \(\lambda_n\in\mathbb R\) need not equal one or be positive. The factor \(1/n\) sets the root count scale, while \(\lambda_n\) describes the regression of complete coverage on observed terminal coverage within the group. Because \(\lambda_n\) is a regression calibration coefficient, a high rank correlation between observed and complete thickness does not require \(\lambda_n=1\). Unequal calibration can attenuate or reverse adjacent slope differences while the complete network identity remains exact. This explains why the overall decline in main \hyperlink{figure-5}{Fig.~5} can coexist with local reversals.

For the same graph, observation and root group, let \(x_e^G=\bar C_G(e)\) and \(\kappa_n=R_G/n>0\), so that Equation (S4) gives \(I_e(0)=\kappa_n x_e^G\). Regressing either response on observed \(x_e\), with identical weights and fitted intercepts, multiplies the residual and total sums of squares by \(\kappa_n^2\). Hence \(R^2(I(0)\sim x\mid n)=R^2(x^G\sim x\mid n)\) whenever defined, even if \(R_S\ne R_G\). This equality follows from the constant scale factor and describes relative scatter around the fitted lines within each group.

Main \hyperlink{figure-4}{Figure~4} pairs foreground observations with black hollow circles for the same edges and unchanged complete network losses. The horizontal coordinates use denominators \(R_ST_S\) and \(R_GT_G\), so horizontal shifts can reflect normalization as well as coverage recovery. Complete points follow equation (S4) within each graph, and pooling different \(R_G\) can produce several slopes within one root class. Dashed rays show \(I_{\mathrm{strict},S}=(R_S/n)\bar C_S\), not fitted lines or identities for complete loss. Display offsets by generation affect only the Growth foreground markers, not complete points or calculations.

Removing the root multiplier gives the coordinate diagnostic \(z_e=R_Sx_e/n(e)=W_S(e)\). Applied to the same edges and unchanged strict losses, this transformation increased both the global linear \(R^2\) and the Spearman correlation in all 92 original synthetic and empirical analysis units (\hyperlink{figure-S6}{Supplementary Fig. 6}). The median global \(R^2\) rose from 0.0270 to 0.511 in Growth, from 0.766 to 0.921 in ER and from 0.768 to 0.977 in BA, and the empirical values rose from 0.896 to 0.998 and from 0.0500 to 0.929. Growth kept substantial residual variation, consistent with its less accurate observed coverage.
\par\begin{figure}[!htbp]\centering\begin{minipage}{\linewidth}\phantomsection\label{fig:S6}\hypertarget{figure-S6}{}\addcontentsline{toc}{subsection}{Supplementary Figure 6. Removing the root count from pair thickness}
{\centering\textbf{Supplementary Figure 6. Removing the root count from pair thickness}\par\smallskip
\includegraphics[width=\linewidth,height=215mm,keepaspectratio]{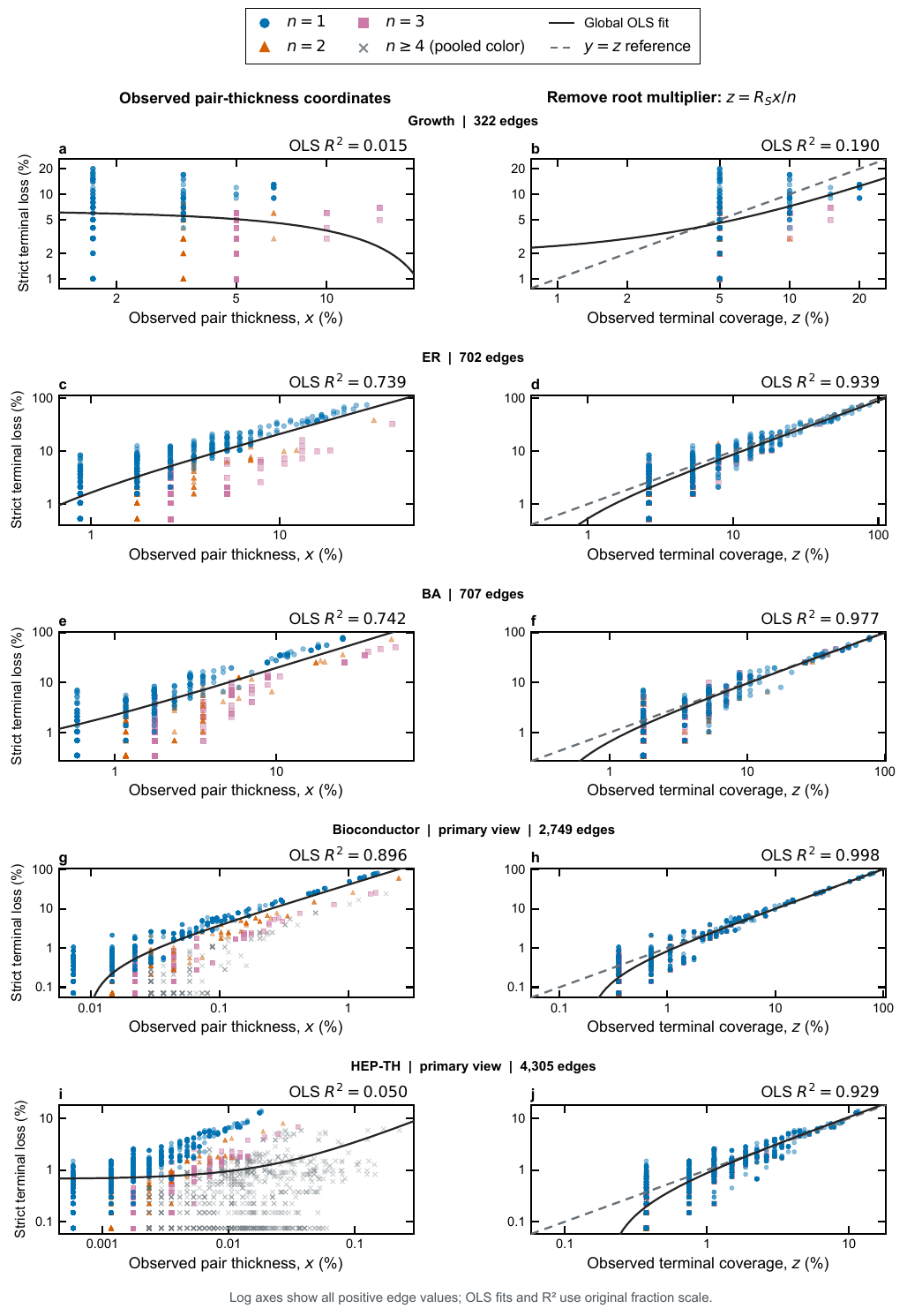}\par}\smallskip
{\fontsize{9}{11}\selectfont Left panels plot strict terminal loss against observed pair thickness; right panels remove the normalized upstream factor to give observed terminal coverage. Colors identify root groups, and the fitted line summarizes the pooled relationship. The same candidate edges and losses appear on both sides, so any change in fit comes from the change in predictor. Axes are logarithmic, but regressions are fitted on the original fraction scale. The first registered synthetic examples are shown, and aggregate conclusions use all 92 analysis units.\par}
\end{minipage}\end{figure}\par
Within any exact root group, the coordinate change is a constant rescaling that leaves fitted values and the residual sum of squares unchanged, so its improvement of the pooled fit identifies a contribution from mixing root count scales. Equality with the fixed line \(I=z\) would also require accurate absolute coverage recovery, which a high \(R^2\) from a line with a fitted intercept does not establish.

\subsection*{3.3 Tolerance separates terminal exposure from failure propagation}

Strict failure reaches all covered terminals, whereas under tolerance only a subset may fail. For \(U_G(e)>0\), let \(q_e(\tau)\) be the fraction of terminals reachable from the target that fail after the cut. Then

\[
I_e(\tau)=W_G(e)q_e(\tau)
=\frac{R_G}{n(e)}\bar C_G(e)q_e(\tau),
\qquad 0\le q_e(\tau)\le1.
\tag{S7}
\]

Exposure \(W_G\) stays fixed while the extent of propagation changes with the support rule, and the corresponding observed group slope becomes

\[
b_n(\tau)=\frac{R_S}{n}
\frac{\operatorname{Cov}\!\left(W_S,W_Gq_e(\tau)\mid n\right)}
{\operatorname{Var}(W_S\mid n)}.
\tag{S8}
\]

If \(q_e(\tau)\) is constant within a group, it rescales the group’s line. Variation correlated with coverage can change slopes and residuals. Topological induction shows that the loss of each cut cannot increase with \(\tau\), because fewer supports fail while failure thresholds cannot decrease. Regression slopes need not follow the same pattern, since they depend on covariance across edges.

The first structural restriction is local. A single cut removes exactly one support from its target, which fails if and only if \(\max\{1,\lceil\tau d_v\rceil\}=1\), equivalently \(\tau\le1/d_v\). Above this value no node fails, because all other supports are initially intact and the DAG has no feedback that could start another failure. At \(\tau=1/2\), only targets with original indegree one or two can initiate a cascade. Once failure starts, downstream nodes respond to combinations of missing supports, including failures that arrive along branches that later reconverge. A large upstream root count by itself does not determine whether those supports are substitutable.

\subsection*{3.4 Terminal critical tolerances determine the complete response}

The loss curve is the tail of a distribution of terminal critical tolerances, which gives an exact structural representation of the propagation factor. For each node \(v\) and cut \(e\), let \(h_v(e)\in(0,1]\) be the greatest tolerance at which the node fails, with the separate symbol \(\bot\) for a node that never fails. Then

\[
I_e(\tau)=\frac{1}{T_G}
\sum_{t\in\mathcal T_G}
\mathbf1\{h_t(e)\ne\bot\text{ and }\tau\le h_t(e)\},
\tag{S9}
\]

where \(\mathcal T_G\) is the designated terminal set. Heterogeneous suppression within a root group corresponds to differences in the distributions of terminal critical values across its edges. Failure holds at equality, and the drop occurs immediately to the right of a critical value. A missing value differs from zero, because every node reached by a strict cascade fails over a nonempty interval starting at zero.

These critical values can be computed exactly in topological order. Original roots have value \(\bot\). Assign support critical value one to the permanently cut input. Every other input \(u\to v\) has support critical value \(h_u(e)\). Discard \(\bot\) values and sort the remaining support values as \(a_{(1)}\ge\cdots\ge a_{(r)}\), where \(r\in\{0,\ldots,d_v\}\) is their number and every retained value lies in \((0,1]\). If \(r=0\), set \(h_v=\bot\); otherwise,

\[
h_v(e)=\max_{1\le j\le r}
\min\left\{a_{(j)},\frac{j}{d_v}\right\}.
\tag{S10}
\]

At least \(j\) supports are unavailable if and only if \(\tau\le a_{(j)}\), and \(j\) losses meet the failure threshold if and only if \(\tau\le j/d_v\). Thus, failure occurs if and only if both inequalities hold for some \(j\). Conversely, whenever a node fails, \(j\) can be chosen as the number of unavailable supports. The union of these failure intervals ends at their largest endpoint, and topological induction yields all node values. The recurrence determines the full response on a known graph; prediction from an incomplete observation also requires information about unobserved downstream support.

The upper endpoint has a simpler path interpretation. At \(\tau=1\), a nonroot node survives exactly when at least one uncut original input comes from a surviving predecessor, which by topological induction is equivalent to reachability from the fixed original roots in \(G-e\). Let \(D_G(e)\in\{0,\ldots,U_G(e)\}\) count designated terminals for which every path from an original root passes through \(e\). Then

\[
I_e(1)=\frac{D_G(e)}{T_G},\qquad
\frac{D_G(e)}{T_G}\le I_e(\tau)\le\frac{U_G(e)}{T_G}.
\tag{S11}
\]

An edge keeps an unchanged terminal loss curve throughout \([0,1]\) if and only if \(D_G(e)=U_G(e)\), that is, when it is a common bottleneck for every terminal it covers. Such an edge can have multiple upstream roots and a downstream region of nodes with many incoming edges, provided all root support to those terminals must traverse it. Among the scanned candidates, every Growth and ER network contained at least one multiroot edge of this kind, with 182 and 267 such edges across the two families; the Bioconductor and HEP-TH views contained 15 and eight, and BA contained none. The binary tree is the limiting case in which every candidate edge keeps its loss at every tolerance.

\subsection*{3.5 The uniform parent special case predicts ensemble thickness}

Once a network is given, the strict identity fixes the slope scale, whereas a generative model predicts downstream coverage before the complete network is observed. We first derive a scalar recursion for the independent uniform parent process used to generate the Growth cohort. Supplementary Note 8.1 extends it to age preference. The scalar form is computationally simpler and supports the larger uniform growth tests.

Starting from \(R_0\ge1\) roots, with integer \(R_0\), each generation \(g\in\{1,\ldots,L\}\) adds \(N_g\) nodes, where the cohort size \(N_g\) is a positive integer. Each new node independently draws a parent count \(J_g\) from \(\eta_g\) and chooses that many distinct parents uniformly from all \(M_{g-1}=R_0+\sum_{h<g}N_h\) earlier nodes. The parent count law \(\eta_g(j)\ge0\) sums to one over integers \(1\le j\le M_{g-1}\), so \(J_g\) is always feasible. Parent choices are independent across nodes, do not depend on existing graph structure and use a pool frozen within each generation.

The last \(N_L=T\) nodes are the designated terminals, and the evaluated specification has \(R_0=3\), \(L=8\), \(N_g=100\) and \(\eta_g(1)=\eta_g(2)=\eta_g(3)=1/3\).

Fix a target \(v\) born at generation \(b<L\). Let \(Z_{g-1}\in\{1,\ldots,M_{g-1}\}\) count \(v\) and all its descendants born before generation \(g\), with \(Z_b=1\), and let the increment \(X_g\in\{0,\ldots,N_g\}\) count new descendants. A new node joins the descendant set if at least one of its parents belongs to it. Conditional on \(Z_{g-1}=z\), its inclusion probability \(p_g(z)\in[0,1]\) and the generation update are

\[
p_g(z)=\sum_j\eta_g(j)
\left[1-\frac{\binom{M_{g-1}-z}{j}}{\binom{M_{g-1}}{j}}\right],
\qquad
X_g\mid Z_{g-1}=z\sim\operatorname{Bin}(N_g,p_g(z)),
\qquad Z_g=Z_{g-1}+X_g .
\tag{S12}
\]

These finite state transitions are exact, because parent choices are independent across new nodes and every earlier subset of the same size has the same intersection probability. Terminal coverage is the final increment \(U=X_L\), not the cumulative descendant count \(Z_L\). With \(\mu_b(z)=\Pr(Z_{L-1}=z\mid b)\) the probability mass on feasible integer states \(z\), the birth-conditioned coverage distribution \(\rho_b(u)\) on integers \(u\in\{0,\ldots,T\}\) and its mean \(\bar u_b\in[0,T]\) are

\[
\rho_b(u)=\sum_z\mu_b(z)\binom{T}{u}
p_L(z)^u[1-p_L(z)]^{T-u},
\qquad
\bar u_b=\mathbb E[U\mid b]
=T\sum_z\mu_b(z)p_L(z).
\tag{S13}
\]

For a final generation target, \(U=1\). We propagate the full distribution of \(Z\) because replacing the expectation of \(p_L(Z)\) by \(p_L(\mathbb E Z)\) would introduce a mean field approximation.

The same reachability recursion, initialized at one fixed root, gives the mean upstream root count of a fixed node born at generation \(a\) as \(\alpha_a=R_0\mathbb E[p_a(Z^{(r)}_{a-1})]\) for \(a>0\), with \(\alpha_0=1\). Here \(\alpha_a\in[1,R_0]\), and \(Z^{(r)}\) is the descendant process initialized at one of the \(R_0\) planted roots, labeled \(r\). Root symmetry and linearity of expectation suffice, so the reachability events of different roots need not be independent. For fixed labels \(u,v\) at integer generations \(0\le a<b\le L\), let \(A_e\) be the event that \(u\to v\) exists, and write \(n(u)\in\{1,\ldots,R_0\}\) for the roots reaching \(u\) and \(U(v)\in\{0,\ldots,T\}\) for the terminals reachable from \(v\). The parent choices that determine \(n(u)\), those that create \(A_e\) and the later choices that determine \(U(v)\) are independent. Consequently,

\[
n(u)\perp U(v)\mid A_e,a,b,\qquad
\mathbb E[C_{\widetilde G}(e)\mid A_e,a,b]=\alpha_a\bar u_b,
\qquad
\mathbb E[U/T\mid n,A_e,a,b]=\frac{\bar u_b}{T}.
\tag{S14}
\]

Here \(\widetilde G\) is the unpruned generated graph, and \(U/T\) is its strict terminal loss fraction. The ancestor closure of all terminals keeps an existing edge exactly when \(U>0\), so its conditional raw thickness mean is \(\alpha_a\bar u_b/\Pr(U>0\mid b)\). Pruning preserves the upstream root set of each retained edge, but the total retained root count \(R_G\) of a graph can vary, so an expectation of normalized thickness cannot replace that random denominator by \(R_0\) without further analysis.

We compared these predictions with 24,000 independently generated graphs across six prespecified fixed edge birth conditions, 4,000 per condition (\hyperlink{figure-S7}{Supplementary Fig. 7}). The simulations generated parent sets directly and conditioned on the forced edge using \(\Pr(J_b=j\mid A_e)=j\eta_b(j)/\mathbb E[J_b]\). Theoretical and simulated mean raw thickness were, for example, 7.21 and 7.25 for a generation-\(0\)-to-\(1\) edge and 1.58 and 1.66 for a generation-\(1\)-to-\(4\) edge, with bootstrap intervals over graphs of [7.09, 7.39] and [1.58, 1.73]. Exact enumeration of small graphs and independent simulations supported the finite state calculation.
\par\begin{figure}[!htbp]\centering\begin{minipage}{\linewidth}\phantomsection\label{fig:S7}\hypertarget{figure-S7}{}\addcontentsline{toc}{subsection}{Supplementary Figure 7. Predicted moments for fixed birth cohorts in the growth ensemble}
{\centering\textbf{Supplementary Figure 7. Predicted moments for fixed birth cohorts in the growth ensemble}\par\smallskip
\includegraphics[width=\linewidth,height=170mm,keepaspectratio]{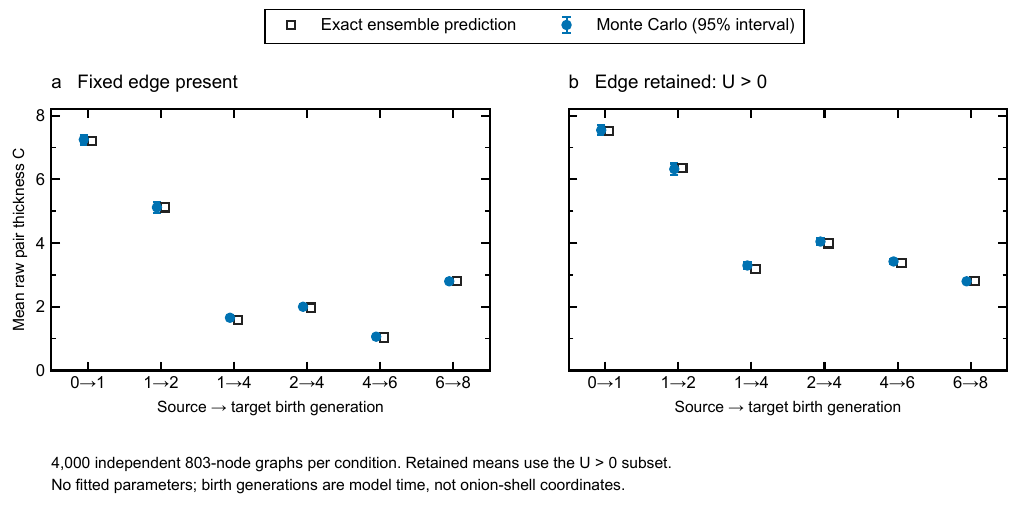}\par}\smallskip
{\fontsize{9}{11}\selectfont Exact ensemble predictions and independent direct simulation for six cohorts defined by the birth times of the focal edges, with 4,000 conditional graphs per cohort. Mean raw pair thickness is compared with the finite state calculation, separately for edges in the full graph and edges retained after backward tracing from terminals. The root count in the raw score is the number of upstream origins reaching the edge tail, and birth generation is original model metadata. These conditional graphs evaluate theory separately from the edge importance test cohort.\par}
\end{minipage}\end{figure}\par
Equation (S14) also separates slope grouping from differences in the average loss of each group. At fixed birth conditions, the expected strict loss does not depend on the upstream root count, even though every realized edge satisfies the root-dependent slope identity. Mixing birth conditions can associate root count with terminal coverage because their conditional means differ. The fixed edge calculation supplies the mean root count and the full terminal coverage distribution. Supplementary Note 7 extends it to the root count distribution and to joint mean counts before and after observation. Within-graph averages include random denominators and therefore target different quantities from the fixed-label expectations and mean counts derived here.

\subsection*{3.6 Terminal sampling converts the growth distribution into an edge predictor}

Let \(1\le m<T\) terminals be sampled uniformly without replacement, independently of the choices that generate the graph. For a fixed existing edge with complete coverage \(U\), the observed downstream count obeys \(k_e\mid U\sim\operatorname{Hypergeometric}(T,U,m)\), and combining this likelihood with Eq. (S13) gives

\[
\Pr(U=u\mid b,k_e,T,m)
=
\frac{\rho_b(u)\binom{u}{k_e}\binom{T-u}{m-k_e}}
{\sum_s\rho_b(s)\binom{s}{k_e}\binom{T-s}{m-k_e}}.
\tag{S15}
\]

Conditioning on edge existence \(A_e\) and on the fixed generation parameters is implicit in Eq. (S15), whose denominator must be positive. Here \(u\in\{k_e,\ldots,T-m+k_e\}\), and summing over \(s=0,\ldots,T\) gives the same result when infeasible binomial coefficients are set to zero. An edge is discovered when \(k_e>0\). The actual positive count already conditions on discovery, so an additional discovery weight would count selection twice. Conditioning the prior on \(U>0\) yields the same posterior because its normalizing constant cancels for a discovered edge.

The sampling law also links the growth ensemble to observation fidelity. For a fixed complete graph and an edge fixed before sampling, let \(W_e=U_e/T\), \(W_{S,e}=K_e/m\) and \(c_m=(T-m)/[m(T-1)]\), with \(T>1\) and \(1\le m\le T\). Uniform terminal sampling gives \(\mathbb E_S W_{S,e}=W_e\) and \(\operatorname{Var}_S(W_{S,e})=c_mW_e(1-W_e)\), moments taken before conditioning on discovery \(K_e>0\). Under the specified formation process, the birth-conditioned growth distribution \(\rho_b\) supplies the distribution of \(U_e\) and hence of the sampling errors of fixed edges.

Edges share the same uniform terminal sample. If \(\mathcal D_e\) and \(\mathcal D_f\) are their complete descendant sets restricted to designated terminals, the inclusion probabilities of uniform sampling without replacement give \(\operatorname{Cov}_S(W_{S,e},W_{S,f})=c_m[|\mathcal D_e\cap\mathcal D_f|/T-W_eW_f]\). This is a covariance over repeated samples of fixed edges, not over edges in one realized view. The fitted slope and scatter of a group therefore depend on terminal set overlap as well as on marginal coverage, root group composition and discovery. Predicting the ensemble distribution of fitted slopes and scatter requires joint structural information beyond the marginal law \(\rho_b\), and neither the marginal predictor nor the mean count law of Supplementary Note 7 supplies these graph level fluctuations.

For the integer hidden coverage \(H_e=U-k_e\in\{0,\ldots,T-m\}\), we use the posterior mean and the corresponding estimates of complete coverage and raw thickness,

\[
\widehat H_e=\mathbb E[U-k_e\mid b,k_e,T,m],
\qquad
\widehat U_e=k_e+\widehat H_e,
\qquad
\widehat C_G(e)=n(e)\widehat U_e.
\tag{S16}
\]

Under strict support, \(\widehat U_e/T\) predicts full terminal loss and \(\widehat H_e/(T-m)\) predicts loss among unsampled terminals. The raw pair estimate keeps the origin multiplier and is evaluated separately in \hyperlink{figure-S5}{Supplementary Fig. 5}. In the primary fresh test its ranking efficiency on unsampled terminals was 0.520, against 0.702 when hidden terminal loss was targeted directly, so a better estimate of pair thickness does not remove the difference between the two counting targets.

The update also has an equivalent growth interpretation. For \(b<L\), conditional on the final-step descendant propensity \(p=p_L(Z_{L-1})\), the sampled and unsampled counts are independent binomial variables with sizes \(m\) and \(T-m\), so \(\widehat H_e=(T-m)\mathbb E[p\mid b,k_e]\), and a larger observed count updates the remaining propensity within its birth-conditioned distribution. In the penultimate generation \(p=2/703\) is fixed for the evaluated model, so the posterior mean of the hidden count is constant across observed counts, and these exact ties are kept in the ranking. For an observed last generation target, \(U=k_e=1\) and hidden coverage is zero.

The posterior uses the summaries \((b,k_e,T,m)\) and the known generation rules. Its conditional mean minimizes expected squared error for a fixed labeled edge. Supplementary Note 2.5 evaluates prediction errors on independent test graphs, averaging over each graph's random candidate population and then assigning equal weight to graphs.

\FloatBarrier\section*{Supplementary Note 4. Extended methods and reproducibility}
\addcontentsline{toc}{section}{Supplementary Note 4. Extended methods and reproducibility}

\subsection*{4.1 Observation procedure}

The indexed predecessor query in Methods returns a complete predecessor list and an explicit empty list for a root. A missing record therefore does not imply zero indegree. Finite-depth replay stops before querying boundary nodes. The acquisition pseudocode is

\begin{quote}\begin{minipage}{\linewidth}\small\begin{verbatim}
V = selected terminal nodes; E = empty set
frontier = selected terminal nodes
while frontier is not empty:
    next_frontier = empty set
    for v in frontier:
        for u in complete_predecessors(v):
            add edge (u, v) to E
            if u is not already in V:
                add u to V and next_frontier
    frontier = next_frontier
return the observed graph (V, E)
\end{verbatim}\end{minipage}\end{quote}

The bounds of \(|V_S|\) queries and \(O(|V_S|+|E_S|)\) processing assume an indexed oracle and constant-time set bookkeeping, and exclude oracle construction, data transfer, canonical sorting and support propagation.

\subsection*{4.2 Planted root recovery and ranking design}

Methods specifies the planted generator, attachment weights and exact edge budget; the final terminal pools contain 760, 1,960 or 3,960 nodes.

Each factorial condition has 60 graphs. Nested root-discovery orders use the nominal fraction grid

\[
\{.001,.0025,.005,.01,.02,.04,.08,.16,.24,.32,.50,.67,.80,.90,.95,1\}.
\]

Pair ranking recovery used 32 orders and the grid

\[
\{.16,.24,.32,.50,.67,.75,.80,.85,.90,.925,.95,.975,.99,1\}.
\]

For a nominal fraction \(f\in(0,1]\), the sample size is \(m=\min\{T,\max(1,\lceil fT\rceil)\}\); coincident integer sizes are evaluated once, and required fractions use \(m/T\). Root and stable ranking success follow Methods, and each graph averages its first successful fractions across its registered orders.

Pair scores use exact endpoint sets, and ranking uses all reference edges with positive counts, scoring unobserved edges zero. Ties receive average ranks. Ordinary correlation is undefined for constant vectors, and an inherited recovery convention assigns one to an exactly identical constant score vector; no such degenerate correlations occurred. One census state per graph closes all its sampling trajectories.

Empirical condition crossings use mean stable success of at least 0.95 with equal graph weights. A 10,000-replicate hierarchical bootstrap, resampling graphs and then whole terminal trajectories within each graph, describes the stability of the crossings.

\subsection*{4.3 OpenAlex finite reference replay}

OpenAlex acquisition and cleaning follow Methods. Each subset is replayed from its own terminals and their immediate predecessors, so its two step query boundary is relative to the sample, and terminal membership follows work identity, not sink status \textsuperscript{\hyperlink{supp-ref:1}{1}}.

The 512 stratified orders produced 5,120 sample rows but only 4,607 distinct information states. In the complete replay, 13,948 components form the backbone with \(Q\ge0.10\), where \(Q\in[0,1]\) is the fraction of designated terminals reachable from a component. Each sample recomputes its boundary roots instead of inheriting those of the complete reference.

Backbone comparisons use work identities, and rankings use stable edges between singleton components, with missing reference edges scored zero in the full-reference comparison. Main \hyperlink{figure-3}{Fig.~3d} compares path and pair counts with restricted or all roots on a fixed set of 44,879 reference edges. Restricted roots reach at least 10\% of the designated terminals and are selected separately in each scored graph.

Top-one-percent diagnostics keep cutoff ties. The planted cutoff uses the full universe of edges with positive counts, whereas OpenAlex uses each view’s stable candidates with positive scores. Top-set recovery requires both precision and recall of at least 0.90, maximum-set recovery requires exact agreement of the tied maxima, and joint tail success follows the planted nesting rule.

\subsection*{4.4 Networks used for pair thickness vulnerability}

Methods, “Vulnerability networks”, specifies all generators, graph boundaries, counts and terminal samples. The Growth analysis graph is the ancestor closure of the final terminals within the 803-node construction.

The binary tree, with one predecessor per nonroot, supplies the structural boundary for tolerance. Main \hyperlink{figure-4}{Fig.~4} and \hyperlink{figure-S1}{Supplementary Fig. 1} use 30 observations of it, and the tolerance scan in main \hyperlink{figure-6}{Fig.~6} and the importance analyses use one fixed observation.

HEP-TH vulnerability and importance analyses use the primary observation. Supplementary Note 12 and \hyperlink{figure-S15}{Supplementary Fig. 15} examine ten nested terminal orders.
\par\begin{figure}[!htbp]\centering\begin{minipage}{\linewidth}\phantomsection\label{fig:S15}\hypertarget{figure-S15}{}\addcontentsline{toc}{subsection}{Supplementary Figure 15. Sampling and root count variation in three empirical references}
{\centering\textbf{Supplementary Figure 15. Sampling and root count variation in three empirical references}\par\smallskip
\includegraphics[width=\linewidth,height=146mm,keepaspectratio]{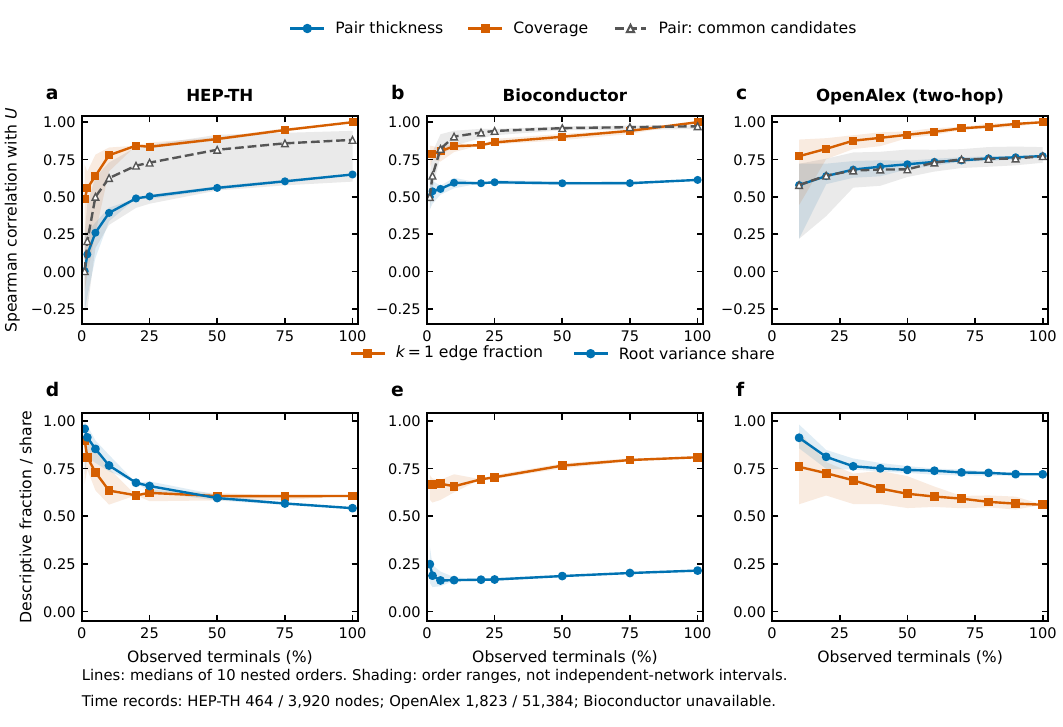}\par}\smallskip
{\fontsize{9}{11}\selectfont \textbf{a to c}, Spearman correlation of sampled pair thickness (blue circles) or sampled coverage (orange squares) with complete terminal coverage \(U\) on discovered edges; gray dashed triangles hold each order's candidates fixed at its smallest budget. \textbf{d to f}, Fraction of discovered edges reaching exactly one sampled terminal, and the descriptive root variance share defined in Supplementary Note 11. Lines show medians across ten nested orders, and shading shows their full ranges within each topology. HEP-TH and Bioconductor orders keep their exact original primary samples. In OpenAlex, acquisition is limited to two steps from the sample, and boundary roots are recomputed for each observation. Complete tracing within the fixed reference is tabulated separately. Each column is one frozen topology. Date availability is given below the panels, and missing or conflicting dates were not imputed. These overall correlations measure a different property from the efficiency of selecting the few highest ranked edges.\par}
\end{minipage}\end{figure}\par
Every vulnerability observation is a complete ancestor closure in which all observed edges are candidates, and each cut is evaluated independently in the corresponding complete analysis graph.

\subsection*{4.5 Exact parameter response and statistical comparisons}

The exact recurrence of Supplementary Note 3.4 keeps predictors, candidate populations and terminal denominators fixed, and Methods gives the five snapshots and the treatment of equality and right limits at critical points.

Cuts entering the same target have identical response curves, so one curve was calculated per target and assigned to every incoming candidate edge, while root counts and thickness remain specific to each edge. The scan covers 30,297 targets and 307,440 combinations of edge and snapshot.

Methods specifies slope estimability, supported groups and the five folds blocked by target. Targets are ordered deterministically by their hashed identities and assigned to folds cyclically. Every fold must contain supported test edges. Group eligibility and fold assignment are fixed before tolerances are compared.

All regressions use the same supported candidates. The pooled cubic uses centering and scaling estimated on training data only and requires four predictor values and full rank. Out-of-fold coefficients pool residual sums before aggregation across graphs, the grouped minus pooled difference is paired within each graph, and negative predictive scores are kept.

An all-zero response has loss zero but undefined \(H_\tau\) and predictive \(R^2\), so BA shape and predictive summaries use only five graphs at the last two tolerances, while loss summaries keep all 30. Synthetic curves report graph medians and interquartile ranges.

The pooled cubic can extrapolate poorly beyond its training predictor range. We retain these failures as extrapolation errors and document the independently checked extreme case in the archive.

\subsection*{4.6 Edge importance scores and evaluation on unsampled terminals}

Methods, under “Structural importance benchmark”, defines the endpoint universes, path weights, betweenness scores, budgets and ties. All methods use the same observed closure and original indegrees. Pair thickness restricts NPPC to roots and designated terminals, whereas the actual sinks used by SPC and SPLC can differ from those terminals \textsuperscript{\hyperlink{supp-ref:3}{3},\hyperlink{supp-ref:4}{4},\hyperlink{supp-ref:5}{5}}.

The rows attributed to this study are pair thickness, root-corrected coverage and the Growth posterior estimates. Under strict support, the observed cascade equals \(W_S=k_e/m\), making it identical to the coverage baseline. Under tolerant rules it is calculated separately, and every static score, including random ranking, receives the same initiation gate.

Besides mean selected loss and oracle efficiency, the retained outcomes include normalized discounted cumulative gain with linear relevance, rank correlation and nonzero-loss frequency, each evaluated separately on all and on unsampled terminals. Target-level selection uses each target’s maximum incoming score (Methods).

Oracle efficiency is undefined when oracle loss is zero, whereas absolute loss stays zero, so at full tolerance BA has five graphs with defined full-terminal efficiency but four for unsampled terminals. The original benchmark intervals are exploratory.

\subsection*{4.7 Growth theory and independent posterior tests}

Equation (S12) propagates the full descendant distribution, giving birth-specific terminal coverage and the mean upstream root count. Six fixed edge cohorts, \(0\to1\), \(1\to2\), \(1\to4\), \(2\to4\), \(4\to6\) and \(6\to8\), each have 4,000 directly generated graphs. To condition on edge existence, simulations use the size-biased target indegree distribution of Note 3.5. They generate parent sets directly rather than drawing from the theoretical transition. Means before and after terminal tracing are assessed separately, and these 24,000 conditional graphs are separate from the importance test cohorts.

The prior \(\rho_b\) was fixed before 100 unconditioned test graphs were generated. Each graph supplies independently sampled, not necessarily nested, observations of 10, 20 and 40 terminals. Predictors use the observed topology, generator birth labels, sampled identities and fixed parameters, and complete coverage is reserved for evaluation.

All five coverage predictors estimate \(H_e\) and set \(\widehat U_e=k_e+\widehat H_e\). The uncorrected extrapolation uses \((T-m)k_e/m\), and the boundary-aware version uses that value for \(b<L\) and zero for the final generation. A uniform prior on \(U=0,\ldots,T\) with the same sampling likelihood gives \((T-m)(k_e+1)/(m+2)\) for \(b<L\), with the known point mass at \(U=1\) as the final generation prior. For the birth and discovery control, define \(q_m(u)=1-\binom{T-u}{m}/\binom{T}{m}\); its hidden count mean is

\[
\widehat H_{\mathrm{birth}}
=
\frac{\sum_u u\rho_b(u)q_m(u)-(m/T)\sum_u u\rho_b(u)}
{\sum_u\rho_b(u)q_m(u)}.
\tag{S17}
\]

This control uses the birth generation and the discovery event for the hidden part, and the observed count is added when its full coverage prediction is constructed. The fifth predictor is the Growth posterior of Eq. (S16). The uniform prior and boundary-aware extrapolation rank unsampled outcomes identically, although their count predictions differ. To give structural methods the same known boundary, their versions evaluated on unsampled terminals score final generation targets zero. The raw posterior pair score \(n\widehat U\) is kept separately. Theoretical ties in the penultimate generation are represented identically, and floating-point ranking scores are rounded to 12 significant digits to remove numerical tie splitting.

The primary outcome was the mean squared error of \(H_e\) at \(m=20\), averaged first over distinct discovered nonterminal targets and then over the 100 graphs. Auxiliary outcomes included count and raw thickness errors, discrete equal-tail interval coverage and width, the discrete continuous ranked probability score (CRPS) and the Brier score for zero hidden coverage. Errors for full and hidden counts coincide, because \(\widehat U-U=\widehat H-H\), so they provide the same error comparison. Proper scores and the birth-resolved probability check were interpreted separately from interval coverage.

Uncertainty used 2,000 graph bootstrap replicates. Three primary paired comparisons tested the posterior against boundary-aware coverage, the uniform prior and birth and discovery. One-sided sign randomization used 100,000 draws and a plus-one estimate, which requires sign exchangeability or symmetry, and all three Holm-adjusted values were \(3.0\times10^{-5}\). Separate exact binomial tests of a win probability above one half excluded ties and used their own Holm adjustment, and the posterior won in 100, 100 and 90 of 100 graphs. Ranking, other sample sizes and birth strata were descriptive.

\FloatBarrier\section*{Supplementary Note 5. Mathematical correspondence with prior methods}
\addcontentsline{toc}{section}{Supplementary Note 5. Mathematical correspondence with prior methods}

\subsection*{5.1 Pair counts, reverse reachability and the strict edge target}

NPPC counts distinct connected endpoint pairs whose paths include an edge \textsuperscript{\hyperlink{supp-ref:3}{3},\hyperlink{supp-ref:4}{4}}. Let \(\operatorname{Anc}^{*}_H(u)\) and \(\operatorname{Desc}^{*}_H(v)\) include the endpoint itself. For any start and end sets \(A,B\subseteq V_H\), the path concatenation argument of Supplementary Note 3.1 shows that the number of pairs connected through \(e=(u,v)\) is \(P_{A,B}(e)=|A\cap\operatorname{Anc}^{*}_H(u)|\,|B\cap\operatorname{Desc}^{*}_H(v)|\). Taking \(A=B=V_H\) gives conventional NPPC, and taking \(A=\mathcal R_H\) and \(B=\mathcal T_H\) gives equation (1), so this endpoint restriction is the precise relation between the two scores.

The restriction can change rankings. On the chain \(r\to a\to b\to t\), with root \(r\) and designated terminal \(t\), NPPC assigns the successive edges scores of \((3,4,3)\) and pair thickness assigns \((1,1,1)\). The latter counts the single origin and outcome relation behind each prerequisite in the chain. In larger graphs pair thickness keeps the multiplicity of origin conditions, which can then be separated from the number of exposed outcomes. The NPPC papers use their weights to analyze citation connectivity and extract main paths, whereas the present analysis connects the root and terminal count to functional outcome loss under the specified dependency rule.

Other measures also distinguish sources from targets. The hourglass analysis of Sabrin and Dovrolis counts source-to-target paths through a vertex and factorizes path centrality into upstream and downstream path counts \textsuperscript{\hyperlink{supp-ref:6}{6}}, with several paths between the same endpoints contributing separately. Intermediacy instead measures the probability that at least one active path between a selected source and target passes through a vertex \textsuperscript{\hyperlink{supp-ref:7}{7}}. These quantities measure path participation or probabilistic connectivity, whereas equation (3) relates distinct endpoint pairs to the loss of designated outcomes after one necessary relation fails.

\textbf{Serial subdivision and process detail.} Replace \(e=(u\to v)\) in \(G\) by \(e_1=(u\to w)\) and \(e_2=(w\to v)\), where \(w\) is a new nonterminal node with indegree and outdegree one, and keep the original root and designated terminal sets in the resulting graph \(G'\). Reachability between original nodes is unchanged, the roots reaching \(w\) are exactly those reaching \(u\), and its designated terminal descendants are exactly those of \(v\). Hence \(C_{G'}(e_1)=C_{G'}(e_2)=C_G(e)\) and \(\bar C_{G'}(e_1)=\bar C_{G'}(e_2)=\bar C_G(e)\), every other original edge keeps its pair score, and this holds for internal designated terminals as well as sinks. Inserting \(w\) between \(a\) and \(b\) in the chain above changes NPPC to \((4,6,6,4)\), while all pair scores and strict terminal losses remain one. Counting all nodes as endpoints makes the added process detail part of the measured quantity, whereas fixing origin and outcome roles preserves the original counting target.

Under strict support, cutting either replacement edge removes the input originally supplied to \(v\) and produces the same failed set among original terminals. In fact \(I^{G'}_{e_1}(\tau)=I^{G'}_{e_2}(\tau)=I^G_e(\tau)\) for every \(0\le\tau\le1\) under equation (2). The new node always requires its sole input, because \(\max\{1,\lceil\tau d_w\rceil\}=1\), and the indegree of \(v\) is unchanged. When another original edge is cut, \(w\) relays the availability of \(u\); when either replacement edge is cut, the corresponding input to \(v\) is unavailable. Topological induction therefore preserves the final states of all original nodes, and the same correspondence holds for complete ancestor closures of an identical terminal sample. This invariance applies to corresponding edge scores and terminal loss responses. Subdivision adds an edge and, when traced, a query, so it can change summaries weighted by edges, fixed percentage intervention budgets and finite-depth observations. The argument uses serial subdivision and the original support rule, with no added delays or independent failure sources.

For a sampled terminal \(t\), define \(\mathrm{RR}(t)=\{v:v\leadsto t\}\), including \(t\). Then \(k_e/m=m^{-1}\sum_{t\in\mathcal T_S}\mathbf1\{v\in\mathrm{RR}(t)\}\) for \(e=(u,v)\), which is the singleton coverage frequency of reverse influence sampling (RIS) with every propagation probability set to one and the sampled target population restricted to designated terminals \textsuperscript{\hyperlink{supp-ref:8}{8},\hyperlink{supp-ref:9}{9}}. The equality holds sample by sample. Strict dependency gives it a failure interpretation because losing a necessary input fails the target, and a failed predecessor propagates failure downstream. All descendants therefore fail, and \(k_e/m\) is also the observed strict terminal loss. Matched RR coverage and corrected coverage would reproduce exactly the same benchmark row.

The deterministic terminal count also specializes weighted influence \textsuperscript{\hyperlink{supp-ref:10}{10}}. Assigning unit weight to designated terminals and zero to other nodes, setting propagation probabilities to one and initializing the failed state at the edge head maps strict failure propagation to a reachability count. The quantity counts outcomes reached from a failed edge head. Marginal edge influence instead compares the influence of a seed set before and after edge deletion.

The complete procedures have different outputs. RIS keeps reverse influence sets for greedy optimization of the joint coverage of a seed set, whereas our tracing keeps the union of queried dependency edges and identifies roots and support relations before evaluating independent edge failures. Both allow local adjacency access and differ in target population, retained information and downstream task. Uniform sampling without replacement, as in the vulnerability experiments, gives a hypergeometric coverage count for a fixed edge, whereas independent repeated target draws give a binomial count. Both designs estimate the same coverage proportion but have different sampling distributions. Supplementary Note 3.6 treats conditioning on edge discovery. Shared queries are an implementation choice. The acquisition bound of Supplementary Note 4.1 counts predecessor queries and graph construction, excluding seed-set optimization and score computation.

Necessary dependency already has a functional interpretation in software. Abate et al. define a strong dependency by necessary coexistence in every healthy installation and rank packages by the size of their impact sets \textsuperscript{\hyperlink{supp-ref:11}{11}}; their score belongs to a component, and its impact population comprises dependent components. Setó-Rey et al. measure package vulnerability through pairs of defect sources and affected packages, counting different defects separately even when they reach the same package \textsuperscript{\hyperlink{supp-ref:12}{12}}. Here the intervention is one required relation, the outcomes are designated terminals, and equation (3) separates distinct upstream origins from distinct exposed outcomes. These established notions of necessary dependency and downstream consequences provide the basis for relating pair thickness to the root groups studied here.

All incoming edges of one target have the same complete strict loss and corrected coverage, while their upstream root counts, and hence pair thicknesses, can differ. This motivates separating origin multiplicity, observation calibration and support substitution when interpreting the thickness and loss response of an edge.

\subsection*{5.2 Threshold aggregation and dominators}

The numerical local operator in equation (S10) is a discrete Sugeno integral \textsuperscript{\hyperlink{supp-ref:13}{13}}. To see this, encode inputs that never fail as zero and use the capacity \(\mu(A)=|A|/d_v\in[0,1]\), where \(A\) is any subset of the original input set; the sorted Sugeno expression in descending order is then \(\max_j\min\{a_{(j)},j/d_v\}\), and padded inputs of value zero do not affect a positive maximum. The flag for never failing must still be kept separately, because a node with no affected supports does not fail even at tolerance zero. This established aggregation form yields the topological recursion and the terminal tail interpretation when applied to the dependency support rule, as proved in Supplementary Note 3.4.

The full-tolerance endpoint is also a standard reachability concept. If a super-root with arcs to every original root is added and edge \(e\) is subdivided by a new vertex \(w_e\), the designated terminals counted by \(D_G(e)\) are precisely those dominated by \(w_e\) in the resulting flowgraph \textsuperscript{\hyperlink{supp-ref:14}{14}}, which supplies the standard bottleneck interpretation.

\subsection*{5.3 Growth laws and prediction targets}

Work on recursive DAGs already treats multiple roots, parent selection without replacement, branching and urn representations and exact or asymptotic reachability distributions \textsuperscript{\hyperlink{supp-ref:15}{15},\hyperlink{supp-ref:16}{16},\hyperlink{supp-ref:17}{17},\hyperlink{supp-ref:18}{18}}. Direction conventions must be aligned, because upstream reachability from a newest vertex and the downstream descendants of a fixed older vertex are different random objects. Our finite state calculation uses a fixed parent pool within each generation and targets the increment of the last generation.

Bayesian finite population prediction is established \textsuperscript{\hyperlink{supp-ref:19}{19}}, with recent approaches using predictive models for the unobserved population \textsuperscript{\hyperlink{supp-ref:20}{20}}. Here the model-specific objects are the final cohort coverage prior and the joint mean count of root multiplicity, full coverage and sampled coverage after mixing birth conditions. The root subset chain and inclusion-exclusion of Supplementary Note 7 resolve root class composition within this ensemble, and the independent tests quantify that composition and its response to parent choice restricted to branches.

The comparison that removes dependence after pooling isolates the role of the birth mixture analytically. The posterior is evaluated separately against coverage and birth and discovery controls. These validations condition on the true generator and the original birth labels.

\subsection*{5.4 Observation models and scope of comparison}

Traceroute sampling, completion from an induced subgraph, reconstruction from noisy interactions and queries that recover the ancestor closure expose different information \textsuperscript{\hyperlink{supp-ref:21}{21},\hyperlink{supp-ref:22}{22},\hyperlink{supp-ref:23}{23},\hyperlink{supp-ref:24}{24},\hyperlink{supp-ref:25}{25}}, and ecological reconstruction is another setting in which the observation model is essential \textsuperscript{\hyperlink{supp-ref:26}{26},\hyperlink{supp-ref:27}{27}}. Related applications link reconstruction to risk in supply networks \textsuperscript{\hyperlink{supp-ref:28}{28},\hyperlink{supp-ref:29}{29}} and rank infrastructure components by failure consequences \textsuperscript{\hyperlink{supp-ref:30}{30},\hyperlink{supp-ref:31}{31}}. Comparisons with these approaches require matching the intervention, outcome population and information available to each method.

\subsection*{5.5 Upstream and downstream factors in classical weights}

On the observed DAG \(H\), for \(e=(u,v)\in E_H\), write \(a_H(u)=\lvert\operatorname{Anc}^{*}_H(u)\rvert\) and \(d_H(v)=\lvert\operatorname{Desc}^{*}_H(v)\rvert\), which include the focal nodes and take integer values from one to \(\lvert V_H\rvert\). Let \(p_H(x,y)\) be the nonnegative integer number of directed paths, with \(p_H(x,x)=1\), and for \(A,B\subseteq V_H\) define \(P_A^{-}(u)=\sum_{x\in A}p_H(x,u)\) and \(P_B^{+}(v)=\sum_{y\in B}p_H(v,y)\). The set \(\mathcal Z_H\) contains actual sinks, whereas designated terminals \(\mathcal T_H\) need not be sinks.

The unnormalized benchmark scores factor as follows \textsuperscript{\hyperlink{supp-ref:3}{3},\hyperlink{supp-ref:4}{4}}, each edge score being the product of its two factors.

\begingroup\small
\setlength{\tableunit}{\dimexpr\linewidth-4\tabcolsep\relax}
\begin{longtable}{@{}>{\raggedright\arraybackslash}p{0.160000\tableunit}>{\raggedright\arraybackslash}p{0.420000\tableunit}>{\raggedright\arraybackslash}p{0.420000\tableunit}@{}}
\toprule
Score & Upstream factor & Downstream factor \\
\midrule\endhead
Pair thickness & \(n_H(e)\) & \(U_H(e)\) \\
NPPC & \(a_H(u)\) & \(d_H(v)\) \\
SPC & \(P_{\mathcal R_H}^{-}(u)\) & \(P_{\mathcal Z_H}^{+}(v)\) \\
SPLC & \(P_{V_H}^{-}(u)\) & \(P_{\mathcal Z_H}^{+}(v)\) \\
SPNP & \(P_{V_H}^{-}(u)\) & \(P_{V_H}^{+}(v)\) \\
Root to terminal path count & \(P_{\mathcal R_H}^{-}(u)\) & \(P_{\mathcal T_H}^{+}(v)\) \\
\bottomrule
\end{longtable}\endgroup

For pair thickness and NPPC, endpoint reachability gives a Cartesian product. For path scores, every path through the edge has a unique prefix ending at \(u\) and suffix starting at \(v\), and conversely every such pair concatenates to a path because acyclicity prevents a repeated vertex, which establishes the factors. Path recursions count a possible endpoint of length zero and then continue to later endpoints, so a path may end at an internal designated terminal or pass through it to another; restricting the suffix to sinks would change that score.

The downstream factor of pair thickness is, by definition, the number of distinct designated outcomes. NPPC counts all descendants, and the other scores can count several paths to the same outcome, so deleting the upstream factor from a path score does not in general produce terminal coverage, although the factors can coincide on particular graphs. SPC and SPLC have exactly the same downstream factor for any fixed graph and endpoint convention.

Forward search path count already uses forward paths to assess subsequent citation influence, with extensions for decay and path length limits \textsuperscript{\hyperlink{supp-ref:32}{32}}. The present comparison tests how upstream weighting affects prediction of strict loss on unsampled designated outcomes. Given complete downstream coverage, the upstream factor does not enter the strict loss formula; under partial observation it may still be associated with unobserved coverage, so removing it need not improve prediction. Supplementary Note 13 reports the empirical comparison and implementation checks.

\FloatBarrier\section*{Supplementary Note 6. Measured TF perturbations and structural prioritization}
\addcontentsline{toc}{section}{Supplementary Note 6. Measured TF perturbations and structural prioritization}

\subsection*{6.1 Independent graph and experimental target}

PPTP-seq measures promoter activity after CRISPR interference of transcription factors in Escherichia coli \textsuperscript{\hyperlink{supp-ref:33}{33}}. We used the glucose-medium responses and the independent, author-organized RegulonDB binding annotation with explicit TF identifiers and operon mapping.

The positive-contribution projection keeps positive annotations between distinct operons, deduplicates directed pairs and preserves TF provenance. Its 694 nodes, 988 edges and 68 roots form a DAG without any response-based selection of edges or cycles; negative, mixed, unknown-sign and within-operon relations are excluded. Roots mark the absence of recorded positive inputs, not all molecular origins, and the projection does not assert strict prerequisites.

Each intervention aggregates only the recorded regulatory links whose TF provenance matches the targeted gene. These links form the TF relation bundle; other links from the same operon are excluded. CRISPRi polarity, knockdown efficiency, reporter context, growth effects and feedback can affect steady-state responses even though feedback edges are excluded, and binding-record counts are not comparable interaction strengths.

\subsection*{6.2 Observation, eligibility and outcome}

Ten uniform observations each sample 154 of 309 candidate promoter or operon outputs and withhold 155, giving closures of 197 to 209 nodes and 235 to 272 edges. All methods share each observed graph and its sampled outputs. Eligible TFs stay in the panel even when none of their relations is discovered, with score zero and discovery flagged separately.

A usable response has a finite FC, the log2 activity fold change relative to negative controls, and at least two replicates. Exact TF matching and at least 20 usable unsampled outputs in every view retain 97 of 119 annotated bundles, with 13 lacking identifiers and nine failing availability, and eligibility ignores response magnitude. A deterministic TF ordering assigns 20 TFs to development and 77 to primary testing, with no fitting to responses, and all 97 form a prespecified sensitivity panel.

The outcome is the mean of \(\max(0,-\mathrm{FC})\) over usable unsampled outputs, which measures an activity decrease rather than a binary terminal failure. Test denominators range from 26 to 152 (median 127) and are identical across methods for each TF. Missing responses are not filled with zeros. The estimand therefore averages the available measured outputs within the set of 155 unsampled candidates. Upward and net changes are retained as direction diagnostics.

\subsection*{6.3 Scores, budgets and ties}

Bundle pair thickness is the union of distinct root and sampled terminal pairs whose paths pass through at least one observed bundle edge. Terminal coverage is the union of sampled terminals reachable from those edge heads, and a separately implemented deterministic RR membership calculation gives exactly the same score. NPPC similarly takes the union of endpoint pairs drawn from all nodes across bundle edges. Each union removes duplicates caused by overlapping bundle edges while retaining the endpoint roles of its score.

The other scores are direct-target counts, root-to-terminal paths and sums of edge betweenness over all endpoints or over root and terminal endpoints. Each TF has one source operon, so a DAG path cannot traverse two of its outgoing bundle edges. Constant random scores use exact tie expectations. The identical deterministic RR coverage verifies the terminal coverage calculation.

Primary selection takes eight of the 77 test TFs (10\%, rounded up), and secondary budgets take four (5\%) and 16 (20\%). Oracle efficiency divides the mean response of the selection by the best mean of the same size and is undefined for a zero denominator. Uniform cutoff ties are handled exactly with rational scores and decimal response means. Spearman correlation uses average ranks, and constant random scores give an undefined correlation. Means and ranges over the ten views are descriptive, without biological confidence intervals or significance tests.

\subsection*{6.4 Results and their scope}

\hyperlink{figure-S8}{Supplementary Fig. 8} reports every distinct registered score and budget. At the primary budget, mean efficiencies are 0.706 for pair thickness, 0.694 for terminal (RR) coverage, 0.688 for NPPC, 0.687 for all-pairs betweenness, 0.683 for the path count between roots and designated terminals, 0.665 for betweenness between roots and designated terminals, 0.644 for the count of direct targets and 0.575 for random selection. Pair thickness exceeds NPPC by 1.78 percentage points, and exceeds NPPC in six of ten views and coverage in seven. Its mean Spearman correlation is only 0.0438, against 0.0340 for coverage and 0.0395 for NPPC, so performance at the selection cutoff coexists with weak ordering across the full TF panel.
\par\begin{figure}[!htbp]\centering\begin{minipage}{\linewidth}\phantomsection\label{fig:S8}\hypertarget{figure-S8}{}\addcontentsline{toc}{subsection}{Supplementary Figure 8. Measured responses to transcription factor perturbations}
{\centering\textbf{Supplementary Figure 8. Measured responses to transcription factor perturbations}\par\smallskip
\includegraphics[width=\linewidth,height=165mm,keepaspectratio]{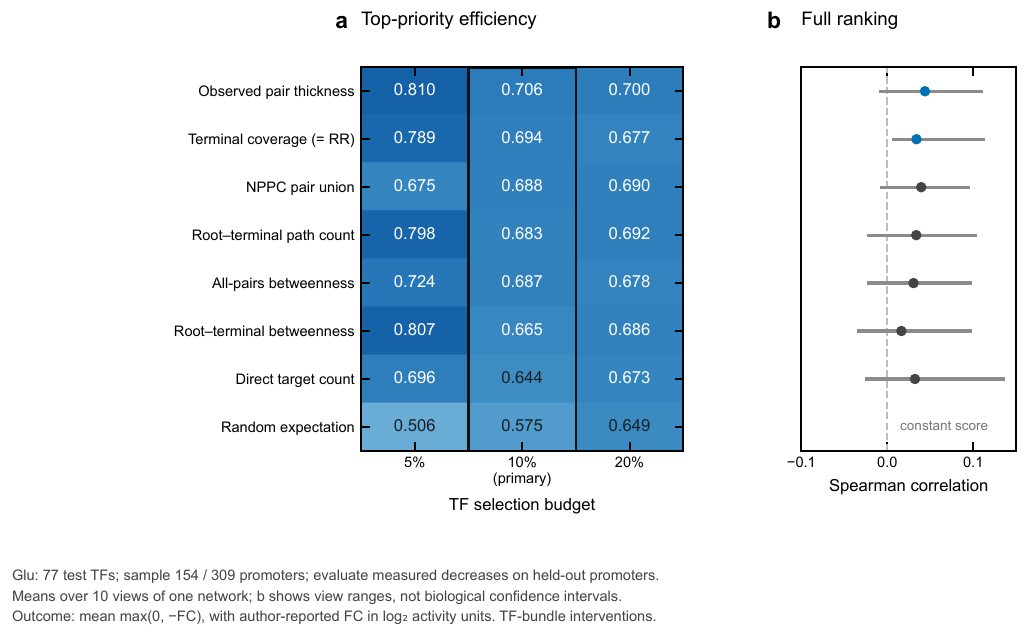}\par}\smallskip
{\fontsize{9}{11}\selectfont \textbf{a}, Mean selection efficiency over ten fixed promoter sampling views for 77 test transcription factors in glucose medium; the primary 10\% budget selects eight factors. \textbf{b}, Ranking correlation with the measured decrease in log2 promoter activity. Dots are view means, and bars span the minimum to maximum across the same ten sampling views of one biological system. Each view samples 154 of 309 promoters and evaluates the 155 unsampled candidates with a finite fold change and at least two replicates. Each factor is evaluated against its mean activity decrease over valid measured outputs, so denominators vary by factor. Each intervention targets one transcription factor and is represented by its associated set of relations. Coverage and deterministic reverse reachability coverage coincide and are shown once, and all registered methods and budgets are kept. The evaluation concerns measured activity decreases following regulator perturbation.\par}
\end{minipage}\end{figure}\par
At 5\%, pair, coverage, NPPC and random efficiencies are 0.810, 0.789, 0.675 and 0.506, and betweenness between roots and designated terminals nearly ties pair thickness at 0.807. At 20\%, the four values are 0.700, 0.677, 0.690 and 0.649, so coverage falls below NPPC, and in the panel of all 97 TFs the 10\% values are 0.688, 0.688, 0.676 and 0.560. The relative performance of the metrics therefore depends on the selection budget.

Across the 770 rows of test TF and view, the mean downward activity burden is 0.0985, the mean upward burden 0.200 and the mean net FC +0.102. Upward responses were substantial and are reported separately from activity decreases. Two pairs of TFs share an operon; flhC and flhD cross the development and test split, whereas gadW and gadX do not. TFs sharing an operon and repeated sampling views are correlated observations from the same biological system. The measured responses provide limited support for using structural scores to prioritize TF perturbations by decreases in promoter activity.

We used fold changes relative to negative controls without subtracting the mean response across promoters to each TF perturbation. Data-use terms for the annotation are distinct from the software license of the repository.

\FloatBarrier\section*{Supplementary Note 7. Prospective composition of root count strata}
\addcontentsline{toc}{section}{Supplementary Note 7. Prospective composition of root count strata}

\subsection*{7.1 Uniform parent root subset evolution and the joint mean count}

The uniform parent process of Supplementary Note 3.5 also admits a scalar recursion for root organization. Together, the descendant and root subset recursions form the uniform special case of the general cohort law in Supplementary Note 8.1. For a fixed subset \(Q\subseteq\mathcal R_G\) of \(q\in\{0,\ldots,R_0\}\) original roots, let \(A_g(Q)\) count all nodes born by generation \(g\) whose entire root ancestry lies within \(Q\); this local subset notation \(Q\) differs from the recovery score of Supplementary Note 4.3. The count \(A_g(Q)\) is an integer between zero and the total population \(M_g=R_0+\sum_{h=1}^gN_h\), with \(A_0(Q)=q\). A newborn has this property precisely when every selected parent does. With the pool of older nodes fixed within each generation, define

\[
h_g(z)=\sum_j\eta_g(j)\frac{\binom zj}{\binom{M_{g-1}}j},\qquad
A_g=A_{g-1}+B_g,\qquad
B_g\mid A_{g-1}=z\sim\operatorname{Bin}(N_g,h_g(z)).
\tag{S18}
\]

Here \(h_g(z)\in[0,1]\) is the probability that every parent belongs to the subset, \(B_g\in\{0,\ldots,N_g\}\) is the number of new nodes with this property and, as elsewhere, \(\binom zj=0\) if \(j>z\). Conditional independence of newborn parent choices gives the binomial update, and propagating its full finite state distribution yields \(f_a(q)=\mathbb E[h_a(A_{a-1}(Q))]\), the probability that the roots of a generation-\(a\) node lie within a given \(q\)-subset. Symmetry among root labels and inclusion-exclusion then give the root count probabilities \(\pi_a(n)\) for integers \(n\in\{1,\ldots,R_0\}\),

\[
\pi_a(n)=\binom{R_0}{n}\sum_{q=0}^n(-1)^{n-q}\binom nq f_a(q),\quad a\ge1;\qquad
\pi_0(n)=\mathbf1\{n=1\}.
\tag{S19}
\]

The calculation allows dependence between individual root reachability events by combining parent selection without replacement with standard inclusion-exclusion over root subsets. Related work on recursive DAGs and urns provides relevant antecedents \textsuperscript{\hyperlink{supp-ref:15}{15},\hyperlink{supp-ref:16}{16},\hyperlink{supp-ref:17}{17}}.

Set \(N_0=R_0\), \(\bar j_b=\mathbb E J_b\) and \(w_{ab}=N_aN_b\bar j_b/M_{b-1}\), the nonnegative expected number of existing edges from birth cohort \(a\) to \(b\). Under uniform choice among labeled parents, the ancestry of the tail of a fixed potential edge is independent of whether that edge is selected. The future terminal coverage of its head depends on later random choices. Equation (S14) therefore extends from moments to the factorization \(\pi_a(n)\rho_b(u)\) conditional on existence. For \(b=L\), \(\rho_L(u)=\mathbf1\{u=1\}\), because each designated terminal covers itself. Uniform sampling of \(m\) terminals gives \(H_m(k\mid u)=\binom uk\binom{T-u}{m-k}/\binom Tm\). Let the nonnegative integer \(N_{n,k,u}\) count generated edges with root multiplicity \(n\), observed coverage \(k\) and full coverage \(u\), where \(u\in\{0,\ldots,T\}\), \(0\le k\le\min(m,u)\) and \(u-k\le T-m\). Linearity of indicator expectations then yields

\[
\mathbb E N_{n,k,u}
=\sum_{0\le a<b\le L}w_{ab}\pi_a(n)\rho_b(u)H_m(k\mid u).
\tag{S20}
\]

Every generated edge contributes, including \(u=0\) and \(k=0\). Restricting to \(k\ge1\) gives the counts of discovered edges, and \((n,k,u)\) maps to raw observed thickness and strict loss \((nk,u/T)\). Shared descendants and terminal samples correlate different edges without altering this expectation. The formula specifies expected counts. In particular, substituting \(R_0\) for the random root count \(R_S\) of the observed closure would not predict the normalization in main \hyperlink{figure-4}{Fig.~4}.

\subsection*{7.2 Independent mean count validation}

\hyperlink{figure-S9}{Supplementary Fig. 9} presents the composition test under uniform growth and the matched control of branch merging. Supplementary Note 8 extends the growth model to signed age preferences.
\par\begin{figure}[!htbp]\centering\begin{minipage}{\linewidth}\phantomsection\label{fig:S9}\hypertarget{figure-S9}{}\addcontentsline{toc}{subsection}{Supplementary Figure 9. Growth and observation predict response group composition under uniform attachment}
{\centering\textbf{Supplementary Figure 9. Growth and observation predict response group composition under uniform attachment}\par\smallskip
\includegraphics[width=\linewidth,height=165mm,keepaspectratio]{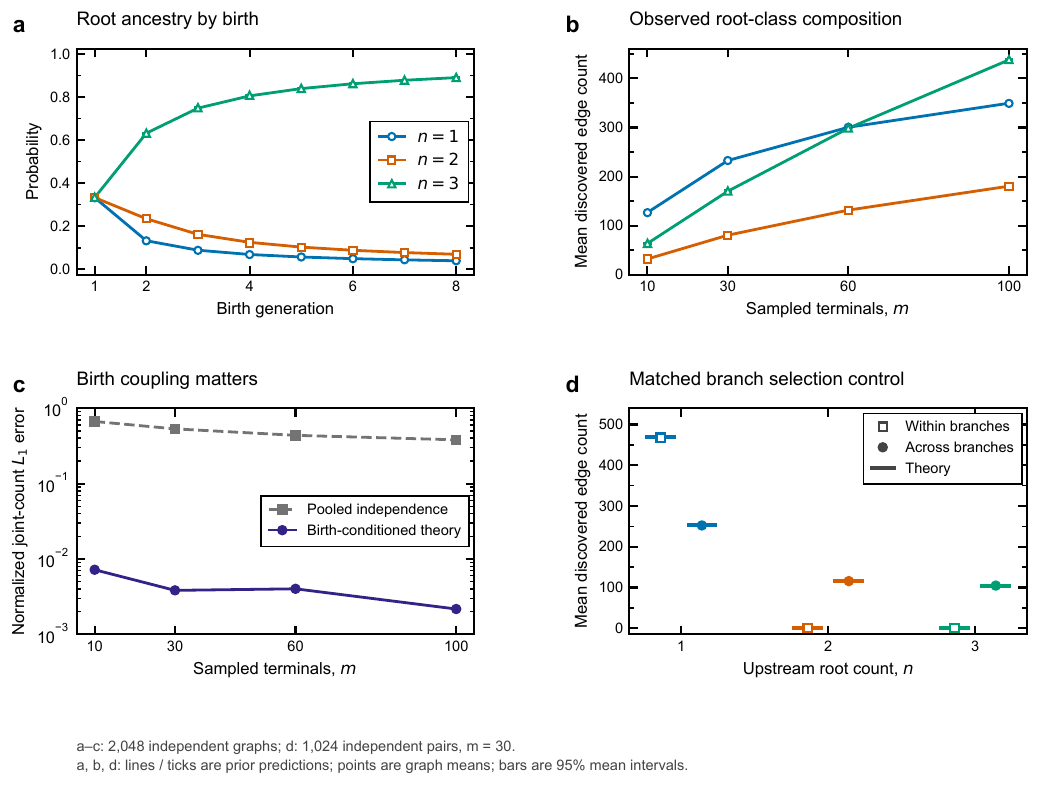}\par}\smallskip
{\fontsize{9}{11}\selectfont \textbf{a}, Probabilities of each root count by birth generation. \textbf{b}, Expected discovered edge counts by upstream root count and terminal sample size. Lines are predictions frozen before 2,048 independent graphs were generated, and symbols are graph means with approximate 95\% t intervals. The process has three roots, eight generations of 100 nodes with the parent pool fixed within each generation, and uniform parent counts on 1,2,3. \textbf{c}, Normalized \(L^1\) error of the ensemble mean discovered \((n,K,U)\) count vector, where \(n\) is the upstream root count and \(K,U\) are the sampled and complete terminal coverage counts. The pooled independence control keeps the exact root count and coverage marginals but removes the association created by mixing birth cohorts, as an analytic diagnostic. \textbf{d}, A separate control of 1,024 graph pairs that share an initial graph with three roots and nine nodes, the input count of each node and the sampled terminal labels. Open squares restrict parents to the same root branch, and filled circles allow parents from different root branches; both arms generate their own futures. Horizontal ticks give theoretical mean counts, and bars are approximate 95\% intervals for graph means. Edges ending in the initial scaffold are excluded, and zero classes are kept. These predictions concern ensemble mean counts in fixed raw coordinates.\par}
\end{minipage}\end{figure}\par
The uniform design uses \(R_0=3\), eight generations of 100 nodes and uniform \(J\in\{1,2,3\}\). Predictions were fixed before explicit parent selection generated 2,048 independent unpruned DAGs, each supplying one uniform terminal order with nested samples of 10, 30, 60 and 100 final generation nodes.

The predicted final generation root proportions for \(n=1,2,3\) are 0.0398, 0.0699 and 0.890, against simulated means of 0.0401, 0.0692 and 0.891, so terminals with multiple roots dominate this ensemble while edge response strata still arise. At \(m=30\), predicted mean discovered counts by root class are 233, 80.6 and 170, against 233, 80.2 and 170 in simulation. The one-root class is predicted to be largest at \(m=10\) and the three-root class at \(m=100\), so growth and discovery together determine the observed composition.

Both \(K\) and \(U\) were binned into 0, 1, 2 to 3, 4 to 7, 8 to 15, 16 to 31, 32 to 63 and 64 to 100. The prespecified diagnostic family contains 27 generation and root proportions, 12 discovered root count means and 768 binned means. Of these 807 entries, 197 have positive variance across graphs and at least 20 nonzero graph observations, making the approximate simultaneous diagnostic applicable. None excludes its prediction at the Bonferroni 99\% threshold. The maximum absolute standardized error is 3.14, below 4.381. Another 468 entries are deterministic boundaries and agree exactly. The remaining 142 sparse or constant entries retain their observed values and undefined intervals and are not counted as validations based on normal intervals. The diagnostic does not assume independence between cells.

To test the role of birth mixing, let \(J(n,u)=\sum_{a<b}w_{ab}\pi_a(n)\rho_b(u)\). An analytical ablation replaces it with \(J_{\rm pool}(n,u)=A(n)B(u)/W\), where \(A(n)=\sum_uJ(n,u)\), \(B(u)=\sum_nJ(n,u)\) and \(W=1600\) is the same expected total edge count, so the two intensities share their root and coverage marginals over all edges. The same hypergeometric observation law preserves total expected discovery, but their predicted root compositions among discovered edges differ, because the ablation removes the association created by mixing birth cohorts. Both marginal distributions are calculated from the growth model.

Normalized \(L^1\) error is the sum of absolute differences between predicted and empirical mean counts in bins of discovered edges, divided by the theoretical expected number of discovered edges. The complete law gives 0.00719, 0.00383, 0.00402 and 0.00217 for \(m=10,30,60,100\), and the ablation 0.666, 0.531, 0.436 and 0.380. At \(m=30\) their absolute error difference is 0.528, with a whole graph bootstrap 95\% interval of [0.521, 0.528] from one thousand resamples that share graph indices across all four nested budgets. These errors assess agreement of ensemble mean count vectors.

\subsection*{7.3 Matched control of merging between root branches}

The control keeps three roots and a shared scaffold in which each root has two single-parent children. Each of the subsequent eight generations contains 99 births, with 33 assigned in advance to each root branch, and paired networks use identical draws \(J\sim\mathrm{Uniform}\{1,2,3\}\) for each node. In arm A, parents are selected uniformly among older nodes of the assigned branch; in arm B, they are selected among all older nodes. The parent pools remain fixed within each generation. Each arm generates its own complete future, with the 99 final births as terminals. One shared order of terminal labels yields samples of 10, 30, 60 and 99, with \(m=30\) primary. Candidate edges have heads born after the scaffold. Its six edges remain in the reachability calculation but are excluded from the outcome and count population.

Initialization is essential. In arm B, the root subset chain starts at \(A_0(Q)=3q\) with pool \(M_0=9\) and subsequent birth counts of 99, and all scaffold tails have \(\pi_0(n)=\mathbf1\{n=1\}\). In arm A, all tails have one root, and the descendant process is computed within a branch with \(M_0=3\) and 33 births per generation, after which the branch expected counts of Equation (S20) are multiplied by three. Both arms use the global sampling law \(H_m(k\mid u)\) with \(T=99\), including arm A, whose \(u\le33\); replacing it by a terminal sample of fixed size within each branch would be incorrect. Each arm has an expected candidate count of \(8\times99\times2=1584\).

Predictions were frozen before 1,024 independent graph pairs were produced. At \(m=30\), arm A has no discovered multiroot edges by construction, whereas arm B has a mean count of 220 and a mean within-graph fraction of 46.4\%, with paired bootstrap intervals of [218, 222] and [46.3\%, 46.6\%] from 2,000 whole-pair resamples. Arm A shows one observed root class and every tested B graph shows all three, an occupancy result for the tested sample. Mean fractions and occupancy probabilities are not obtained by dividing or thresholding theoretical mean counts.

Predicted and measured discovered counts in arm A are 469 and 468, and the predicted root class means in arm B, 253, 115 and 104, compare with measured means of 252, 116 and 105. All 32 registered root class and total mean checks across arms and budgets meet the prespecified \(4\mathrm{SE}+10^{-8}\) tolerance, with a maximum standardized error of 1.39. This control bins each coverage count into 0, 1, 2 to 4, 5 to 9, 10 to 19, 20 to 39, 40 to 69 and 70 to 99, unlike the preceding experiment. Of 1,536 retained joint bins, all 146 with expectation at least one satisfy the secondary diagnostic \(\max(5\mathrm{SE},0.5)\), a Monte Carlo consistency criterion, not a calibrated familywise significance test. The total discovered count differs by 4.78 between arms, with paired interval [1.51, 8.12].

Both arms start from the same roots and scaffold, match each node’s birth schedule and input count, and generate their own subsequent dependencies and outcomes. The comparison tests the predicted change in group composition when parent choices extend across root branches.

These predictions use the specified finite growth model with independent parent choices from a fixed pool within each generation. Supplementary Note 8 tests calibrated age preference within that family.

\FloatBarrier\section*{Supplementary Note 8. Signed age preference and predictive stratification}
\addcontentsline{toc}{section}{Supplementary Note 8. Signed age preference and predictive stratification}

\subsection*{8.1 A law for cohort counts includes preferences for older and recent parents}

The age-dependent attachment of main equation (8) uses a fixed pool in each generation, independent parent choices by different newborns and sequential weighted sampling without replacement within a parent set. Put \(N_0=R_0\). Birth cohort \(h\in\{0,\ldots,g-1\}\) has eligible size \(N_h\), and the total earlier population is \(M_{g-1}=\sum_{h<g}N_h\). The dimensionless age parameter \(\beta\in\mathbb R\) gives a node in that cohort the positive weight \(\omega_h=\exp(\beta h)>0\), after canceling the common factor \(\exp(-\beta g)\). Thus equally available nodes one cohort apart have weights in the ratio \(\exp\beta\). Aggregate cohort choice also depends on cohort populations. Negative beta favors older nodes at each draw. This preference differs from the longer opportunity for selection that an old node already has under uniform growth.

Assume \(\eta_g\) is supported on \(1\le j\le\sum_{h<g}N_h\), so every parent count is feasible. Let \(Z_h\in\{0,\ldots,N_h\}\) be the number of eligible nodes of a specified subset in cohort \(h\), and let \(c_h(t)\in\{0,\ldots,N_h\}\) count selections from that cohort in the first \(t\in\{0,\ldots,j\}\) draws of an ordered sequence \(h_1,\ldots,h_j\) with entries in \(\{0,\ldots,g-1\}\); bold symbols collect the counts across all earlier cohorts. With total weight \(W=\sum_{h<g}N_h\omega_h>0\), the probability \(A_{g,\beta}(\mathbf Z;\mathbf N)\in[0,1]\) of selecting all parents within the subset is

\[
A_{g,\beta}(\mathbf Z;\mathbf N)=
\sum_j\eta_g(j)\sum_{h_1,\ldots,h_j<g}
\prod_{t=1}^{j}
\frac{[Z_{h_t}-c_{h_t}(t-1)]\omega_{h_t}}
{W-\sum_{h<g}c_h(t-1)\omega_h}.
\tag{S21}
\]

Terms containing an unavailable selection have probability zero. The denominators retain the order of the draws, distinguishing this sequential law from the alternative that assigns an unordered set a probability proportional to the product of its weights. When \(\beta=0\), equation (S21) reduces to \(\sum_j\eta_g(j)\binom{\sum Z_h}{j}/\binom{\sum N_h}{j}\), and the total subset size suffices. At nonzero beta, two subsets of equal size but different cohort composition generally have different selection probabilities, which is why a vector state replaces the scalar uniform recursion.

For a fixed subset of \(q\) roots, let \(Z_h\) count nodes whose entire root ancestry lies in that subset; starting from \(Z_0=q\), each newborn qualifies precisely when all its parents qualify. For a focal node born in \(b\), let instead \(D_h\in\{0,\ldots,N_h\}\) count its descendants in cohort \(h\), including itself, with \(D_b=1\) and earlier counts zero; a later newborn is a descendant when at least one of its parents is. Thus

\[
X_g\mid\mathbf Z\sim\operatorname{Bin}\{N_g,A_{g,\beta}(\mathbf Z;\mathbf N)\},
\qquad
Y_g\mid\mathbf D\sim\operatorname{Bin}\{N_g,1-A_{g,\beta}(\mathbf N-\mathbf D;\mathbf N)\}.
\tag{S22}
\]

Here \(X_g,Y_g\in\{0,\ldots,N_g\}\) count new nodes confined to the chosen root subset and new descendants of the focal node, respectively, and each new count is appended to the relevant cohort state after every generation. Conditional binomial sampling follows from independent newborn choices from the same frozen pool, not from independence of their unconditional ancestries. Root exchangeability and the inclusion-exclusion of equation (S19) give \(\pi_a(n;\beta)\), and the final increment \(Y_L\) gives \(\rho_b(u;\beta)\), with \(\rho_L(u;\beta)=\mathbf1\{u=1\}\). Exact finite state propagation enumerates all states of positive probability, and numerical evaluation uses floating-point arithmetic without mean field closure, probability clipping or empirical calibration of those probabilities.

Let \(w_{ab}(\beta)\) be \(N_b\) times the expected number of cohort-\(a\) parents per newborn in cohort \(b\), obtained by summing the same ordered draws with their cohort multiplicities. The uniform expression \(N_aN_b\mathbb E J_b/(\sum_{h=0}^{b-1}N_h)\) is recovered only at zero beta, and no approximation by weights with replacement is used. Within a cohort, a specified edge is selected independently of its tail ancestry. The head’s future descendants depend on its birth and on later random choices, rather than on its incoming ancestry. Conditional on endpoint births and edge existence, the root and terminal laws therefore factorize, and Equation (S20) becomes

\[
\mathbb E N_{n,k,u}(\beta)=
\sum_{a<b}w_{ab}(\beta)\pi_a(n;\beta)\rho_b(u;\beta)
\frac{\binom uk\binom{T-u}{m-k}}{\binom Tm}.
\tag{S23}
\]

The law predicts ensemble mean counts, including zero coverage. Shared descendants and root discovery couple the edges within each graph, so graph-to-graph variation in fitted slopes depends on that joint structure. Substituting mean counts into the nonlinear \(S_{\mathrm{root}}\) does not give its expectation. Degree preference, hidden fitness, sequential growth within cohorts and feedback require different state descriptions. The model is defined for every finite real \(\beta\). Numerical validation uses the parameter grid specified below.

\subsection*{8.2 Fixed definition and validation of root group predictive gain}

For each discovered edge, the predictor is \(x=nK/(R_Sm)\) and the strict response is \(y=U/T\). Main equation (10) compares grouped and pooled fits with the target-blocked folds of Note 4.5. A group needs at least ten targets, three predictor values and two training predictor values in every fold, and all models use the same eligible edges, selected without reference to responses.

Fitted lines include intercepts. Test residuals are concatenated over folds and divided by a common sum of squares of the supported responses. The prespecified pooled cubic uses centering and scaling from training data only, four training values and full rank, and diagnostics of predictor ranges identify extrapolation failures.

The gain \(S_{\mathrm{root}}\in\mathbb R\) measures the predictive information supplied by root labels; it does not depend on axis units, can be negative and requires nonconstant responses for its denominator. We report grouped accuracy, the number of eligible groups, the supported fraction and the zero-response fraction alongside it, and a single group, inadequate support or constant responses leave the gain undefined.

The diagnostic controls use a fixed 40-point predictor grid on \([0.05,1]\) in each of three groups, with 120 distinct target identifiers. The common line response is \(y=0.05+0.35x\) and the branch response \(y=0.45x/n\). For each of 128 independent response datasets, independent noise values \(z\), each uniform on \([-1,1]\), multiply the response by \(1+az\), with noise amplitude \(a\in\{0,0.25,0.5,0.75,1\}\) and group labels \(n\in\{1,2,3\}\); the same noise vector is reused across amplitudes and geometries. A label null permutes root labels once per dataset, preserving predictor and response pairs and group sizes, and no response is clipped. Bootstrap intervals from 2,000 resamples of whole datasets condition on this fixed design. The controls test sensitivity to constructed response noise, not to the correlated hypergeometric noise produced by terminal sampling.

Noiseless gains are 0 for a shared line and 0.502 for distinct branches; at maximum noise their means are \ensuremath{-}0.0309 and 0.243, and shuffled labels give \ensuremath{-}0.0354. Rescaling x by 19 and y by seven leaves the statistic unchanged to numerical precision, and reversing training and test relationships gives \ensuremath{-}0.408. Single-group, all-zero, constant-positive and insufficient-support controls remain undefined. The controls validate how the predictive diagnostic responds to the supplied group labels.

\subsection*{8.3 Prespecified signed beta scan and finite law validation}

The prespecified scan used \(R_0=3\), eight cohorts of 100 nodes and independent uniform \(J\in\{1,2,3\}\), with nine beta values equally spaced from \(-\log4\) to \(\log4\) and 128 independently generated complete graphs per value. One uniform terminal permutation per graph gives nested samples of 30, 60 and 100, and the 30-terminal view is primary. Candidate edges satisfy \(K>0\), and every sample recomputes discovery and predictor support under the unchanged metric definition, so comparisons across sample fractions include changes in the candidate population and supported predictor range as well as in coverage estimation. All 3,456 records of graph observations are kept, including undefined scores.

Supplementary Table 4 reports the primary scan. Means condition on the stated estimable graphs, while the denominator for estimability and root occupancy is always 128. Intervals use 2,000 resamples of whole graph records and keep paired identities for comparisons across samples within graphs. A case with only one estimable graph has no interpretable confidence interval for the graph population, and the degenerate interval returned by the raw bootstrap is kept in the analysis record but not drawn. Hollow markers flag fewer than ten estimable graphs as a display caution, without changing eligibility or discarding conditions.

\par\begin{minipage}{\linewidth}
\textbf{Supplementary Table 4. Primary signed preference scan (30 of 100 terminals).} Mean predictive scores use only estimable graphs and are not unconditional network means.\par\smallskip

\begingroup\small
\setlength{\tableunit}{\dimexpr\linewidth-6\tabcolsep\relax}
\begin{tabular}{@{}>{\raggedright\arraybackslash}p{0.160000\tableunit}>{\raggedright\arraybackslash}p{0.280000\tableunit}>{\raggedright\arraybackslash}p{0.280000\tableunit}>{\raggedright\arraybackslash}p{0.280000\tableunit}@{}}
\toprule
Beta / log(2) & Estimable graphs / 128 & Mean gain & Mean grouped R squared \\
\midrule
-2 & 1 & 0.504 & 0.519 \\
-1.5 & 5 & 0.386 & 0.395 \\
-1 & 16 & 0.465 & 0.502 \\
-0.5 & 78 & 0.519 & 0.562 \\
0 & 126 & 0.567 & 0.686 \\
0.5 & 128 & 0.503 & 0.801 \\
1 & 128 & 0.461 & 0.878 \\
1.5 & 128 & 0.441 & 0.919 \\
2 & 128 & 0.447 & 0.935 \\
\bottomrule
\end{tabular}\endgroup
\end{minipage}\par

At \(\beta=-\log4\), every graph contains all three observed root classes, with mean observed edge proportions of 0.695, 0.137 and 0.168 for one, two and three roots. Yet the two-root group has fewer than three observed predictor values in all 128 graphs, and the three-root group in 127, so the undefined gain reflects insufficient variation within groups, not missing roots or an absence of branches. At full observation, 47 of these graphs become estimable. At \(\beta=\log4\), all primary views are estimable and the three-root edge fraction averages 0.586.

Under uniform choice, the primary mean gain is 0.567 (95\% interval 0.555 to 0.579), with grouped predictive R squared 0.686 and pooled R squared 0.119; at \(\beta=\log4\) these become 0.447 (0.443 to 0.450), 0.935 and 0.488. The squared correlation between observed and complete thickness improves from 0.416 to 0.873 over all 128 graphs in each condition, whereas the gain is conditional on estimability. The gain and coverage recovery have different denominators and prediction targets. Better calibration can improve both grouped and pooled predictions while narrowing the difference between them, so it need not increase the gain.

Within uniform growth graphs, increasing observation from 30 to 100 terminals changes the gain by \ensuremath{-}0.235 (paired 95\% interval \ensuremath{-}0.247 to \ensuremath{-}0.223; 126 paired estimable graphs). With complete observation, all 1,048 estimable graphs have grouped R squared equal to one within floating-point precision. This exact fit follows from the strict identity and provides a consistency check on the calculation. The records for each graph retain the cubic sensitivity analysis and predictor range diagnostics. Large gains against a cubic that extrapolates beyond the training predictor range are not interpreted as stronger stratification.

A separate finite law check uses three roots, five cohorts of 20 nodes, beta values \(-\log4,\allowbreak-\log2,\allowbreak0,\allowbreak\log2,\allowbreak\log4\) and 128 new graphs per value. Before these graphs were generated, the exact cohort recursion froze the joint mean edge counts at \((n,U)\), including zero coverage. With coverage bins of 0, 1, 2, 3 to 5, 6 to 10 and 11 to 20, the normalized absolute errors of the mean binned counts are 0.0146, 0.0303, 0.0220, 0.0246 and 0.0176, finite sample discrepancies with a shared theoretical expected edge count denominator of 200. The score scan in the larger networks uses a different configuration and does not apply this count prediction. Pointwise bootstrap bands for sparse cells are diagnostics, not a simultaneous acceptance criterion, and their coverage fraction is not used as a success rate.

Predictions and analysis choices were fixed before the formal graphs were generated.

\subsection*{8.4 Historical calibration and independent hidden coverage prediction}

\hyperlink{figure-S10}{Supplementary Fig. 10} uses \(\beta=0,\log2,\log4\), with 128 historical and 512 disjoint test graphs per condition (Methods), and each test observes six terminals. The fitted growth, empirical, uniform-model and known-parameter priors are fixed before testing and share the hypergeometric likelihood.
\par\begin{figure}[!htbp]\centering\begin{minipage}{\linewidth}\phantomsection\label{fig:S10}\hypertarget{figure-S10}{}\addcontentsline{toc}{subsection}{Supplementary Figure 10. Calibrated age preference predictions and controls with the same information}
{\centering\textbf{Supplementary Figure 10. Calibrated age preference predictions and controls with the same information}\par\smallskip
\includegraphics[width=\linewidth,height=150mm,keepaspectratio]{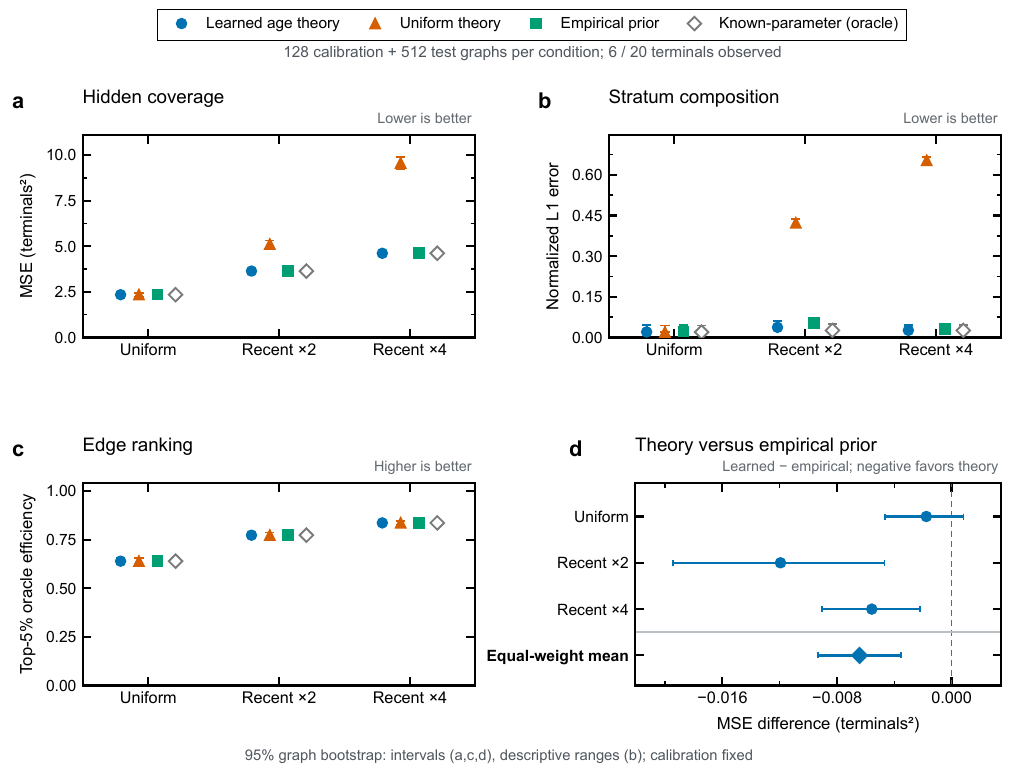}\par}\smallskip
{\fontsize{9}{11}\selectfont \textbf{a}, Hidden count mean squared error over distinct discovered nonterminal targets. \textbf{b}, Normalized absolute error of the ensemble mean discovered count vector. \textbf{c}, Top 5\% oracle efficiency on unsampled terminals. \textbf{d}, Paired difference in prediction error, fitted theory minus the empirical prior, where negative values favor fitted theory. Each preference uses 128 calibration and 512 independent test graphs, with three roots, five cohorts of 20 nodes and six observed terminals. The empirical prior receives the same calibration graphs and test information, and theory with known parameters is a reference using the true model. Bars in a, c and d are conditional 95\% bootstrap intervals resampling whole test graphs; those in b are descriptive bootstrap percentile ranges for the nonlinear discrepancy. Fitted theory and the empirical prior have identical registered top 5\% efficiency in every test graph.\par}
\end{minipage}\end{figure}\par
The estimated values of \(\beta\) are 0.00297, 0.715 and 1.38 for the three conditions, respectively. They are obtained from the likelihood of unordered multisets of parent birth generations, because the order of parent selection is unobserved. Methods specifies the known schedule, empirical pseudocount and terminal generation boundary. Known-parameter theory uses additional information and serves as a reference.

The primary endpoint averaged squared hidden count error over distinct discovered nonterminal targets within each graph and then weighted graphs equally, and the overall comparison weighted the three conditions equally. Fitted theory had errors of 2.34, 3.64 and 4.61 and the empirical prior 2.35, 3.65 and 4.62, a mean difference of \ensuremath{-}0.00642 (conditional 95\% bootstrap interval over graphs \ensuremath{-}0.00932 to \ensuremath{-}0.00352), or a 0.182\% error reduction. Uncertainty used 2,000 paired resamples of whole test graphs conditional on the frozen calibration fit and excludes uncertainty from the calibration sample.

The uniform model had errors of 2.34, 5.12 and 9.56. In the two conditions favoring recent parents, normalized absolute errors of the mean binned discovered \((n,K,U)\) counts were 0.0374 and 0.0264 for fitted theory, 0.422 and 0.653 for uniform theory and 0.0535 and 0.0315 for the empirical prior, with coverage bins of 0, 1, 2 to 3, 4 to 7, 8 to 15 and 16 to 20 and the shared empirical mean count of discovered edges as denominator. The displayed bootstrap ranges for this nonlinear discrepancy are descriptive, not confidence intervals for the true ensemble discrepancy.

At the top 5\% budget, fractional ties were allocated after scores were rounded to 12 decimal places. Fitted theory, the empirical prior and known-parameter theory had identical efficiencies on unsampled terminals in each of the 1,536 test graphs, with condition means of 0.639, 0.773 and 0.836 and a common mean advantage of 2.20 percentage points over boundary-aware extrapolation. The theoretical and empirical priors therefore share the observed ranking gain. The test assesses calibrated prediction within one family with age preference at one graph size.

Supplementary Figs. 4, 5, 7 and 9 keep the uniform special case with separate graph cohorts and sample sizes.

\FloatBarrier\section*{Supplementary Note 9. Analytical prediction of population response strata}
\addcontentsline{toc}{section}{Supplementary Note 9. Analytical prediction of population response strata}

\hyperlink{figure-S11}{Supplementary Fig. 11} presents the constructed controls for \(S_{\mathrm{root}}\), which is calculated within each finite graph using the protocol in Supplementary Note 8.2. We now derive a population prediction from the cohort law. This separates the information carried by root labels across the growth ensemble from fitting error and support selection within individual graphs.
\par\begin{figure}[!htbp]\centering\begin{minipage}{\linewidth}\phantomsection\label{fig:S11}\hypertarget{figure-S11}{}\addcontentsline{toc}{subsection}{Supplementary Figure 11. Validation of predictive gain from root grouping}
{\centering\textbf{Supplementary Figure 11. Validation of predictive gain from root grouping}\par\smallskip
\includegraphics[width=\linewidth,height=150mm,keepaspectratio]{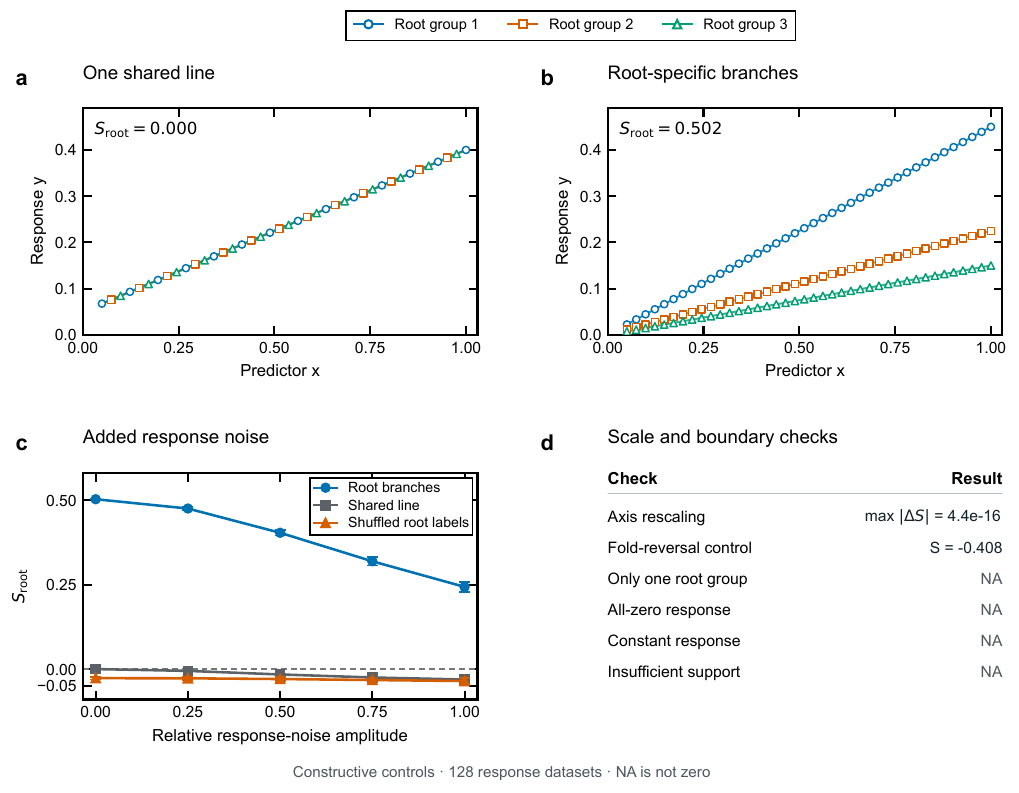}\par}\smallskip
{\fontsize{9}{11}\selectfont \textbf{a,b}, The same 120 targets and predictor values receive either a common linear response with an intercept (a) or three response branches, one per root group (b). Coincident points in a appear as interleaved markers without coordinate jitter, and the complete input is kept. \textbf{c}, Mean \(S_{\mathrm{root}}\) across 128 independent response datasets at each prespecified noise amplitude. The response is multiplied by \(1+az\), where \(a\ge0\) is the prescribed noise amplitude and \(z\) is uniform on \([-1,1]\); each dataset reuses its noise vector across amplitudes, and the null control permutes root group labels while preserving each predictor and response pair. Lines join the sampled conditions, and intervals are conditional 95\% bootstrap intervals from 2,000 resamples of whole response datasets; the noiseless distinct branch and shared line controls are deterministic. This diagnostic tests constructed response noise. \textbf{d}, Positive rescaling of either axis preserves the score to numerical precision, and a prescribed fold reversal example retains a negative score; degenerate comparisons are undefined (NA). \(S_{\mathrm{root}}\) is the difference between five fold cross-validated \(R^2\) from separate group lines and from a pooled line on the same supported candidates, with edges sharing a target kept in one fold. These constructed controls confirm the intended behavior of the measure.\par}
\end{minipage}\end{figure}\par
\subsection*{9.1 From mean counts to conditional responses}

Write \(\mathcal J_{\beta,m}(n,k,u)=\mathbb E N_{n,k,u}(\beta)\) for equation (S23), including zero terminal coverage and undiscovered edges, so that its sum is the expected number of generated edges. Restriction to \(k>0\) defines discovery, with expected count \(Z_{\beta,m}\), and for \(Z_{\beta,m}>0\) the associated population probability law is

\[
P^+_{\beta,m}(n,k,u)=\frac{\mathcal J_{\beta,m}(n,k,u)}{Z_{\beta,m}},
\qquad
Z_{\beta,m}=\sum_{n,k\ge1,u}\mathcal J_{\beta,m}(n,k,u).
\tag{S24}
\]

This ratio of expected counts is the limit of pooling discovered edges from independently generated networks, which differs from choosing a network uniformly and then an edge uniformly. Correlated edges and shared terminal samples leave the law of expected counts valid but matter for simulation uncertainty and for fluctuations in individual networks.

For a supported pair \((n,k)\), \(k>0\), condition equation (S24) to obtain

\[
\begin{aligned}
p_{n,k}(u)&=\frac{\mathcal J_{\beta,m}(n,k,u)}{\sum_v\mathcal J_{\beta,m}(n,k,v)},
\qquad \mu_n(k)=\sum_u\frac{u}{T}p_{n,k}(u),\\
\sigma_n^2(k)&=\sum_u\left[\frac{u}{T}-\mu_n(k)\right]^2p_{n,k}(u).
\end{aligned}
\tag{S25}
\]

Here \(p_{n,k}(u)\in[0,1]\) is a probability mass function on \(u\in\{k,\ldots,T-m+k\}\), \(\mu_n(k)\in[0,1]\) is the conditional mean loss and \(\sigma_n^2(k)\ge0\) its conditional variance, all obtained without fitting test losses. The conditional means need not lie on a straight line. For probability level \(\alpha\in(0,1)\), the quantiles \(q_\alpha=\min\{u/T:\sum_{v\le u}p_{n,k}(v)\ge\alpha\}\) give the central interval \([q_{0.05},q_{0.95}]\), whose discrete endpoints can raise its actual probability above 0.90. The interval describes the spread of individual edge outcomes in the network population weighted by expected edge counts. We obtain these analytical predictions by evaluating the finite state recursion and finite sums numerically.

\subsection*{9.2 Population slopes and the contribution of root labels}

The random predictor is \(X=nK/m\in[0,R_0]\) and the response \(Y=U/T\in[0,1]\). This coordinate omits the random denominator of observed roots \(R_S\) used in \hyperlink{figure-4}{Fig.~4}, so the two sets of slopes convert by a fixed factor only when \(R_S\) is constant across the population, and predicting the original random normalization would require its joint law.

For each nonempty root group, with positive predictor variance under \(P^+\), the optimal slope \(b_n\in\mathbb R\) and intercept \(a_n\in\mathbb R\) are

\[
b_n=\frac{\operatorname{Cov}(X,Y\mid n,K>0)}{\operatorname{Var}(X\mid n,K>0)},
\qquad a_n=\mathbb E[Y\mid n,K>0]-b_n\mathbb E[X\mid n,K>0].
\tag{S26}
\]

Groups with constant predictors use the optimal intercept alone and have undefined slopes, and empty groups carry no population weight. Let \(L_{\mathrm{pooled}}\ge0\) and \(L_{\mathrm{group}}\ge0\) be the minimum mean squared prediction errors of a pooled line and of lines fitted separately by root count. For nonconstant \(Y\), define

\[
S_{\mathrm{pop}}=\frac{L_{\mathrm{pooled}}-L_{\mathrm{group}}}{\operatorname{Var}(Y)}.
\tag{S27}
\]

Nested optimal predictors ensure \(0\le S_{\mathrm{pop}}\le1\). The quantity measures how much root labels improve population linear prediction. It does not measure geometric cluster separation and is not the expectation of the cross-validated \(S_{\mathrm{root}}\), which can be negative because of finite sample fitting. At full observation, \(K=U\) and \(Y=X/n\), so grouped risk is zero, which checks the strict counting identity. More observation need not increase the relative gain. At \(\beta=0\), the predicted grouped \(R^2\) rises from 0.304 to 0.529 to 1 for \(m=2,6,20\), while \(S_{\mathrm{pop}}\) is 0.182, 0.523 and 0.460, because the pooled predictor improves as well.

An untruncated benchmark isolates sampling attenuation. Retain the \(K=0\) outcomes, fix \(n\), and put \(M=\mathbb E[Y\mid n]\in[0,1]\) and \(V=\operatorname{Var}(Y\mid n)\ge0\). Uniform sampling gives \(\mathbb E[K/m\mid Y]=Y\) and a mean conditional sampling variance \(B\ge0\), so

\[
B=\frac{T-m}{m(T-1)}\mathbb E[Y(1-Y)\mid n],
\qquad
b_n^{\mathrm{all}}=\frac{1}{n}\frac{V}{V+B},
\qquad a_n^{\mathrm{all}}=M(1-nb_n^{\mathrm{all}}).
\tag{S28}
\]

This follows from \(\operatorname{Cov}(X,Y\mid n)=nV\) and \(\operatorname{Var}(X\mid n)=n^2(V+B)\). Equation (S28) requires positive predictor variance and applies before conditioning on discovery, whereas slopes among discovered edges use equation (S26). For \(T=1\), the only positive sample is complete and \(B=0\).

\subsection*{9.3 Root information after conditioning on birth}

At fixed \(\beta\) and head birth \(b\), edge existence and tail ancestry depend on choices up to cohort \(b\), whereas head coverage depends on later choices. With attachment based only on age, the future coverage law \(\rho_b\) is independent of the earlier ancestry and of the head indegree, and selecting an incoming edge weights heads by indegree without changing their future coverage law. Independent uniform terminal sampling therefore gives

\[
\Pr(U=u\mid b,n,K=k,\beta)
=\frac{\rho_b(u;\beta)H_m(k\mid u)}{\sum_v\rho_b(v;\beta)H_m(k\mid v)}
=\Pr(U=u\mid b,K=k,\beta).
\tag{S29}
\]

Thus \(n\) changes the mixture of head births in equation (S25), but adds no coverage information once the true birth, sampled count and model parameter are fixed. Conditioning on a positive \(k\) already accounts for discovery. The result allows dependence between edges while conditioning on age-based attachment, known birth labels, independent uniform terminal sampling and a fixed value of \(\beta\).

\subsection*{9.4 Prospective validation and sparse support}

\hyperlink{figure-S12}{Supplementary Fig. 12} uses the five preferences \(-\log4,\allowbreak-\log2,\allowbreak0,\allowbreak\log2,\allowbreak\log4\), with 512 independent networks each and 2,560 overall, three roots, five cohorts of 20 nodes and uniform parent counts from one to three. Nested samples contain \(m=2,6,20\) final generation terminals. Predictions precede test generation, and \(\beta=0,m=6\) is the prespecified primary display.
\par\begin{figure}[!htbp]\centering\begin{minipage}{\linewidth}\phantomsection\label{fig:S12}\hypertarget{figure-S12}{}\addcontentsline{toc}{subsection}{Supplementary Figure 12. Conditional response forecasts across all age preferences}
{\centering\textbf{Supplementary Figure 12. Conditional response forecasts across all age preferences}\par\smallskip
\includegraphics[width=\linewidth,height=192mm,keepaspectratio]{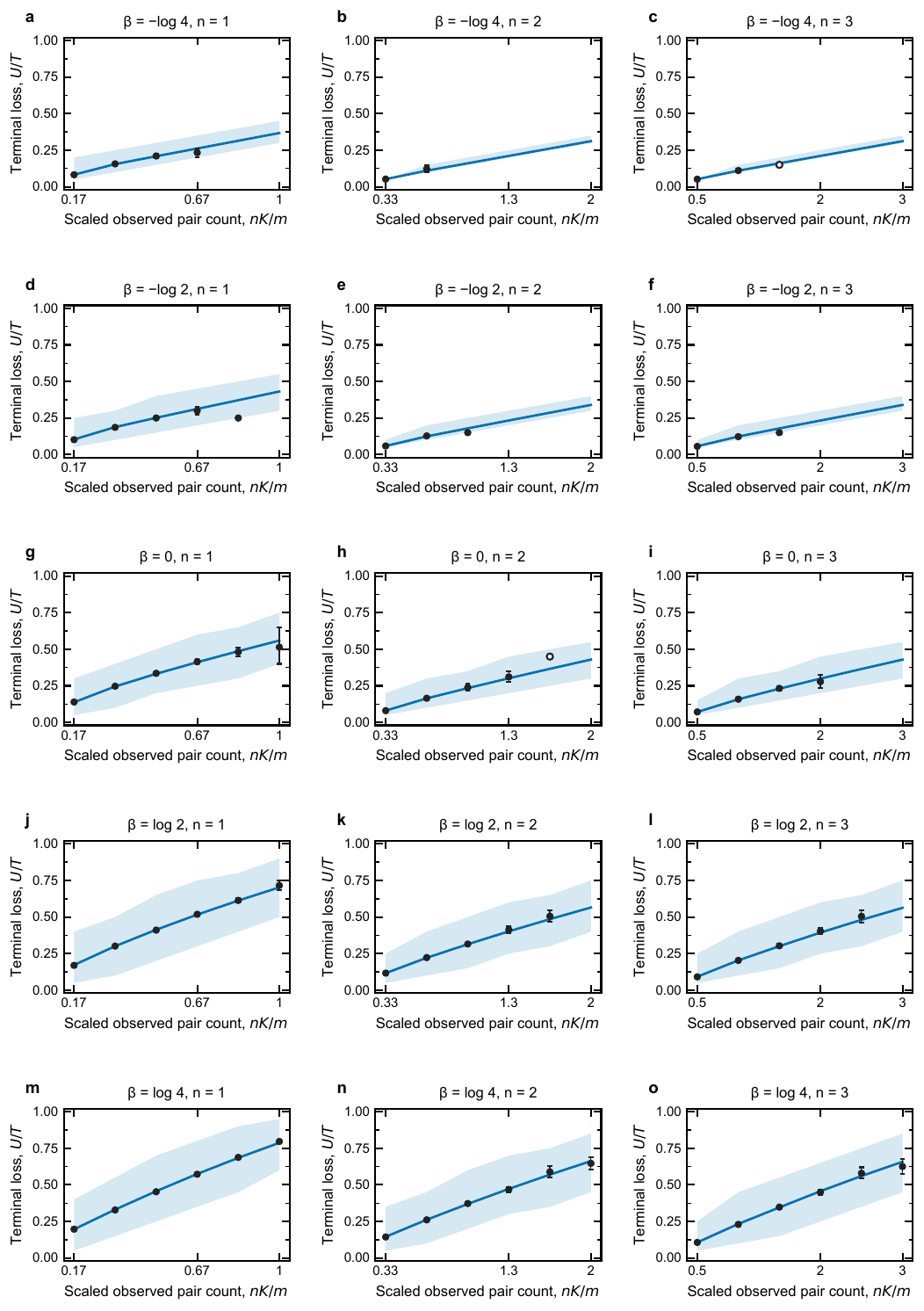}\par}\smallskip
{\fontsize{9}{11}\selectfont Rows show \(\beta=-\log4,-\log2,0,\log2,\log4\) and columns one, two and three upstream roots, at \(m=6\) of \(T=20\) terminals, with 512 independent graphs per preference. Lines and shaded bands are exact conditional means and discrete central 90\% prediction intervals, and markers with black whiskers are simulated means with pointwise 95\% bootstrap intervals resampling whole graphs; hollow markers have one supporting graph. The full theoretical support is shown, including cells without simulated observations. Here \(\beta\in\mathbb R\) sets the age preference, \(m,T\) are the sampled and total terminal counts, \(n\) is the upstream root count and \(K\) counts covered sampled terminals. The coordinate \(X=nK/m\) uses the sampled terminal count as its denominator.\par}
\end{minipage}\end{figure}\par
Uncertainty used 1,000 whole-network bootstrap draws within each preference, keeping the same sampled graphs across observation levels, and means, variances, slopes and interval coverage were computed from pooled counts. Pointwise intervals are descriptive and carry no simultaneous coverage interpretation. All cells and support counts are kept, including empirical absences and undefined slopes.

At \(m=6\), predicted \(S_{\mathrm{pop}}\) values were 0.320, 0.409, 0.523, 0.579 and 0.597, against empirical values of 0.338, 0.409, 0.523, 0.584 and 0.599, and all 15 primary root group slopes lay within their pointwise simulation intervals. Across all preference and sample conditions, normalized absolute discrepancies in joint mean counts ranged from 0.00951 to 0.0422, with the expected count of all 200 generated edges as denominator. Under partial observation, discrete interval masses were 94.7 to 98.0\% in theory and 94.8 to 97.9\% empirically, and the largest difference in any condition was 0.524 percentage points. These aggregate coverage checks do not establish precision in every rare conditional cell.

Sparse support limits the empirical assessment of the most extreme predictions. At \(\beta=-\log4,m=2\), the higher predictor value \(K=2\) occurred in only two networks for \(n=2\) and one for \(n=3\), and slopes were estimable in 874 and 642 of 1,000 bootstrap draws. Their conditional bootstrap intervals missed the rare population tail, limiting their value for slope validation. Hollow plotted means mark cells supported by one network. No networks or preferred conditions were added after these discrepancies were seen.

The population law is weighted by expected edge counts (Note 9.1). To predict the distributions of slopes or \(S_{\mathrm{root}}\) across individual networks, we would also need to describe dependence through shared descendants and root discovery. Ensemble means alone do not supply this information.

\FloatBarrier\section*{Supplementary Note 10. Prediction with limited historical calibration}
\addcontentsline{toc}{section}{Supplementary Note 10. Prediction with limited historical calibration}

The growth distribution also supplies a prior for estimating unsampled terminal coverage from sampled counts and target birth generation (Supplementary Note 3.6). Historical calibration adjusts this prior to the observed formation process (Supplementary Note 8.4). Here we ask whether the fitted model improves on an empirical prior given the same limited history and observations.

\subsection*{10.1 Matched information and independent blocks}

\hyperlink{figure-S13}{Supplementary Fig. 13} compares growth-informed and empirical priors when history and terminal observations are sparse, using the same five-cohort model with \(\beta=-\log2,0,\log2\). For each preference, 16 independent blocks each contained 128 calibration networks and 128 fresh test networks, giving 6,144 of each overall. The first 16 calibration networks formed a nested smaller budget, and the test samples of two and six terminals were also nested.
\par\begin{figure}[!htbp]\centering\begin{minipage}{\linewidth}\phantomsection\label{fig:S13}\hypertarget{figure-S13}{}\addcontentsline{toc}{subsection}{Supplementary Figure 13. Predictive value of formation priors under scarce history and observation}
{\centering\textbf{Supplementary Figure 13. Predictive value of formation priors under scarce history and observation}\par\smallskip
\includegraphics[width=\linewidth,height=128mm,keepaspectratio]{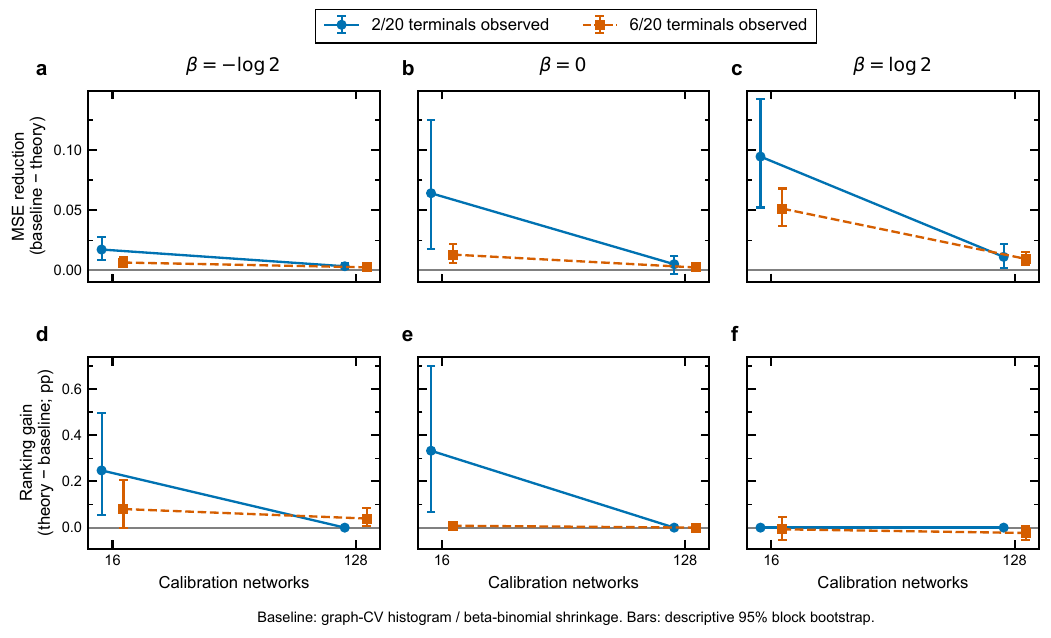}\par}\smallskip
{\fontsize{9}{11}\selectfont Fitted theory compared with an empirical prior given the same information, with shrinkage between a histogram and a beta-binomial distribution selected by cross-validation over graphs. Columns show preference for older parents, uniform choice and preference for recent parents. The horizontal axis compares histories of 16 and 128 calibration networks, and colors distinguish the two terminal observation budgets. Top panels show the absolute reduction in mean squared hidden count error (baseline minus theory, in squared terminal count units), and bottom panels the change in top 5\% oracle efficiency on unsampled terminals, in percentage points. Every preference has 16 independently calibrated blocks with 128 fresh test graphs per block. Histories and observations are nested, and all methods share test graphs. Points and descriptive 95\% intervals use 10,000 paired bootstrap draws resampling blocks, with equal weight for preferences in aggregate summaries and no adjustment for multiplicity. Shared ordinate scales allow visual comparison across history sizes, and zero ranking gains and unresolved intervals are shown. With the least history and sparsest observation, the aggregate reduction in count error is 1.32\% and the efficiency gain 0.194 percentage points, and only 22 of 6,144 test graphs change fractional selections (Supplementary Table 5).\par}
\end{minipage}\end{figure}\par
All nonoracle methods received the same historical networks, birth schedule, parent count law, exact birth of each test target and observed \((K,m,T)\), but no test coverage. Fitted theory estimated \(\beta\) within fixed bounds \([-3,3]\) by the likelihood of unordered parent birth multisets, summing admissible draw orders. A histogram prior was constructed separately for each birth cohort from all historical nodes, including those with zero coverage, with a total uniform pseudocount of one. The stronger comparator mixed this histogram with a fitted beta-binomial distribution. Its mixture weight was selected from seven fixed values by four-fold cross-validation over whole historical graphs, maximizing the coverage log score on withheld graphs. All priors imposed \(U=1\) for final cohort targets and used the same hypergeometric likelihood. Known-parameter theory served as a reference with additional information, and boundary-aware coverage extrapolation as a diagnostic comparator.

Fitted priors were fixed within each block before its test graphs were generated. Beta-binomial fitting used two prescribed L-BFGS-B starts and a candidate at the binomial limit. After inspecting some calibration data, we added a Powell fallback for cases in which both starts failed. This change preceded test generation and left the experimental design and endpoints unchanged.

\subsection*{10.2 Endpoints, uncertainty and results}

The co-primary endpoints were the hidden count MSE for \(H=U-K\), averaged over distinct discovered nonterminal targets within each graph, and the top 5\% single-edge ranking efficiency among discovered edges. Ranking used \(\lceil0.05|E_S|\rceil\) candidates with fractional allocation among tied scores at the budget boundary, and efficiency compared the summed hidden losses of individual edges with the oracle sum at the same budget, not the effect of deleting edges simultaneously. Empty target sets and zero oracle losses were left non-estimable, not scored as zero performance.

Metrics were averaged within independent calibration and test blocks. Descriptive 95\% intervals used 10,000 paired bootstrap draws of whole blocks, resampling blocks separately within each preference and preserving nested budgets and samples, with equal weight for the three preferences in aggregates. Relative MSE reductions were recomputed as ratios inside each draw. Test graphs sharing a fitted prior were not treated as independent calibration replicates, and intervals were not adjusted for multiple comparisons.

\par\begin{minipage}{\linewidth}
\textbf{Supplementary Table 5. Prediction informed by growth relative to the empirical shrinkage prior selected by cross-validation over graphs.} Brackets give descriptive 95\% intervals from bootstrap resampling of blocks. MSE reductions are percentages, and efficiency differences are percentage points.\par\smallskip

\begingroup\small
\setlength{\tableunit}{\dimexpr\linewidth-6\tabcolsep\relax}
\begin{tabular}{@{}>{\raggedright\arraybackslash}p{0.160000\tableunit}>{\raggedright\arraybackslash}p{0.280000\tableunit}>{\raggedright\arraybackslash}p{0.280000\tableunit}>{\raggedright\arraybackslash}p{0.280000\tableunit}@{}}
\toprule
Historical networks & Observed terminals & Relative MSE reduction (\%) & Ranking efficiency gain (pp) \\
\midrule
16 & 2/20 & 1.32 [0.827, 1.86] & 0.194 [0.0721, 0.337] \\
16 & 6/20 & 0.964 [0.734, 1.21] & 0.0271 [\ensuremath{-}0.00640, 0.0748] \\
128 & 2/20 & 0.149 [0.0530, 0.244] & 0 [0, 0] \\
128 & 6/20 & 0.197 [0.118, 0.279] & 0.00540 [\ensuremath{-}0.0103, 0.0235] \\
\bottomrule
\end{tabular}\endgroup
\end{minipage}\par

With 16 histories and two observed terminals, efficiency gains were 0.248, 0.333 and zero percentage points for older, uniform and recent preferences. Fractional selections changed in 22 of 6,144 tests, 21 with different efficiencies, and with 128 histories and two terminals the selections were identical. Count prediction can therefore improve without changing which edges enter the selection budget. In this comparison, the clearly resolved ranking benefit occurs with the smaller information budget.

The comparison evaluates calibration and prediction within the specified age-dependent family.

\FloatBarrier\section*{Supplementary Note 11. Birth structure and sampling in the original Growth cohort}
\addcontentsline{toc}{section}{Supplementary Note 11. Birth structure and sampling in the original Growth cohort}

\subsection*{11.1 Frozen graphs and observation design}

The 30 original Growth graphs keep their primary samples and complete terminal ancestor closures. Generator labels give the tail birth \(a\) and head birth \(b\), integers with \(0\le a<b\le8\), and every reference edge has positive coverage \(U\in\{1,\ldots,100\}\) and \(n\in\{1,2,3\}\).

One fixed random order per graph placed a permutation of its original 20 sampled terminals before a permutation of the rest, so prefixes with \(m\in\{5,10,20,50,100\}\) gave nested observations that recover the original sample exactly at \(m=20\). Complete tracing from each prefix determined the observed graph, and sampled coverage \(k\in\{1,\ldots,m\}\) is positive on discovered edges. The primary curves use currently discovered edges, and a separate comparison fixes the candidates to those discovered at \(m=5\); \hyperlink{figure-S14}{Supplementary Fig. 14} shows both populations.
\par\begin{figure}[!htbp]\centering\begin{minipage}{\linewidth}\phantomsection\label{fig:S14}\hypertarget{figure-S14}{}\addcontentsline{toc}{subsection}{Supplementary Figure 14. Birth structure and sampling in the original Growth cohort}
{\centering\textbf{Supplementary Figure 14. Birth structure and sampling in the original Growth cohort}\par\smallskip
\includegraphics[width=\linewidth,height=166mm,keepaspectratio]{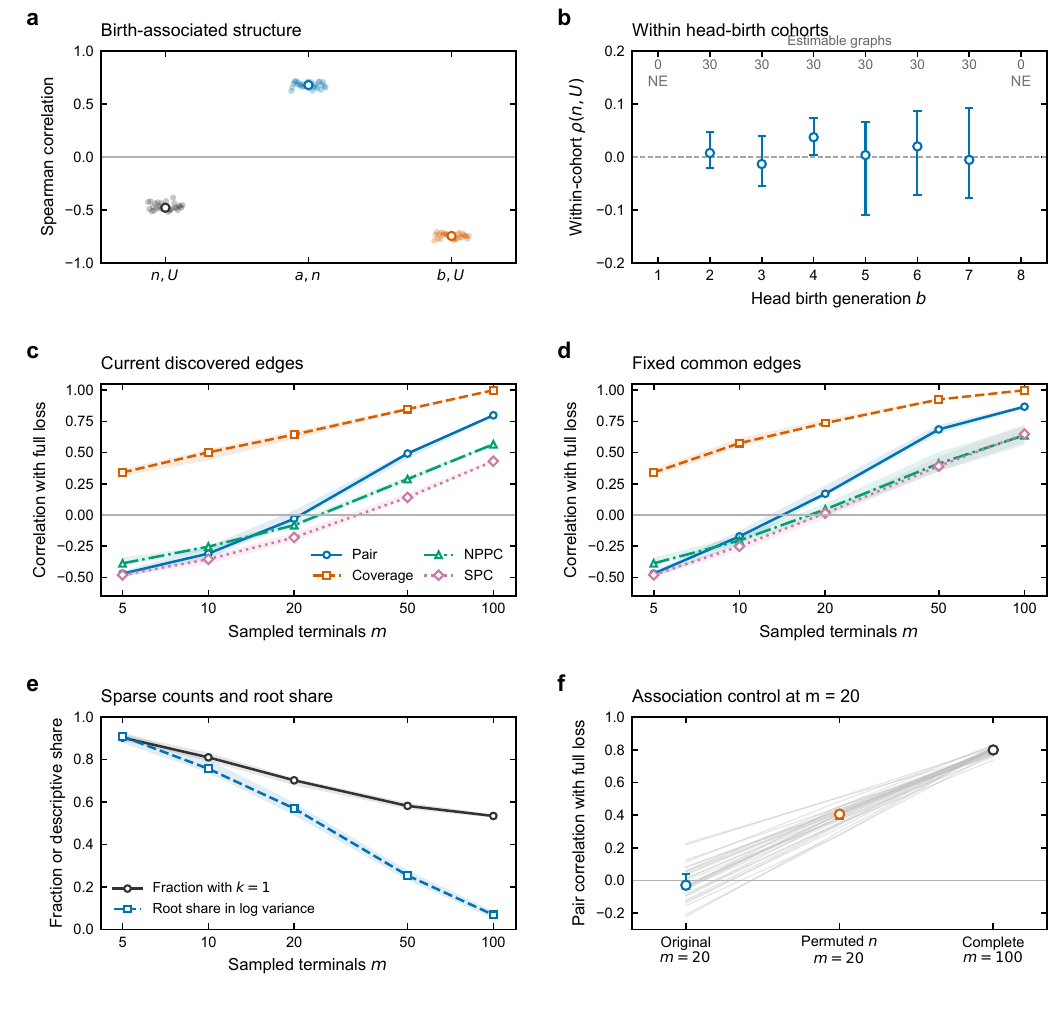}\par}\smallskip
{\fontsize{9}{11}\selectfont \textbf{a}, Correlations among upstream root count \(n\), complete terminal coverage \(U\), tail birth generation \(a\) and head birth generation \(b\) on positive-coverage reference edges; faint points are individual graphs. \textbf{b}, Correlations of \(n\) and \(U\) within head birth generations, with counts of estimable graphs; NE marks an unestimable correlation caused by a constant variable. \textbf{c}, Correlations of four observed scores with complete strict loss among edges discovered at each budget. \textbf{d}, The same comparisons restricted throughout to edges discovered at \(m=5\). \textbf{e}, Fraction of discovered edges with \(k=1\), and the descriptive share of \(\log(nk)\) variance assigned to \(\log n\) when covariance is split equally. \textbf{f}, Original correlations at \(m=20\), within-graph median correlations from 200 permutations of \(n\) at fixed \(k,U\), and correlations under complete observation; thin lines connect statistical comparisons within the same graph. Markers are medians across 30 independent graphs, and bars and shading are descriptive 95\% percentile intervals from 2,000 graph bootstrap replicates, unadjusted for multiple comparisons. Lines between registered budgets guide the eye. Nested terminal prefixes preserve each original sample of 20 terminals, and counts and birth ranges are defined in Supplementary Note 11.\par}
\end{minipage}\end{figure}\par
NPPC was the product of observed ancestor counts at the tail and descendant counts at the head, each including its endpoint, and SPC counted paths through an edge from observed roots to observed sinks. Pair thickness, coverage, NPPC and SPC used identical observed graphs and candidates, and complete terminal loss served only for evaluation. Correlations were computed within graphs. Conditional analyses kept both head birth strata and strata defined by both endpoint births, with edge counts, distinct values and non-estimable states; constant variables did not receive zero correlation, and group slopes included fitted intercepts.

All log variances and covariances used population moments. Splitting the covariance term equally gives the root component the descriptive share \(\{\operatorname{Var}(\log n)+\operatorname{Cov}(\log n,\log k)\}/\operatorname{Var}(\log(nk))\), defined when the denominator is positive, which can fall outside \([0,1]\) and is not a causal attribution. At \(m=20\), 200 permutations shuffled \(n\) among the discovered edges of each graph while keeping \(k\), \(U\) and the candidate set. This statistical control need not correspond to a realizable network. We first summarized the permutation correlations by their median within each graph, then aggregated those summaries across graphs. Unless stated otherwise, aggregates are medians across equally weighted graphs, with descriptive 95\% percentile intervals from 2,000 graph bootstrap replicates, unadjusted for multiple comparisons. Edges, nested observations and permutations are not independent network units.

\subsection*{11.2 Associations with formation time}

Across positive-coverage reference edges, median correlations were \ensuremath{-}0.479 between \(n\) and \(U\), 0.677 between \(a\) and \(n\) and \ensuremath{-}0.745 between \(b\) and \(U\). Within head generations 2 to 7, median correlations between \(n\) and \(U\) ranged from \ensuremath{-}0.0130 to 0.0370, and the generation 4 estimate stayed weakly positive, with a descriptive 95\% interval of 0.00300 to 0.0730. Generation 1 had constant \(n\) and generation 8 constant \(U\), so neither conditional correlation was estimable. Conditioning greatly reduced the pooled association in this cohort. The exact conditional independence in Supplementary Note 3.5 concerns the specified ensemble at fixed edge birth conditions, whereas the empirical correlations describe individual realized graphs.

The root groups of the full reference also differed in coverage. For \(n=1,2,3\), the medians of within-graph mean coverage were 5.16, 1.92 and 1.57 terminals, the median fractions with \(U=1\) were 0.275, 0.586 and 0.723, and the median edge fractions were 0.367, 0.193 and 0.450; these separately computed medians need not sum to one. Source Data keep all per-graph means, upper quantiles, maxima and supports. These summaries describe positive-coverage edges in this growth cohort.

\subsection*{11.3 Sampling, candidate composition and the pooled correlation}

At \(m=20\), a median 70.3\% of discovered edges had \(k=1\), and the descriptive root variance share was 0.570. Median correlation with complete strict loss was \ensuremath{-}0.0290 for pair thickness, \ensuremath{-}0.0810 for NPPC, \ensuremath{-}0.179 for SPC and 0.645 for sampled coverage, and the correlation for pair thickness reached 0.799 under complete observation. On the fixed candidates first discovered at \(m=5\), correlations for pair thickness were 0.171 at \(m=20\) and 0.868 under complete observation. Both measurement resolution and the selection of discovered edges therefore shape the primary curve. The across-graph median of within-graph permutation medians was 0.406. The association of root counts with coverage helps explain the original low correlation. The shuffle compares correlations after reassigning root counts across the same edges, with coverage and outcomes held fixed.

Median calibration coefficients for one, two and three roots were 0.577, 0.443 and 0.503, and each graph’s slopes satisfy \(b_n=3\lambda_n/n\). Full-precision values, observations and independent reconstruction checks are kept in the archive and Source Data.

\FloatBarrier\section*{Supplementary Note 12. Temporal associations in fixed empirical references}
\addcontentsline{toc}{section}{Supplementary Note 12. Temporal associations in fixed empirical references}

\subsection*{12.1 Boundaries, recorded times and sampling}

We reused the exact HEP-TH, Bioconductor and OpenAlex reference boundaries, treating each as one fixed topology. Ten nested terminal orders kept the original primary samples of 267 HEP-TH and 283 Bioconductor terminals, with additional budgets at the rounded-up fractions 0.01, 0.02, 0.05, 0.10, 0.25, 0.50, 0.75 and 1 of their designated terminal populations. For OpenAlex, ten existing orders stratified by publication year supplied samples of 5 to 50 terminals in steps of five. The primary replay kept acquisition to two steps relative to each sample and recomputed boundary roots, with complete tracing within the fixed reference as a separate sensitivity analysis, so equality of observed and reference root counts is not imposed on the two step observations.

HEP-TH times came from unique recorded source dates, grouped by calendar month; of 503 nodes with date records, 39 had conflicting dates and were treated as missing. Identifier months, which orient the frozen topology (Methods), were examined separately and not substituted for absent source dates. OpenAlex used recorded publication years, with equal years tied and missing boundary dates not imputed. Bioconductor supplied a release snapshot without reliable package birth dates, so its temporal associations were not estimable. The prespecified expectations were that root count increases with tail time and coverage decreases with head time, without assuming universally monotone trends.

Spearman correlations compared sampled pair thickness and sampled coverage with complete terminal coverage on currently discovered edges, on candidates already discovered at the smallest budget and, with undiscovered edges scored zero, over the whole reference, where all edges had positive designated-terminal coverage. Time correlations used only available dates and kept their exact support. Head time strata required at least three edges and variation in both quantities, and stratum correlations were not averaged into an independence test. Variance shares used the population moments and covariance split of Supplementary Note 11. At the original HEP-TH and Bioconductor samples, 200 root count permutations kept sampled and complete coverage; for OpenAlex, which has no unique historical primary observation, this control used a prespecified representative order at ten terminals, separately for each acquisition boundary. \hyperlink{figure-S15}{Supplementary Fig. 15} reports ten-order medians and full ranges, which describe sampling variation within each topology.

\subsection*{12.2 Temporal evidence and its support}

HEP-TH had unique source dates for 464 of 3,920 nodes, and OpenAlex had recorded publication years for 1,823 of 51,384 nodes. On the available reference edges, tail time correlated with root count at 0.617 and 0.636, supported by 1,079 and 11,292 edges, and head time with complete coverage at \ensuremath{-}0.458 and \ensuremath{-}0.497, supported by 1,839 and 87,161 edges. The very different head and tail support in OpenAlex reflects which publication records were queried within the frozen reference.

Across all reference edges, root count and coverage correlated at \ensuremath{-}0.0500 in HEP-TH and \ensuremath{-}0.00500 in OpenAlex, and associations within head time strata were mixed. Eighteen of 32 HEP-TH strata and 39 of 68 OpenAlex strata were estimable, with correlations from \ensuremath{-}0.299 to 0.579 and from \ensuremath{-}0.186 to 0.307, respectively. The dated subsets agree with the expected directions. Missing dates and mixed conditional associations leave the historical mechanism for the whole network and conditional independence unresolved.

\subsection*{12.3 Sampling differences do not equate to ranking efficiency}

At the original primary samples, pair thickness correlated with complete coverage at 0.490 in HEP-TH and 0.592 in Bioconductor, and at 0.650 and 0.614 under complete observation, with primary root variance shares of 0.676 and 0.167; permuting root counts gave median correlations of 0.529 and 0.639. These empirical views therefore did not reproduce the near-zero pooled correlation of the Growth cohort. In OpenAlex, median correlations across the ten orders rose from 0.640 at ten sampled terminals to 0.773 at all fifty under acquisition to two steps, and complete tracing within the fixed reference at ten terminals gave 0.703.

These diagnostics describe overall rank associations within the fixed references. The HEP-TH and Bioconductor importance benchmark instead evaluates losses from independently cutting the highest-ranked edges. Source Data keep strata, undefined states, fixed-candidate sensitivities and permutation ranges, and no new temporal metadata were collected.

\FloatBarrier\section*{Supplementary Note 13. Removing upstream factors from edge scores}
\addcontentsline{toc}{section}{Supplementary Note 13. Removing upstream factors from edge scores}

\subsection*{13.1 Design and evaluation}

This prespecified secondary analysis uses the frozen observations of 30 graphs per synthetic family, the tree, Bioconductor, HEP-TH and ten 40-terminal Rio Grande views. Each product and its downstream factor use identical observed candidates, and full references serve only for evaluation. Note 5.5 defines the six products and five distinct downstream factors, with SPC and SPLC sharing one, and exact integer path recursions continue beyond designated terminals that have outgoing edges.

The primary endpoint was oracle efficiency at the 5\% candidate edge budget for strict loss among unsampled terminals, with each selected edge cut independently from the intact reference. The hidden loss is \((U_G(e)-k_e)/(T_G-m)\), with the counts defined in the main text and \(m<T_G\). For budget fraction \(f\in\{0.01,0.05,0.10\}\), the number selected is the integer \(B_f=\max\{1,\lceil f|E_S|\rceil\}\). Every method used expected loss under uniform tie breaking and the same oracle. Zero oracle loss leaves efficiency undefined, whereas a selection causing zero loss against a positive oracle has efficiency zero. Budgets of 1\% and 10\%, evaluation on all terminals and selection over distinct targets were kept as sensitivities, with the target analysis using each target's largest incoming score and a budget over distinct targets.

Synthetic summaries weight graphs equally, with descriptive 95\% percentile intervals from 2,000 paired graph bootstrap resamples, unadjusted for multiplicity, and differences within \(10^{-12}\) count as unchanged. The tree, Bioconductor and HEP-TH results are single cases, and river means and ranges summarize ten views of one reference.

An auxiliary analysis correlates the upstream factor with a normalized rank difference. This difference is the product midrank minus the hidden-loss midrank, divided by the candidate count minus one. Positive differences denote overranking, and constant inputs leave the correlation undefined. The rank-difference statistic contains the product score, so its descriptive association with the upstream factor partly reflects their shared construction. Source Data keep all ranks and sensitivities.

\subsection*{13.2 Results and interpretation}

\hyperlink{figure-S16}{Supplementary Fig. 16} compares each full score with its own downstream factor, and the distinct downstream quantities with sampled terminal coverage.
\par\begin{figure}[!htbp]\centering\begin{minipage}{\linewidth}\phantomsection\label{fig:S16}\hypertarget{figure-S16}{}\addcontentsline{toc}{subsection}{Supplementary Figure 16. Removing upstream factors from multiplicative edge scores}
{\centering\textbf{Supplementary Figure 16. Removing upstream factors from multiplicative edge scores}\par\smallskip
\includegraphics[width=\linewidth,height=185mm,keepaspectratio]{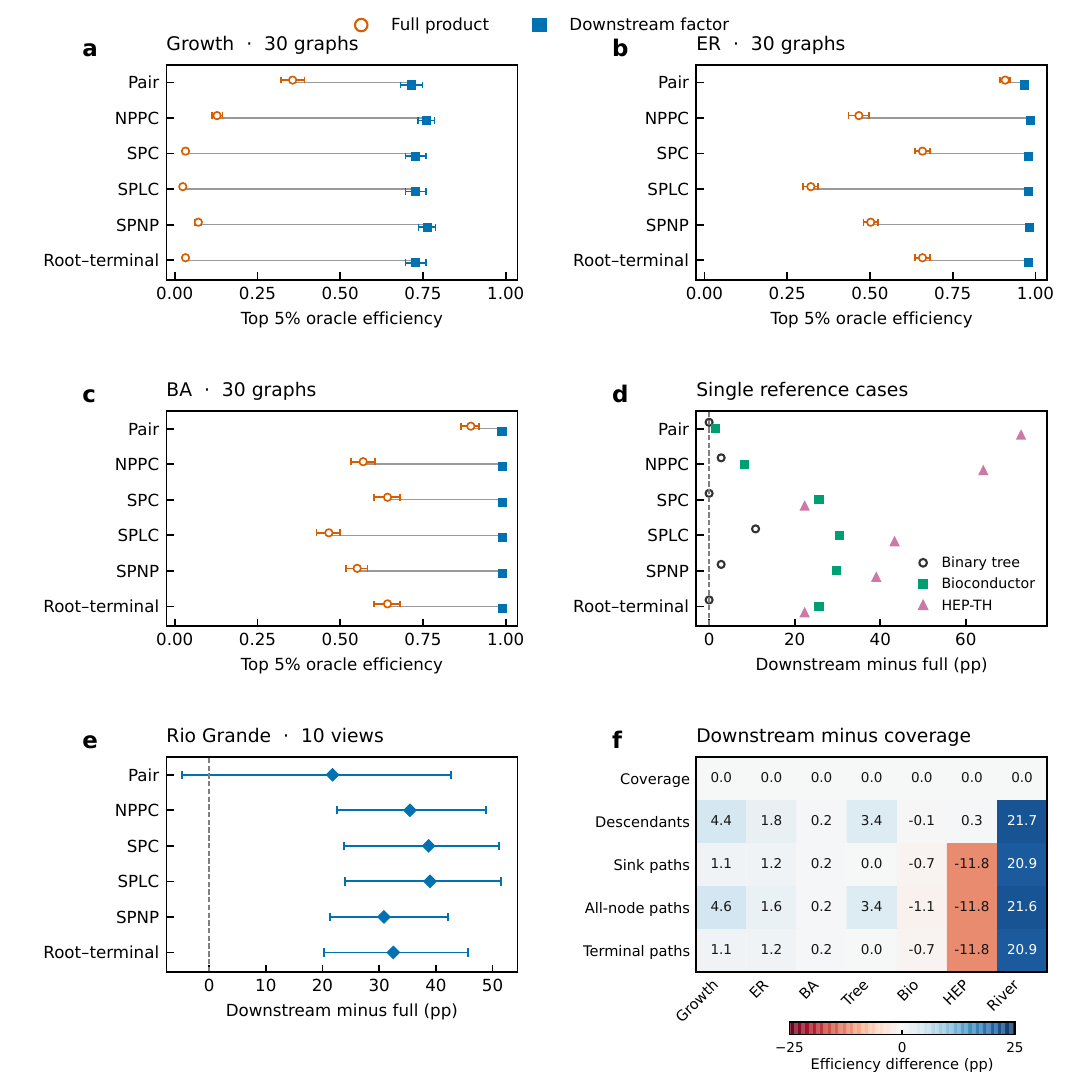}\par}\smallskip
{\fontsize{9}{11}\selectfont Panels a to c show mean oracle efficiency for selecting the highest ranked 5\% of observed edges in 30 Growth, 30 ER and 30 BA graphs, with orange open circles for the full product and blue squares for its own downstream factor. Bars are descriptive 95\% intervals from 2,000 whole graph bootstrap resamples, and gray segments connect paired method summaries. Each selected edge is cut independently under strict dependency, and losses are evaluated on unsampled designated terminals using the same candidate set, budget and exact expected loss under uniform tie breaking. Panel d shows downstream minus full differences for the single tree, Bioconductor and HEP-TH cases, and panel e the means and ranges of paired differences across ten fixed views of one Rio Grande reference. Panel f shows mean differences between the five defined downstream quantities and sampled terminal coverage, where positive values favor the listed downstream quantity and pp denotes percentage points. SPC and SPLC use the same downstream method, their shared sink path suffix. Source Data give all budgets and distinct target sensitivities.\par}
\end{minipage}\end{figure}\par
At the primary budget, removing the upstream factor increased efficiency for each of the six comparisons in every synthetic graph, although the five distinct downstream quantities differed in predictive value. In Growth, distinct descendant counts reached a mean efficiency of 0.760 and paths to all nodes 0.762, against 0.716 for sampled terminal coverage, with paired gains over coverage of 0.0440 (descriptive 95\% interval [0.0225, 0.0647]) and 0.0459 ([0.0235, 0.0697]). Sink and designated terminal path counts both reached 0.727, a paired gain of 0.0109 [\ensuremath{-}0.00570, 0.0291], leaving uncertainty about some positive mean differences.

Descendant counts also exceeded coverage in mean efficiency in ER and BA, at 0.984 versus 0.966 and 0.991 versus 0.989, and in BA all four noncoverage downstream scores had paired intervals spanning zero. In HEP-TH, descendant counts reached 0.938 against 0.935 for coverage, whereas sink paths, designated terminal paths and paths to all nodes stayed near 0.817, and in Bioconductor coverage exceeded all four alternative downstream scores. River view means were 0.928 for descendant counts and 0.711 for coverage, but root correction of pair thickness reduced efficiency by 0.0470 in one view. The tree kept unchanged pair, SPC and root-to-terminal path rankings after ablation, and at the 1\% budget pair correction reduced efficiency by 0.0357 in one Growth graph. Source Data keep these exceptions and all sensitivity results.

These comparisons show how separating upstream participation from downstream exposure changes edge prioritization. Descendant and path counts can carry information about unsampled coverage beyond that captured by sampled terminal coverage. Each score combines its choice of endpoints with its treatment of multiple paths, and the comparison evaluates these combinations as a whole. Their relative performance varies across networks.

\FloatBarrier\section*{Supplementary References}
\addcontentsline{toc}{section}{Supplementary References}

\begin{samepage}\noindent\hypertarget{supp-ref:1}{}1. Priem, J., Piwowar, H. \& Orr, R. OpenAlex: A fully-open index of scholarly works, authors, venues, institutions, and concepts. \emph{26th International Conference on Science, Technology and Innovation Indicators (STI 2022)} (2022). \href{https://doi.org/10.5281/zenodo.6936227}{doi:10.5281/zenodo.6936227}.\par\vspace{-6pt}\end{samepage}
\begin{samepage}\noindent\hypertarget{supp-ref:2}{}2. Taube, J. C., Miller, P. B. \& Drake, J. M. An open-access database of infectious disease transmission trees to explore superspreader epidemiology. \emph{PLOS Biology} \textbf{20}, e3001685 (2022). \href{https://doi.org/10.1371/journal.pbio.3001685}{\begingroup\urlstyle{same}\nolinkurl{10.1371/journal.pbio.3001685}\endgroup}.\par\vspace{-6pt}\end{samepage}
\begin{samepage}\noindent\hypertarget{supp-ref:3}{}3. Hummon, N. P. \& Doreian, P. Connectivity in a citation network: The development of DNA theory. \emph{Social Networks} \textbf{11}, 39–63 (1989). \href{https://doi.org/10.1016/0378-8733(89)90017-8}{doi:10.1016/0378-8733(89)90017-8}.\par\vspace{-6pt}\end{samepage}
\begin{samepage}\noindent\hypertarget{supp-ref:4}{}4. Batagelj, V. \emph{Efficient Algorithms for Citation Network Analysis}. arXiv:cs/0309023, version 1 (2003). \href{https://arxiv.org/abs/cs/0309023v1}{Original preprint}.\par\vspace{-6pt}\end{samepage}
\begin{samepage}\noindent\hypertarget{supp-ref:5}{}5. Brandes, U. On variants of shortest-path betweenness centrality and their generic computation. \emph{Social Networks} \textbf{30}, 136–145 (2008). \href{https://doi.org/10.1016/j.socnet.2007.11.001}{doi:10.1016/j.socnet.2007.11.001}.\par\vspace{-6pt}\end{samepage}
\begin{samepage}\noindent\hypertarget{supp-ref:6}{}6. Sabrin, K. M. \& Dovrolis, C. The hourglass effect in hierarchical dependency networks. \emph{Network Science} \textbf{5}, 490–528 (2017). \href{https://doi.org/10.1017/nws.2017.22}{\begingroup\urlstyle{same}\nolinkurl{10.1017/nws.2017.22}\endgroup}.\par\vspace{-6pt}\end{samepage}
\begin{samepage}\noindent\hypertarget{supp-ref:7}{}7. Šubelj, L., Waltman, L., Traag, V. \& van Eck, N. J. Intermediacy of publications. \emph{Royal Society Open Science} \textbf{7}, 190207 (2020). \href{https://doi.org/10.1098/rsos.190207}{\begingroup\urlstyle{same}\nolinkurl{10.1098/rsos.190207}\endgroup}.\par\vspace{-6pt}\end{samepage}
\begin{samepage}\noindent\hypertarget{supp-ref:8}{}8. Borgs, C., Brautbar, M., Chayes, J. \& Lucier, B. Maximizing Social Influence in Nearly Optimal Time. \emph{Proceedings of the Twenty-Fifth Annual ACM-SIAM Symposium on Discrete Algorithms}, 946–957 (2014). \href{https://doi.org/10.1137/1.9781611973402.70}{\begingroup\urlstyle{same}\nolinkurl{10.1137/1.9781611973402.70}\endgroup}.\par\vspace{-6pt}\end{samepage}
\begin{samepage}\noindent\hypertarget{supp-ref:9}{}9. Tang, Y., Xiao, X. \& Shi, Y. Influence Maximization: Near-Optimal Time Complexity Meets Practical Efficiency. \emph{Proceedings of the 2014 ACM SIGMOD International Conference on Management of Data}, 75–86 (2014). \href{https://doi.org/10.1145/2588555.2593670}{\begingroup\urlstyle{same}\nolinkurl{10.1145/2588555.2593670}\endgroup}.\par\vspace{-6pt}\end{samepage}
\begin{samepage}\noindent\hypertarget{supp-ref:10}{}10. Kempe, D., Kleinberg, J. \& Tardos, É. Maximizing the spread of influence through a social network. In \emph{Proceedings of the Ninth ACM SIGKDD International Conference on Knowledge Discovery and Data Mining} 137–146 (ACM, 2003). \href{https://doi.org/10.1145/956750.956769}{\begingroup\urlstyle{same}\nolinkurl{10.1145/956750.956769}\endgroup}.\par\vspace{-6pt}\end{samepage}
\begin{samepage}\noindent\hypertarget{supp-ref:11}{}11. Abate, P., Boender, J., Di Cosmo, R. \& Zacchiroli, S. Strong dependencies between software components. In \emph{2009 3rd International Symposium on Empirical Software Engineering and Measurement} 89–99 (IEEE, 2009). \href{https://doi.org/10.1109/ESEM.2009.5316017}{\begingroup\urlstyle{same}\nolinkurl{10.1109/ESEM.2009.5316017}\endgroup}.\par\vspace{-6pt}\end{samepage}
\begin{samepage}\noindent\hypertarget{supp-ref:12}{}12. Setó-Rey, D., Santos-Martín, J. I. \& López-Nozal, C. Vulnerability of Package Dependency Networks. \emph{IEEE Transactions on Network Science and Engineering} \textbf{10}, 3396–3408 (2023). \href{https://doi.org/10.1109/TNSE.2023.3260880}{\begingroup\urlstyle{same}\nolinkurl{10.1109/TNSE.2023.3260880}\endgroup}.\par\vspace{-6pt}\end{samepage}
\begin{samepage}\noindent\hypertarget{supp-ref:13}{}13. Marichal, J.-L. On Sugeno integral as an aggregation function. \emph{Fuzzy Sets and Systems} \textbf{114}, 347–365 (2000). \href{https://doi.org/10.1016/S0165-0114(98)00116-X}{\begingroup\urlstyle{same}\nolinkurl{10.1016/S0165-0114(98)00116-X}\endgroup}.\par\vspace{-6pt}\end{samepage}
\begin{samepage}\noindent\hypertarget{supp-ref:14}{}14. Lengauer, T. \& Tarjan, R. E. A fast algorithm for finding dominators in a flowgraph. \emph{ACM Transactions on Programming Languages and Systems} \textbf{1}, 121-141 (1979). \href{https://doi.org/10.1145/357062.357071}{\begingroup\urlstyle{same}\nolinkurl{10.1145/357062.357071}\endgroup}.\par\vspace{-6pt}\end{samepage}
\begin{samepage}\noindent\hypertarget{supp-ref:15}{}15. Janson, S. The number of descendants in a random directed acyclic graph. \emph{Random Structures \& Algorithms} \textbf{64}, 768–803 (2024). \href{https://doi.org/10.1002/rsa.21195}{\begingroup\urlstyle{same}\nolinkurl{10.1002/rsa.21195}\endgroup}.\par\vspace{-6pt}\end{samepage}
\begin{samepage}\noindent\hypertarget{supp-ref:16}{}16. Burghart, F. Ancestries in random d-DAGs. \emph{arXiv preprint} (2026). arXiv:2606.30475v1. \href{https://arxiv.org/abs/2606.30475}{arXiv record}.\par\vspace{-6pt}\end{samepage}
\begin{samepage}\noindent\hypertarget{supp-ref:17}{}17. Mailler, C. \& Steiner, R. Multi-drawing Pólya urns via labelled random DAGs. \emph{arXiv preprint} (2025). arXiv:2508.20592v1. \href{https://doi.org/10.48550/arXiv.2508.20592}{\begingroup\urlstyle{same}\nolinkurl{10.48550/arXiv.2508.20592}\endgroup}.\par\vspace{-6pt}\end{samepage}
\begin{samepage}\noindent\hypertarget{supp-ref:18}{}18. Janson, S. \& Lo, T. Y. Y. The number of descendants in a preferential attachment graph. \emph{arXiv preprint} (2024). arXiv:2412.13975v1. \href{https://doi.org/10.48550/arXiv.2412.13975}{\begingroup\urlstyle{same}\nolinkurl{10.48550/arXiv.2412.13975}\endgroup}.\par\vspace{-6pt}\end{samepage}
\begin{samepage}\noindent\hypertarget{supp-ref:19}{}19. Ericson, W. A. Subjective Bayesian Models in Sampling Finite Populations. \emph{Journal of the Royal Statistical Society: Series B (Methodological)} \textbf{31}, 195–224 (1969). \href{https://doi.org/10.1111/j.2517-6161.1969.tb00782.x}{\begingroup\urlstyle{same}\nolinkurl{10.1111/j.2517-6161.1969.tb00782.x}\endgroup}.\par\vspace{-6pt}\end{samepage}
\begin{samepage}\noindent\hypertarget{supp-ref:20}{}20. Rodríguez, C. E., Walker, S. G. \& Mena, R. H. A finite population parametric Bayesian bootstrap for quick counts. \emph{The Annals of Applied Statistics} \textbf{19}, 3394–3415 (2025). \href{https://doi.org/10.1214/25-AOAS2066}{\begingroup\urlstyle{same}\nolinkurl{10.1214/25-AOAS2066}\endgroup}.\par\vspace{-6pt}\end{samepage}
\begin{samepage}\noindent\hypertarget{supp-ref:21}{}21. Dall’Asta, L., Alvarez-Hamelin, I., Barrat, A., Vázquez, A. \& Vespignani, A. Exploring networks with traceroute-like probes: Theory and simulations. \emph{Theoretical Computer Science} \textbf{355(1)}, 6–24 (2006). \href{https://doi.org/10.1016/j.tcs.2005.12.009}{\begingroup\urlstyle{same}\nolinkurl{10.1016/j.tcs.2005.12.009}\endgroup}.\par\vspace{-6pt}\end{samepage}
\begin{samepage}\noindent\hypertarget{supp-ref:22}{}22. Achlioptas, D., Clauset, A., Kempe, D. \& Moore, C. On the bias of traceroute sampling: Or, power-law degree distributions in regular graphs. \emph{Journal of the ACM} \textbf{56(4)}, 21:1–21:28 (2009). \href{https://doi.org/10.1145/1538902.1538905}{\begingroup\urlstyle{same}\nolinkurl{10.1145/1538902.1538905}\endgroup}.\par\vspace{-6pt}\end{samepage}
\begin{samepage}\noindent\hypertarget{supp-ref:23}{}23. Kim, M. \& Leskovec, J. The Network Completion Problem: Inferring Missing Nodes and Edges in Networks. \emph{Proceedings of the 2011 SIAM International Conference on Data Mining}, 47–58 (2011). \href{https://doi.org/10.1137/1.9781611972818.5}{\begingroup\urlstyle{same}\nolinkurl{10.1137/1.9781611972818.5}\endgroup}.\par\vspace{-6pt}\end{samepage}
\begin{samepage}\noindent\hypertarget{supp-ref:24}{}24. Peixoto, T. P. Reconstructing Networks with Unknown and Heterogeneous Errors. \emph{Physical Review X} \textbf{8}, 041011 (2018). \href{https://doi.org/10.1103/PhysRevX.8.041011}{\begingroup\urlstyle{same}\nolinkurl{10.1103/PhysRevX.8.041011}\endgroup}.\par\vspace{-6pt}\end{samepage}
\begin{samepage}\noindent\hypertarget{supp-ref:25}{}25. Peixoto, T. P. Network Reconstruction via the Minimum Description Length Principle. \emph{Physical Review X} \textbf{15}, 011065 (2025). \href{https://doi.org/10.1103/PhysRevX.15.011065}{\begingroup\urlstyle{same}\nolinkurl{10.1103/PhysRevX.15.011065}\endgroup}.\par\vspace{-6pt}\end{samepage}
\begin{samepage}\noindent\hypertarget{supp-ref:26}{}26. Young, J.-G., Valdovinos, F. S. \& Newman, M. E. J. Reconstruction of plant–pollinator networks from observational data. \emph{Nature Communications} \textbf{12}, 3911 (2021). \href{https://doi.org/10.1038/s41467-021-24149-x}{\begingroup\urlstyle{same}\nolinkurl{10.1038/s41467-021-24149-x}\endgroup}.\par\vspace{-6pt}\end{samepage}
\begin{samepage}\noindent\hypertarget{supp-ref:27}{}27. Peel, L., Peixoto, T. P. \& De Domenico, M. Statistical inference links data and theory in network science. \emph{Nature Communications} \textbf{13}, 6794 (2022). \href{https://doi.org/10.1038/s41467-022-34267-9}{\begingroup\urlstyle{same}\nolinkurl{10.1038/s41467-022-34267-9}\endgroup}.\par\vspace{-6pt}\end{samepage}
\begin{samepage}\noindent\hypertarget{supp-ref:28}{}28. Reisch, T., Heiler, G., Diem, C., Klimek, P. \& Thurner, S. Monitoring supply networks from mobile phone data for estimating the systemic risk of an economy. \emph{Scientific Reports} \textbf{12}, 13347 (2022). \href{https://doi.org/10.1038/s41598-022-13104-5}{\begingroup\urlstyle{same}\nolinkurl{10.1038/s41598-022-13104-5}\endgroup}.\par\vspace{-6pt}\end{samepage}
\begin{samepage}\noindent\hypertarget{supp-ref:29}{}29. Fessina, M. et al. Inferring firm-level supply chain networks with realistic systemic risk from industry sector-level data. \emph{Scientific Reports} \textbf{16}, 19848 (2026). \href{https://doi.org/10.1038/s41598-026-47883-y}{\begingroup\urlstyle{same}\nolinkurl{10.1038/s41598-026-47883-y}\endgroup}.\par\vspace{-6pt}\end{samepage}
\begin{samepage}\noindent\hypertarget{supp-ref:30}{}30. Hamedmoghadam, H., Jalili, M., Vu, H. L. \& Stone, L. Percolation of heterogeneous flows uncovers the bottlenecks of infrastructure networks. \emph{Nature Communications} \textbf{12}, 1254 (2021). \href{https://doi.org/10.1038/s41467-021-21483-y}{\begingroup\urlstyle{same}\nolinkurl{10.1038/s41467-021-21483-y}\endgroup}.\par\vspace{-6pt}\end{samepage}
\begin{samepage}\noindent\hypertarget{supp-ref:31}{}31. Chulahwat, A., Hassan, E. M., Tariverdi, M., Nunez-del-Prado, M. \& Mahmoud, H. Identifying influential assets in higher order interdependent infrastructure networks through population impact. \emph{Scientific Reports} \textbf{15}, 33351 (2025). \href{https://doi.org/10.1038/s41598-025-15824-w}{\begingroup\urlstyle{same}\nolinkurl{10.1038/s41598-025-15824-w}\endgroup}.\par\vspace{-6pt}\end{samepage}
\begin{samepage}\noindent\hypertarget{supp-ref:32}{}32. Jiang, X. \& Zhuge, H. Forward search path count as an alternative indirect citation impact indicator. \emph{Journal of Informetrics} \textbf{13}, 100977 (2019). \href{https://doi.org/10.1016/j.joi.2019.100977}{\begingroup\urlstyle{same}\nolinkurl{10.1016/j.joi.2019.100977}\endgroup}.\par\vspace{-6pt}\end{samepage}
\begin{samepage}\noindent\hypertarget{supp-ref:33}{}33. Han, Y., Li, W., Filko, A., Li, J. \& Zhang, F. Genome-wide promoter responses to CRISPR perturbations of regulators reveal regulatory networks in Escherichia coli. \emph{Nature Communications} \textbf{14}, 5757 (2023). \href{https://doi.org/10.1038/s41467-023-41572-4}{\begingroup\urlstyle{same}\nolinkurl{10.1038/s41467-023-41572-4}\endgroup}.\par\vspace{-6pt}\end{samepage}
\end{document}